\documentclass[12pt,twoside]{report}

\usepackage{suthesis-2e}
\usepackage{float}
\usepackage{graphicx}

\usepackage{amsmath}
\usepackage{amssymb}

\usepackage[small,bf]{caption2}
 \abovecaptionskip
\belowcaptionskip

\usepackage[rightcaption]{sidecap}
\sidecaptionvpos{figure}{t}
\usepackage{rotating}
\usepackage{lscape}

\newcommand{\rot}[1]{\nabla\times\mathbf{#1}}
\newcommand{\divg}[1]{\nabla\cdot\mathbf{#1}}
\newcommand{\dt}[2]{\frac {\partial^{#2}{#1}} {\partial{t}^{#2}} }

\begin{document}

\title{Magnetization dynamics using ultrashort magnetic field pulses}
\author{Ioan Tudosa}

\principaladviser{Joachim St\"{o}hr}
\firstreader{Theodore Geballe}
\secondreader{Ian Fisher}
\dept{Applied Physics}
\submitdate{August 2005}
\copyrightyear{2005}

\beforepreface
\prefacesection{Abstract}
Very short and well shaped magnetic field pulses can be generated
using ultra-relativistic electron bunches at Stanford Linear
Accelerator. These fields of several Tesla with duration of
several picoseconds are used to study the response of magnetic
materials to a very short excitation. Precession of a magnetic
moment by 90 degrees in a field of 1 Tesla takes about 10
picoseconds, so we explore the range of fast switching of the
magnetization by precession.

Our experiments are in a region of magnetic excitation that is not
yet accessible by other methods. The current table top experiments
can generate fields longer than 100 ps and with strength of 0.1
Tesla only.

Two types of magnetic were used, magnetic recording media and
model magnetic thin films. Information about the magnetization
dynamics is extracted from the magnetic patterns generated by the
magnetic field. The shape and size of these patterns are
influenced by the dissipation of angular momentum involved in the
switching process.

The high-density recording media, both in-plane and perpendicular
type, shows a pattern which indicates a high spin momentum
dissipation. The perpendicular magnetic recording media was
exposed to multiple magnetic field pulses. We observed an extended
transition region between switched and non-switched areas
indicating a stochastic switching behavior that cannot be
explained by thermal fluctuations.

The model films consist of very thin crystalline Fe films on GaAs.
Even with these model films we see an enhanced dissipation
compared to ferromagnetic resonance studies. The magnetic patterns
show that damping increases with time and it is not a constant as
usually assumed in the equation describing the magnetization
dynamics. The simulation using the theory of spin-wave scattering
explains only half of the observed damping. An important feature
of this theory is that the spin dissipation is time dependent and
depends on the large angle between the magnetization and the
magnetic field.

\prefacesection{Acknowledgements}
I would like to thank all the people who supported me along the
way to PhD: my parents whose efforts started all this, my sister
who encouraged me, my professors who taught me new ideas and my
friends whom I commiserated with through the grad school :).
Essential for the work done in this thesis were Christian Stamm,
Hans Siegmann and the wonderful people who provided the magnetic
samples. My thanks go also to Alexander Kashuba and Alex Dobin for
their simulations of the magnetic media. Finally, the support of
my advisor Joachim St\"{o}hr was essential in getting my PhD.

\afterpreface

\chapter{Introduction}

\begin{figure}[!h]
  \begin{center}
  \includegraphics[]{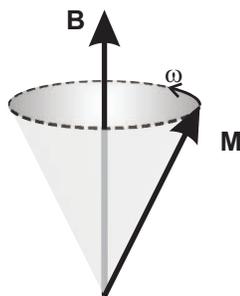}
  \caption[Magnetic Precession]{The precession around the magnetic field direction is the
           most important effect of a magnetic field on a magnetic moment.}
  \label{Fig:Precession}
  \end{center}
\end{figure}

If one wants to study magnetic phenomena at short time scales a
method is to apply a very short pulse magnetic field and study the
response to it. The peak amplitude of the magnetic field has to be
high to obtain observable effects. The reason is that any magnetic
moment precesses in the magnetic field, this is the main effect of
a magnetic field on a magnetic moment, see figure
\ref{Fig:Precession}. Since the precession speed is proportional
to the magnetic field intensity ($\mathbf{\omega}=\gamma
\mathbf{B}$), in order to have reasonable precession angles in
very short time, one needs a big magnetic field.

It is difficult to make very short magnetic field pulses with
conventional solid-state techniques when the electric current used
in generating the magnetic field has to overcome inevitable
capacitive and inductive effects. However, if one makes an
electric current with relativistic electrons in vacuum, those
effects are much more reduced. A short current pulse can be made
more easily because the relativistic electrons can be packed more
efficiently (relativistic contraction). The drawback is one has to
use an accelerator but this method can produce pulses on the order
of tens of femtoseconds, with the required high intensity of the
magnetic field.

This thesis explores the precessional switching in magnetic
recording media and model magnetic thin films using the magnetic
field associated with the electron bunch of the Stanford Linear
Accelerator. The perpendicular magnetic media samples show there
is an additional random mechanism beside the thermal fluctuations
that affects the switching. All magnetic samples present enhanced
damping, only part of which could be explained by spin wave
scattering of the uniform mode using micromagnetics simulation.

After presenting the characteristics of the magnetic field of the
electron bunches we show the experimental setup and the imaging
methods used for viewing the magnetic patterns. Then the equation
governing the magnetization dynamics is introduced and the main
results of the experiments follow. Appendices contain supporting
information and related new ideas triggered by thinking about the
experiments.

\chapter{Electromagnetic Effects of an Electron Pulse}
\label{Ch:EMeffects}
\chaptermark{Electromagnetic Effects}

In this chapter we will calculate the electromagnetic field of an
ultra-relativistic electron beam and understand what are some of
the effects of these fields on different materials. Since the
electron beam is very short we have to look at the relevant time
and length scales and how they arise in Maxwell's equations.

\section{Why relativistic electron beam?}
\sectionmark{Relativistic Electrons}
Very energetic electron beams can be produced at high energy
physics facilities, like linear accelerators. Typically they are
used to probe the matter at ultra small time and length scales in
particle collisions. However one can also use the electromagnetic
field associated with the electron beam to do condensed matter
physics because, being long range, the electric and magnetic field
extend far enough (hundreds of microns) to be useful.

As an example, the Stanford Linear Accelerator facility can
deliver an electron pulse with the energy around 28 GeV. The pulse
normally has a gaussian shape in all 3 directions, for example in
one experiment we had the longitudinal $\sigma_z=700 \mu m$ and
transversal $\sigma_x=5 \mu m$ and $\sigma_y=3 \mu m$. The number
of electrons in the beam is around $n_e=10^{10}$ electrons. For
this energy the relativistic factor $\gamma=1/\sqrt{1-\beta^2}
\approx 5.5\times 10^4,\quad \beta=v/c,\quad v$ being the electron
velocity.

\begin{figure}[!hbt]
  \begin{center}
  \includegraphics[]{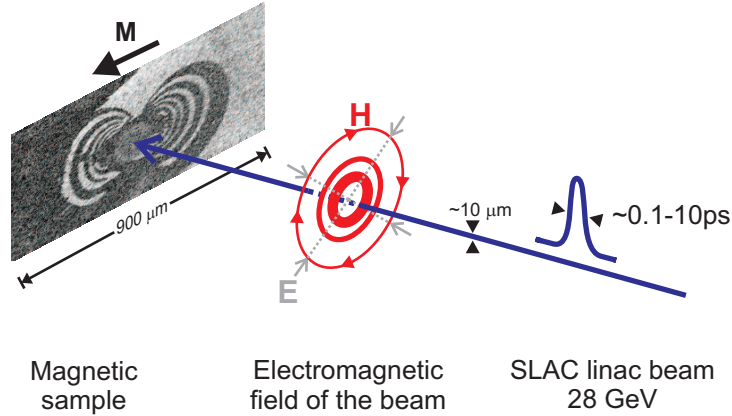}
  \caption[A Typical Experimental Setup for Magnetic Thin Film Samples]
            {A typical experimental setup for magnetic thin film samples,
            shown here with an initially in-plane magnetized sample.
            The electric and magnetic fields associated with the
            electron bunch influence the sample's magnetic
            and electric state. The effect of the magnetic field
            after one pulse has gone through is a magnetic pattern.}
  \label{Fig:ExperimentSetup}
  \end{center}
\end{figure}

An illustration of how this electromagnetic field can be used is
shown in figure \ref{Fig:ExperimentSetup}, where we shoot
electrons through a thin film sample. As the electron beam passes
through the sample it damages an area roughly twice the section of
the beam. The small amount of heat generated in that small area
can only propagate with the phonon speed, on the order of the
sound speed, which is slow compared to magnetic processes of
interest to us. Since the damage is localized there is a chance to
observe at larger distances the effects of the electric an
magnetic fields associated with the pulse. Hence the requirement
to have a well focused beam.

The small spatial extent of the electron beam is possible because
the electromagnetic force between electrons is reduced as their
velocity approach that of light. To see how this comes about we
consider two electrons moving with the same velocity $\mathbf{v}$
in the same direction. The force exerted by one electron on
another is given by the Lorentz formula
\begin{equation}
    \mathbf{F}=e(\mathbf{E}+\mathbf{v}\times\mathbf{B})
    \label{Eq:LorentzFormula}
\end{equation}
where $\mathbf{E,B}$ are the electric and magnetic field generated
by one electron at the location of the other, $e$ being the
electric charge of the electron.

\begin{figure}[!hbt]
  \begin{center}
  \includegraphics[]{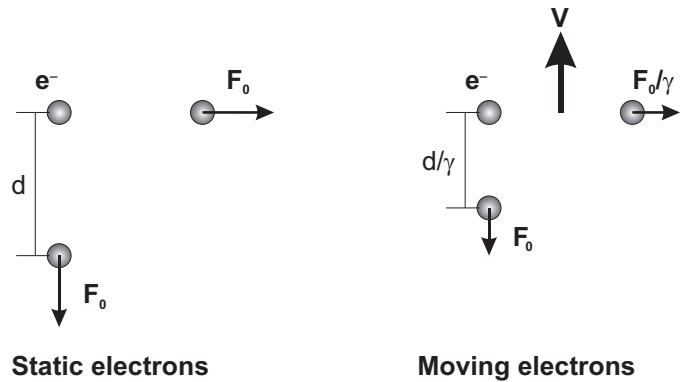}
  \caption[Comparison of Forces Acting between Static Electrons and Moving
        Electrons]{Comparison of forces acting between static electrons and moving electrons of an electron bunch. In a direction parallel
        with the velocity the electromagnetic force remains unchanged; notice that the distance between electrons is reduced by relativistic contraction.
        Perpendicular to the velocity direction, the distance measured in
        the lab frame doesn't change although the force, due to both
        electric and magnetic field, is reduced.
        The reason is that the magnetic Lorentz force acts in the opposite
        sense to the electric force. (The two parallel running electrons
        are similar to two parallel electric currents that attract
        each other and two parallel charged lines that repel each other.)}
  \label{Fig:ThreeElectrons}
  \end{center}
\end{figure}

It turns out that the force can be expressed as gradient of a
potential, called the convective potential (see appendix
\ref{UniformMotion}).
\begin{equation}
    \mathbf{F}=-\mathbf{\nabla}\psi,\qquad where\qquad
    \psi=\frac{e^2(1-v^2/c^2)}{4\pi\epsilon_0s}
\end{equation}
with $s=r\sqrt{1-v^2/c^2\sin^2\theta}$ defined in
\ref{Eq:sDefinition}, $r$ being the distance between electrons and
$\theta$ the angle $r$ makes with $v$. As an application, the
figure \ref{Fig:ThreeElectrons} illustrates how the measured force
between electrons is modified when the electrons move.
The force between moving electrons in the lab frame decreases as
$1/\gamma$ in transversal direction
\cite{chao:book}. The repulsion between electrons (space charge
effect) is then reduced, making possible a better focusing.
Moreover, the relativistic contraction makes the pulse look
shorter in the longitudinal direction in the lab frame.

\section{Electromagnetic fields of an electron beam}
\sectionmark{Electromagnetic fields}
Because space and time are so closely related in the relativistic
regime a fast moving electron, i.e. with high velocity =
space/time, has significantly different electromagnetic field than
a static electron.

\begin{figure}[!hbt]
  \begin{center}
  \includegraphics[]{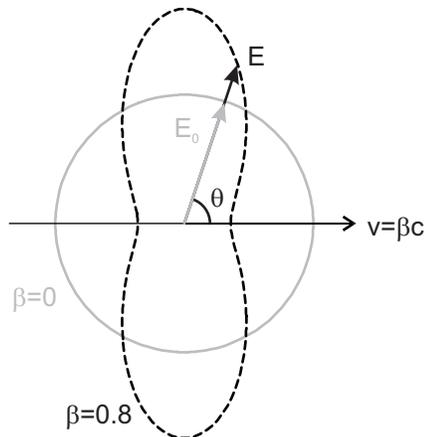}
  \caption[Angular Dependence of the Electric Field Intensity for a Relativistic Electron]
  {Polar plot of angular dependence of the electric field intensity for a relativistic electron
  ($E_0$ is the electric field of a static electron). The electric field as measured in the lab
  frame is enhanced in the perpendicular direction to the motion and reduced in the parallel
  direction.}
  \label{Fig:ElectronFields}
  \end{center}
\end{figure}

The electric field of an electron seen in the lab frame are
plotted below for different values of $\beta=v/c$ in a polar plot
of the intensity with respect to the angle $\theta$ made by the
field with the velocity direction (figure
\ref{Fig:ElectronFields}).

The formula describing the electric field is \cite{purcell:book}
\begin{equation}
    \mathbf{E}(\mathbf{r})=\frac {e\mathbf{\hat{r}}} {4\pi\epsilon_0 r^2}
        \frac {1-\beta^2} {(1-\beta^2\sin^2\theta)^{3/2}}
    \label{Eq:RelativisticElectricField}
\end{equation}
where $r$ is the distance between present position of the electron
and the point of observation. The electric field is radial, but it
is isotropic only for $\beta=0$. Along the direction of motion
($\theta=0,\pi$), the field strength is smaller by a factor of
$\gamma^{-2}$, while in transverse directions ($\theta=\pi/2$) it
is larger by a factor of $\gamma$.

\begin{figure}[!hbt]
  \begin{center}
  \includegraphics[]{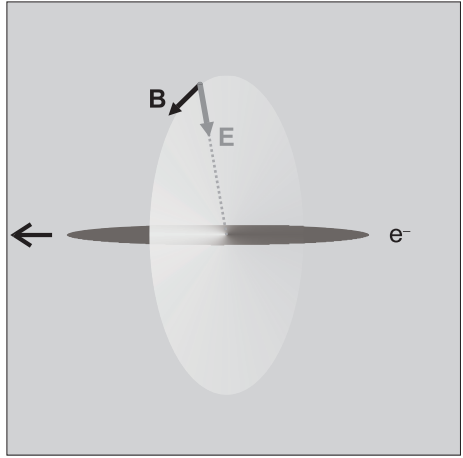}
  \caption[Relativistic Contraction of Electromagnetic Fields]{Electromagnetic
  fields are significant only in a plane perpendicular to the
  direction of propagation.}
  \label{Fig:ElectronBunch}
  \end{center}
\end{figure}

The moving electron is equivalent to an electric current, so it
creates also a magnetic field given by
\begin{equation}
    \mathbf{B}(\mathbf{r})=\frac{\boldsymbol{\beta}\times\mathbf{E}} {c}
    \label{Eq:RelativisticMagneticField}
\end{equation}
An artistic view of the direction of the electric and magnetic
fields is illustrated in figure \ref{Fig:ElectronBunch}.

As the speed approaches the speed of light the electric field is
concentrated in a thin disk perpendicular to the direction of
motion with an angular spread of $2/\gamma$. In the
ultra-relativistic limit $v\sim c$, the thickness of the disk
$2r/\gamma$ is very small and we can approximate it using the
Dirac delta function $2r/\gamma\rightarrow 1/\delta(x-ct)$, $x$
being the component of the position vector of the electron
parallel with its velocity. We can then use the Gauss law with a
very thin box to find the electric field close to the electron
\begin{equation}
    \oint\mathbf{E}d\mathbf{S}=\frac{q}{\epsilon_0}\quad\Rightarrow\quad
    \mathbf{E}_\perp=\frac{q}{2\pi\epsilon_0 r}\delta(x-ct)
    \label{Eq:UltrarelativisticElectricField}
\end{equation}
\begin{figure}[!hbt]
  \begin{center}
  \includegraphics[]{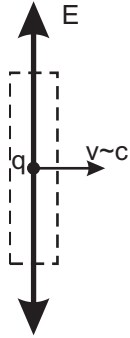}
  \caption[Electric Field of an Ultra-Relativistic Electron]{As the electric field is very much confined
  in a disc perpendicular to the velocity one can use the Gauss law in
  calculating the electric field of an ultra-relativistic electron. Here we enclose
  the charge with a thin cylindrical box. Only the side surface
  with area $2\pi r\ 1/\delta(x-ct)$ contributes to the total surface integral.}
  \label{Fig:GaussBox}
  \end{center}
\end{figure}
The formula for the electric field of an ultra-relativistic
electron beam is just a summation (convolution) of
\ref{Eq:UltrarelativisticElectricField} over all the electrons. If
the bunch has a charge linear density $\lambda(x)$ the electric
field is
\begin{equation}
    E_\perp=\frac{\lambda(x-ct)} {2\pi\epsilon_0 r}
    \label{Eq:BunchElectricField}
\end{equation}

\begin{figure}[!hbt]
  \begin{center}
  \includegraphics[height=3in]{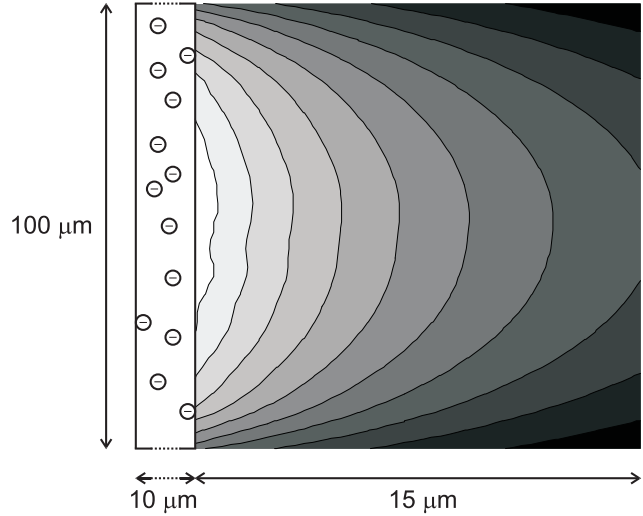}
  \caption[Equipotential Lines of Electric Field]{Equipotential lines of electric field for $10^4$ electrons
  uniform randomly distributed in a long rectangular line. Note that
  the vertical dimension is scaled down, so the variations in that
  direction are enhanced. Even so, the potential lines look rather smooth
  and for almost all purposes the random discrete distribution can be replaced by
  a continuous average charge distribution.}
  \label{Fig:RandomElectrons}
  \end{center}
\end{figure}

Although in the actual beam the charge distribution is discrete
and the random position of one electron inside the electron bunch
follows a gaussian distribution the electromagnetic field becomes
smooth very quickly farther away from the bunch center. As an
illustration, figure \ref{Fig:RandomElectrons} shows the potential
from a uniform random distribution of $10^4$ static electrons
inside a line with length $100\mu m$ and cross section $10\mu m
\times 10\mu m$.
As long as we are not interested in the effects very close to the
beam we can neglect the randomness in the position of the
electrons as only the average counts in the long range.

\begin{figure}[!hbt]
  \begin{center}
  \includegraphics[]{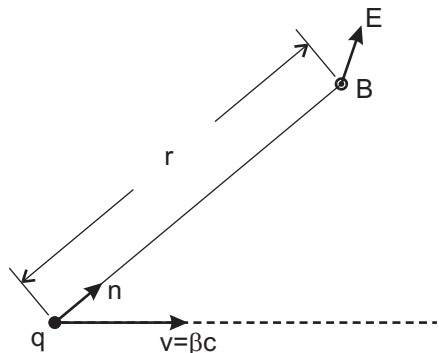}\\
  \caption[A Relativistic Electric Charge]{A relativistic electric charge $q$ in an arbitrary
  motion $\mathbf{v}$ generates at a point $r\mathbf{n}$ away
  electric and magnetic fields described by the Li\'{e}nard-Wiechert
  formula.}
  \label{Fig:LienardWiechert}
  \end{center}
\end{figure}

In general a charge has both velocity and acceleration and
different types of electromagnetic fields are associated with
them. A charge moving with velocity
$\mathbf{v}=\mbox{\boldmath$\beta$\unboldmath} c$ and acceleration
$\mathbf{a}=d\mbox{\boldmath$\beta$\unboldmath}/dt\,c$ will have
magnetic and electric fields that are described, at some point
$\mathbf{r}=r\mathbf{n}$ (see figure \ref{Fig:LienardWiechert}),
by the following Li\'{e}nard-Wiechert formula \cite{jackson:book}.
\begin{eqnarray}
  \mathbf{E}(\mathbf{r},t) &=& \frac{q} {4\pi\epsilon_0}
  \left [
  \frac{\mathbf{n}-\boldsymbol{\beta}} {\gamma^2(1-\boldsymbol{\beta}\cdot\mathbf{n})^3r^2}+
  \frac{\mathbf{n}\times{(\mathbf{n}-\boldsymbol{\beta})\times\boldsymbol{\dot{\beta}}}} {c(1-\boldsymbol{\beta}\cdot\mathbf{n})^3r}
  \right ]_{ret} \label{Eq:LienardWiechert}\\
  \mathbf{B} &=& [\mathbf{n}\times \mathbf{E}]_{ret} \nonumber
\end{eqnarray}
Since any electromagnetic interaction propagates with the speed of
light, we have to take into account the time required for the
electromagnetic field to propagate to the measured point. That is
why the square bracket is evaluated at a time when the
electromagnetic field was leaving the charge, the so called
retarded time.

The first term in the bracket of equation
\eqref{Eq:LienardWiechert} is associated with the speed and is
called a "velocity field". Having a $1/r^2$ dependence, it is
short range and this is the electromagnetic field accompanying the
electron bunch. The second term is associated with acceleration.
The "acceleration field" is generated only when a a charge is
accelerated. This is also called radiation. Because of the $1/r$
long-range dependence, only radiation can contribute to a
non-vanishing flow of energy through a remote spherical surface
enclosing the charge.

Although a charge in uniform motion does not radiate, the electric
and magnetic field of an ultra-relativistic charge resembles that
of a plane wave moving in the same direction, that is $B\perp E$
and $E=cB$.

This fact is exploited by a method of calculating the radiation
emitted by highly relativistic charged particles in interaction
with the condensed media, the pseudo-photon method
\cite{mikaelian:book,fermi:1924}. This method considers
\textit{the effect of a charged particle equivalent to that of a
set of photons} of various frequencies, obtained by Fourier
transform. The radiation emitted is just scattered pseudo-photons
and one can sum over all their interaction cross-sections to
obtain the interaction cross section for the particle.
\begin{equation}
    \sigma_{particle}=\int{n_\omega\sigma(\omega)d\omega}
    \label{Eq:Pseudophoton}
\end{equation}

Calculation of the interactions can be done either in real space
or in the Fourier space. The choice between them is a matter of
ease of calculation or an intuitive picture of phenomena. The
magnetic field pulse associated with a relativistic electron bunch
is equivalent to a superposition of frequency components up to a
cutoff given by the inverse of the pulse duration. The intensity
of these components can be quite high if measured reasonably
($100\mu m$) close to the bunch.

\begin{figure}[!hbt]
  \begin{center}
  \includegraphics[width=\textwidth]{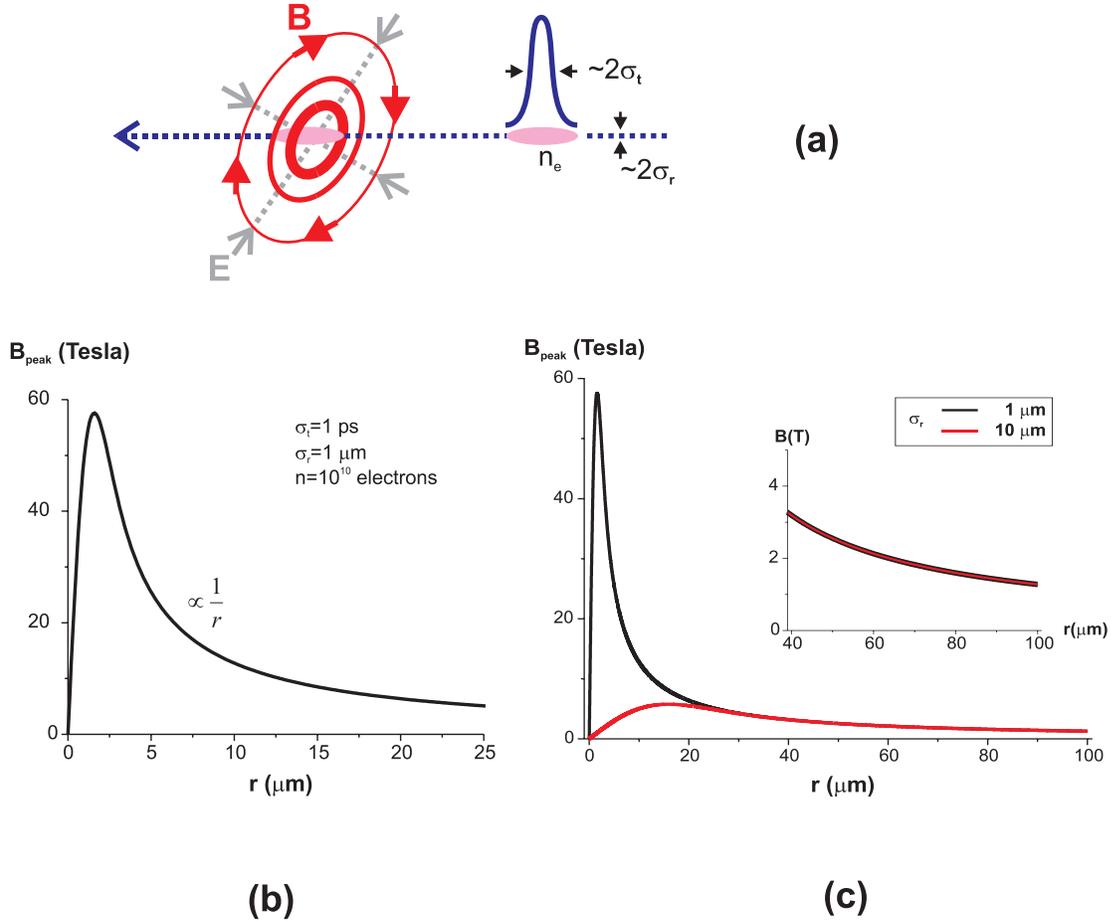}\\
  \caption[Spatial Distribution of the Magnetic Field of a Typical
  Electron Bunch]{Spatial distribution of the magnetic field of a typical
  electron bunch. (a)The important parameters are the number of
  electrons in the bunch $n$, its transversal size $\sigma_r$
  and its duration $\sigma_t$. (b) In the usable space the magnetic
  field has a $1/r$ dependence. (c) Focusing enables high magnetic fields
  near the beam but it is not important at larger distances. }
  \label{Fig:MagneticField}
  \end{center}
\end{figure}

As an example let's take an electron bunch that is gaussian in all
3 directions. It has $n$ electrons, $\sigma_t$ duration and
transversal dimensions $\sigma_x=\sigma_y=\sigma_r$. The magnetic
field generated by the bunch moving with near the speed of light
can be found making use of the Ampere's law:
\begin{equation}
    B(r,t)=\frac{\mu_0 n e }{\sqrt{2\pi}\sigma_t}
    \frac{1-e^{-\frac{r^2}{2\sigma_r^2}}}{2\pi r}
    e^{-\frac{t^2}{2\sigma_t^2}}
    \label{Eq:BunchMagneticField}
\end{equation}
At distances $r$ larger than the diameter of the bunch the
magnetic field is almost the magnetic field around a straight
current wire and has a dependence of $1/r$.
\begin{equation}
    B(r,t)=\frac{\mu_0 I(t)}{2\pi r}=
    \frac{\mu_0 \frac{n e}{\sqrt{2\pi}\sigma_t}
    e^{-\frac{t^2}{2\sigma_t^2}}}{2\pi r}=
    B_{peak}(r) e^{-\frac{t^2}{2\sigma_t^2}}
    \label{Eq:WireMagneticField}
\end{equation}
$B_{peak}(r)$ is the peak intensity of the magnetic field pulse at
a distance $r$ from the center of the bunch. A plot of the
magnetic field for a beam with parameters
$\sigma_t=1ps,\quad\sigma_r=1\mu m,\quad n=10^{10}$. High
intensity magnetic fields are possible near the electron beam if
this is focused well. At larger distances the focusing does not
matter much, see figure \ref{Fig:MagneticField}

\begin{figure}[!hbt]
  \begin{center}
  \includegraphics[]{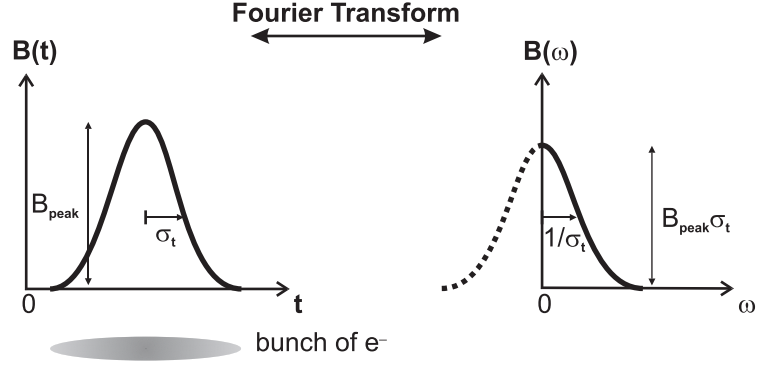}\\
  \caption[Frequency Components of a Gaussian Magnetic Field Pulse]
  {Frequency components of a gaussian magnetic field pulse with
  peak intensity $B_{peak}$ and standard deviation $\sigma_t$. Only
  positive frequencies are meaningful.}
  \label{Fig:FourierTransformPulse}
  \end{center}
\end{figure}

The frequency content of the pulse is easy to calculate. A
magnetic field with a gaussian shape has the Fourier transform
also a gaussian, as in figure \ref{Fig:FourierTransformPulse}.
\begin{equation}
    B(r,t)=B_{peak}(r) e^{-\frac{t^2}{2\sigma_t^2}} \rightarrow
    B(r,\omega)=B_{peak}(r) \sigma_t e^{-\frac{\omega^2}{2(1/\sigma_t)^2}}
    \label{Eq:FourierTransformMagneticField}
\end{equation}
If the bunch is made shorter, keeping the same number of
electrons, higher frequency components are added. By making the
bunch shorter we get a high intensity field over a short period of
time.

The Fourier component of the magnetic field is directly related to
the number of pseudo-photons of that frequency. The energy of the
electromagnetic field is stored as $n(\omega)$ quanta of energy
$\hbar\omega$ per unit volume and frequency interval,
see\eqref{Eq:QuantaMagneticField}.
\begin{align}
    \frac{B(r,\omega)^2}{2\mu_0}d\omega &= n(\omega)(\hbar\omega)d(\omega)
    \quad\Rightarrow\nonumber\\
    n(\omega) = \frac{B(r,\omega)^2}{2\mu_0\hbar\omega}&=
    \frac{(B_{peak}(r)\sigma_t)^2}{2\mu_0\hbar\omega}
    e^{-\frac{\omega^2}{(1/\sigma_t)^2}}
    \label{Eq:QuantaMagneticField}
\end{align}

You may have noticed that the quantity $B_{peak}\sigma_t$ keeps
appearing in formulas. From equation \ref{Eq:WireMagneticField} we
see that $B_{peak}\sigma_t$ is proportional to $q=n e$, the total
amount of charge in the bunch, which is usually constant in our
experiments.

\section{Time scales of electromagnetic phenomena}
\sectionmark{Time Scales}
As the time scale of the experiments become shorter, one has to
understand what are the relevant processes involved. One such
process is interaction of electromagnetic field with matter.
Maxwell's equations govern these processes and in SI units they
are:
\begin{eqnarray}
  \rot{H} &=& \mathbf{j} + \dt{\mathbf{D}}{} \label{Eq:rotH}\\
  \rot{E} &=& - \dt{\mathbf{B}}{} \\
  \divg{B} &=& 0 \\
  \divg{D} &=& \rho_e
  \label{Eq:GaussLaw}
\end{eqnarray}
where the vectors and the scalars are
\begin{description}
    \item[\textbf{H}], magnetic field strength
    \item[\textbf{B}], magnetic flux density
    \item[\textbf{D}], electric flux density
    \item[\textbf{E}], electric field strength
    \item[\textbf{j}], free current density
    \item[$\rho_e$], volume density of free electric charges
\end{description}
Three more equations are required for a general solution. These
equations are material dependent.
\begin{eqnarray}
  \mathbf{j} &=& \sigma\mathbf{E} \label{Eq:OhmLaw}\\
  \mathbf{B} &=& \mu\mathbf{H} \\
  \mathbf{D} &=& \epsilon\mathbf{E}
\end{eqnarray}
with
\begin{description}
    \item[$\sigma$], electric conductivity
    \item[$\mu$], magnetic permeability
    \item[$\epsilon$], dielectric constant
\end{description}

If there is no time dependence we have simply have a static field
\begin{eqnarray}
  \rot{H} &=& \mathbf{j} \\
  \rot{E} &=& 0
\end{eqnarray}
The phenomenon of electromagnetic waves arises from the coupling
of the laws of Faraday and Amp\`{e}re through the time dependent
terms $-\dt{\mathbf{B}}{}$ and $\dt{\mathbf{D}}{}$. The
quasistatic laws are obtained are obtained by neglecting either of
them in Maxwell's equations \cite{haus:book}.
\begin{table}[!hbt]
\begin{center}
\begin{tabular}{|c|c|}
  \hline
  \textbf{Electroquasistatic} & \textbf{Magnetoquasistatic} \\
  \hline
  $\rot{E} = - \dt{\mathbf{B}}{} \approx 0$ & $\rot{E} = - \dt{\mathbf{B}}{}$ \\
  $\rot{H} = \mathbf{j} + \dt{\mathbf{D}}{}$ & $\rot{H} = \mathbf{j} + \dt{\mathbf{D}}{} \approx \mathbf{j}$\\
  $\divg{D} = \rho_e$ & $\divg{D} = \rho_e$ \\
  $\divg{B} = 0$ & $\divg{B} = 0$ \\
  \hline
\end{tabular}
\end{center}
\caption[Electro- and Magnetoquasistatic Equations]{Electro- and
magnetoquasistatic equations.}
\end{table}
The electromagnetic wave propagation is neglected in either set.
Quasistatic here means the finite speed of light is neglected and
fields are considered to considered to be determined by the
instantaneous distribution of sources. The magnitude of the fields
can be used to classify these equations.
\begin{table}[!hbt]
\begin{center}
\begin{tabular}{|c|c|}
  \hline
  \textbf{Electroquasistatic} & \textbf{Magnetoquasistatic} \\
  \hline
  \multicolumn{2}{|c|}{Dominant Equations}\\
  $\rot{E} = 0$ & $\rot{H} = \mathbf{j}$ \\
  $\divg{D} = \rho_e$ & $\divg{B} = 0$ \\
  \hline
  \multicolumn{2}{|c|}{Residual Equations}\\
  $\rot{H} = \mathbf{j} + \dt{\mathbf{D}}{}$ & $\rot{E} = - \dt{\mathbf{B}}{}$ \\
  $\divg{B} = 0$ & $\divg{D} = \rho_e$ \\
  \hline
\end{tabular}
\end{center}
\caption[Classification of Quasistatic Equations]{Classification
of quasistatic equations according to the magnitude of the fields}
\end{table}

The quasistatic approximation is justified only when the magnitude
of the neglected terms is small compared to the other retained
terms. For example, consider a system of \textit{free space and
perfect conductors}. If the system has a typical length scale $L$
and the excitation has a characteristic time $\tau$ then we can
approximate $\nabla\rightarrow 1/L,\frac{\partial}{\partial
t}\rightarrow 1/\tau$. It turns out \cite{haus:book} that the
error in the approximated fields are
\begin{equation}
    \frac{E_{error}}{E}=\frac{\mu_0\epsilon_0L^2}{\tau^2}\hspace{1em},\hspace{1em}
    \frac{B_{error}}{B}=\frac{\mu_0\epsilon_0L^2}{\tau^2}
    \label{Eq:ErrorQuasistaticField}
\end{equation}
so the approximation is good when we have sufficiently slow time
variation (large $\tau$) and sufficiently small dimensions $L$
\begin{equation}
    \frac{\mu_0\epsilon_0L^2}{\tau^2}\ll1 \Rightarrow \frac{L}{c}\ll\tau
    \label{Eq:SmallQuasistaticError}
\end{equation}
This is valid when an electromagnetic wave can propagate the
characteristic length of the system in a time shorter than the
time of interest. To determine which quasistatic approximation to
use it helps to consider which fields are retained in the static
limit ($\tau\rightarrow\infty$). For a given characteristic time,
systems can often be divided in smaller regions small enough to be
quasistatic, but dynamically interacting through their boundaries.

\begin{figure}[!hbt]
  \begin{center}
  \includegraphics[]{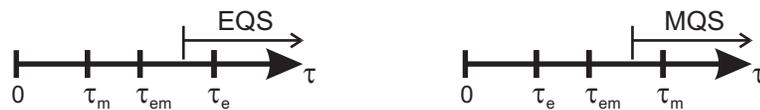}\\
  \caption[Characteristic Times of the Electromagnetic
  Phenomena]{Ordering of the characteristic times of the electromagnetic
  phenomena on the excitation time axis. Left diagram describes system that
  can be treated with an electro-quasistatic approximation, while the right
  diagram shows the ordering for systems amenable to magneto-quasistatic
  approximation. $\tau_e$ is the charge relaxation time, $\tau_m$ magnetic field
  diffusion time, $\tau_{em}$ electromagnetic transit time.}
  \label{Fig:TimeOrdering}
  \end{center}
\end{figure}

In the case of real systems with finite $\epsilon,\sigma,\mu$ we
have to refine the condition \ref{Eq:SmallQuasistaticError}. Let's
again consider a system with a typical length $l$ and an
excitation with a characteristic time $\tau$. We normalize the
quantities in the Maxwell's equations to their typical values.
\begin{align}
  \epsilon(\mathbf{r})=\epsilon\underline{\epsilon}\quad &,\quad
  \sigma(\mathbf{r})=\sigma\underline{\sigma}\quad &,\quad
  \mu(\mathbf{r})=\mu\underline{\mu} \nonumber\\
  \mathbf{r}=l\underline{\mathbf{r}}\quad &,\quad
  t=\tau \underline{t}\quad &,\quad
  \nonumber\\
  \mathbf{E}=\mathcal{E}\underline{\mathbf{E}}\quad &,\quad
  \mathbf{H}=\mathcal{E}\sqrt{\frac{\epsilon}{\mu}}\underline{\mathbf{H}}\quad &,\quad
  \rho=\frac{\epsilon\mathcal{E}}{l}\underline{\rho}
  \label{Eq:NormalizedQuantities}
\end{align}
Then the Maxwell's equations, with the help of the material
dependent equations, become
\begin{equation}
\begin{aligned}
  \underline{\nabla}\cdot\underline{\epsilon}\underline{\mathbf{E}} &= \underline{\rho}
  \\\\
  \underline{\nabla}\times\underline{\mathbf{H}} &= \frac{\tau_{em}}{\tau}\left(\frac{\tau}{\tau_e}\underline{\mathbf{E}}+\frac{\partial\underline{\epsilon}\underline{\mathbf{E}}}{\partial\underline{t}}\right)
  \\\\
   &= \frac{\tau_m}{\tau_{em}}\underline{\mathbf{E}}+\frac{\tau_{em}}{\tau}\frac{\partial\underline{\epsilon}\underline{\mathbf{E}}}{\partial\underline{t}}
  \\\\
  \underline{\nabla}\times\underline{\mathbf{E}} &= -\frac{\tau_{em}}{\tau}\frac{\partial\underline{\mathbf{H}}}{\partial\underline{t}}
  \\\\
  \underline{\nabla}\cdot\underline{\mu}\underline{\mathbf{H}} &= 0
\end{aligned}
\label{Eq:NormalizedMaxwellEquations}
\end{equation}
The times involved here represent charge relaxation time $\tau_e$,
magnetic diffusion time $\tau_m$ and the transit time for an
electromagnetic wave $\tau_{em}$
\begin{equation}
    \tau_e\equiv\frac{\epsilon}{\sigma};\hspace{2em}
    \tau_m\equiv\mu\sigma l^2;\hspace{2em}
    \tau_{em}\equiv\frac{l}{c}=l\sqrt{\mu\epsilon}
    \label{Eq:CharacteristicTimes}
\end{equation}
As an example in copper, with $\sigma=6\times
10^7\;\Omega^{-1}m^{-1}$, $l=0.3\;mm$, the times are
$\tau_{em}=10^{-12}\;s$, $\tau_m=7\times10^{-6}\;s$,
$\tau_e=1.5\times 10^{-19}\;s$. The electromagnetic transit time
is the geometric mean of the other two
$\tau_{em}=\sqrt{\tau_e\tau_m}$, that is, it lies between $\tau_e$
and $\tau_m$. The importance of various terms in Maxwell's
equations is given by the following ratios
\begin{equation}
    \frac{\tau_e}{\tau};\hspace{2em}
    \frac{\tau_m}{\tau};\hspace{2em}
    \frac{\tau_{em}}{\tau}
    \label{Eq:CharacteristicTimesRatios}
\end{equation}
The picture \ref{Fig:TimeOrdering} gives the domain of validity
for the quasistatic approximation

As the excitation time becomes shorter, dynamical processes
associated with charge relaxation, electromagnetic wave
propagation and magnetic diffusion appear.


\section{Conductive boundary}
\begin{figure}[!hbt]
  \begin{center}
  \includegraphics[width=\textwidth]{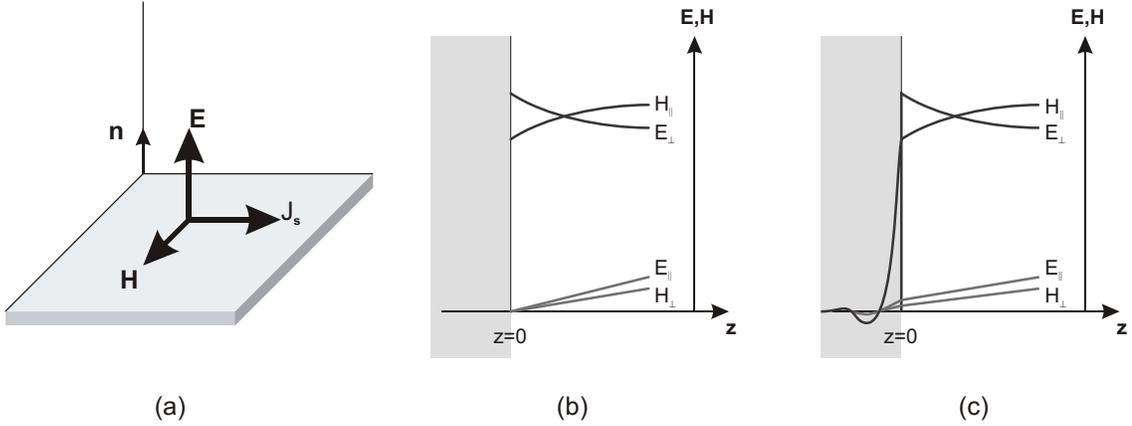}\\
  \caption[Fields Traversing a Conductive Boundary]{Fields traversing a conductive boundary. (a),(b) A perfect
  conductor has only normal electric field and tangential magnetic field
  just at the surface. (c) A good, but not perfect conductor allows the
  tangential magnetic field to penetrate inside over a skin depth distance.
  (The magnetic field plotted here sinusoidal time variation, hence the
  oscillatory behavior.)}
  \label{Fig:BoundaryFields}
  \end{center}
\end{figure}

Most magnetic media samples are metallic and therefore good
conductors. The conduction electrons, being highly mobile, will
tend to screen any electromagnetic fields inside a conductor. In a
perfect conductors they move instantly in response to changes in
the fields, no matter how rapid, to produce the correct surface
charge density and surface current to screen the electric and
magnetic field inside. The boundary conditions for a perfect
conductor are
\begin{align}
    \mathbf{n\cdot D} &= \Sigma \nonumber\\
    \mathbf{n\times H} &= \mathbf{J_s} \nonumber\\
    \mathbf{n\cdot(B-B_c)} &= 0 \nonumber\\
    \mathbf{n\times(E-E_c)} &= 0
    \label{Eq:PerfectBoundary}
\end{align}
where $\mathbf{n}$ is the unit normal of the surface directed
outward, $\Sigma$ the surface charge, $\mathbf{J_s}$ the surface
current and the subscript $c$ refers to the fields inside
 the conductor. Just outside the surface only normal $\mathbf{E}$
 and tangential $\mathbf{H}$ fields can exist, while inside the
 fields drop abruptly to zero as shown in figure
 \ref{Fig:BoundaryFields}.

The finite conductivity of a good conductor that is not perfect
implies that the surface is not as effective in screening the
magnetic field. In fact the penetration of the magnetic field is
described by a diffusion type equation.

From Ampere's equation \ref{Eq:rotH}, eliminating the electric
field, we can deduce the following differential equation for
magnetic field:
\begin{equation}
    \epsilon\mu\dt{H_c}{2}+\sigma\mu\dt{H_c}{}=\triangle{H_c}
    \label{Eq:MagneticFieldEquation}
\end{equation}
In good conductors $\sigma$ is large and the conduction term
dominates over the displacement term. The equation then becomes a
classical diffusion equation.
\begin{equation}
    \triangle{H_c}-\sigma\mu\dt{H_c}{}=0
    \label{Eq:MagneticFieldDiffusion}
\end{equation}
For a tangential sinusoidal (frequency $\omega$) magnetic field
$\mathbf{H_{\parallel}}$ outside the solution for $\mathbf{H_c}$
inside the conductor is
\begin{equation}
    \mathbf{H_c}=\mathbf{H_{\parallel}}e^{-z/\delta}e^{iz/\delta}
    \label{Eq:ParallelMagneticField}
\end{equation}
where $\delta=\sqrt{2/(\mu\omega\sigma)}$ is the skin depth. The
Faraday law requires that there is also a small electric field
associated with this magnetic field given by equation
\eqref{Eq:ParallelElectricField}.
\begin{equation}
    \mathbf{E_c}=\sqrt{\frac{\mu\omega}{2\sigma}}(1-i)
    (\mathbf{n\times H_{\parallel}})e^{-z/\delta}e^{iz/\delta}
    \label{Eq:ParallelElectricField}
\end{equation}
Using boundary condition for the electric field implies a similar
small tangential electric field $\mathbf{E_{\parallel}}$.

The above solutions show an exponential decay and their amplitude
is significant only on lengths a few times the skin depth. The
magnetic field is much larger than the electric field as shown in
\ref{Fig:BoundaryFields}. Because of the finite conductivity any
perpendicular magnetic field will also penetrate the conductor,
however, an evaluation using the Faraday law gives
$\mathbf{B_{\perp}}$ on the same order of magnitude as
$\mathbf{E_{\parallel}}$.

\begin{figure}[!hbt]
  \begin{center}
  \includegraphics[]{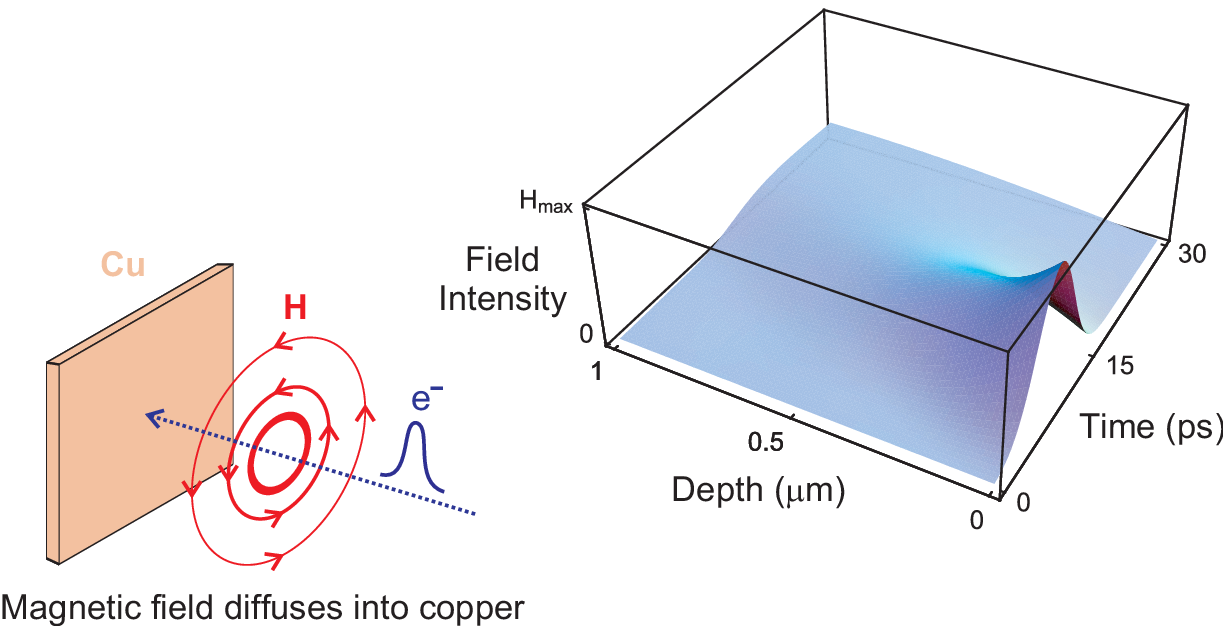}\\
  \caption[Magnetic Field Diffusion]{The magnetic field associated with an electron bunch diffuses into
  a conductive sample, a copper slab. If one is interested only in a thin layer
  at the surface (much thinner than $1\mu m$) the magnetic diffusion can be neglected
  and the magnetic field can be considered constant across its thickness.}
  \label{Fig:MagneticFieldDiffusion}
  \end{center}
\end{figure}

In first order one can approximate a gaussian pulse with a half
period of a sinusoidal oscillation, but a numerical solution for
the diffusion of a gaussian pulse associated with a relativistic
electron beam is shown in figure \ref{Fig:MagneticFieldDiffusion}
The skin depth in a typical metal like copper is of the order of
microns for magnetic field pulse of picoseconds duration. As one
moves deeper under the surface, the magnetic field pulse is spread
more and the maximum intensity is lower. However the time integral
of the magnetic field remains the same, being proportional to the
amount of charge in the beam.

\section{Electric charge relaxation}
A conductor reacts to any electric field and redistributes its
charge to screen the it, eventually relaxing into an equilibrium
state. The the electric charge relaxation equation (which can be
derived from equations
\ref{Eq:rotH},\ref{Eq:GaussLaw},\ref{Eq:OhmLaw}) and its solution
are
\begin{equation}
    \dt{\rho_e}{} +
    \frac{\sigma}{\epsilon}\rho_e = 0\hspace{1em},\hspace{1em}
    \rho_e = \rho_{e0}e^{-\frac{t}{\tau_e}}
    \label{Eq:ChargeEquation}
\end{equation}
There is one correction to the above equations in the case of
conductors. The time $\tau_e\approx 10^{-19}s$ for copper seems
very short and nonphysical, after all, light travels only
$0.3\mbox{\AA}$ during that time. Ohm's law used in free electric
charge equation \ref{Eq:ChargeEquation} is valid only for times
larger than the time between collisions $\tau_c$. The assumption
of a good conductor is reasonable only if $\tau\gg\tau_c$. For
copper $\tau=5\times 10^{-14}\;s$, so the above considerations
hold at least up to frequencies in infrared spectrum.

The simplest model describing charge relaxation can be derived
from the Drude model of electrical conduction. According to this
model the motion of conduction electrons is described by the
equation:
\begin{equation}
    m\mathbf{\ddot{r}}=-e\mathbf{E}-m\mathbf{\dot{r}}/\tau_c
\end{equation}
that has a solution
\begin{equation}
\mathbf{r}\sim\mathbf{e}^{-\frac{t}{2\tau_c}\pm\mathbf{i}\omega_pt}
\label{Eq:ChargeCarrierDynamics}
\end{equation}
with $\tau_c$ the time between collisions and $\omega_p$ the
plasma frequency($\omega_p=\sqrt{\frac{ne^2}{m\epsilon}}
=\sqrt{\frac{\sigma}{\tau\epsilon}}$). Any deviation from
neutrality within the conductor will be restored on the time scale
of $\tau_c$ in a damped oscillatory process (\cite{ohanian:1983}).

\begin{figure}[hbt]
  \begin{center}
  \includegraphics[]{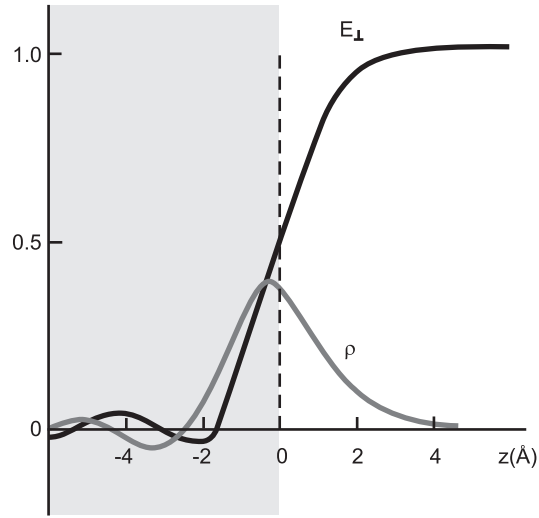}\\
  \caption[Surface Charge and Electric Field]{Surface electrical charge
  density $\rho(Z)$ and normal electric field $E_{\perp}(z)$
  distributions at the surface of a conductor confined to $z<0$.}
  \label{Fig:SurfaceCharge}
  \end{center}
\end{figure}

The electric field is screened very well inside metal thin film
samples. This process takes place very fast with the screening
charge oscillating with plasma frequency and being damped in times
comparable to the collision time of the electrons, see eq.
\ref{Eq:ChargeCarrierDynamics}. The surface charge is located in a
very thin layer of about $\pm 2$ \AA at the surface of the metal
\cite{kenner:1972}, as in figure \ref{Fig:SurfaceCharge}.
.

In the case of insulators and semiconductors the electric field
polarizes the matter and for a high enough field the phenomenon of
dielectric breakdown can occur.The electric field generated by the
electron beam has an amplitude related to the amplitude of the
magnetic field $\mathbf{B}=\frac{\boldsymbol{\beta}}{c}\times
\mathbf{E}$. So, for a magnetic field of 1 Tesla and $\beta\simeq
1$, the electric field has a value of $3\times 10^8 V/m$. For
comparison the electric breakdown in air occurs at an electric
field of $\approx 3\times 10^6 V/m$ while for GaAs it occurs at
$\approx 4\times 10^7 V/m$.

As one makes the electron bunch shorter the charge linear density
increases and so does the electric field, see
\ref{Eq:BunchElectricField}. As a consequence, the electric field
exceeds the breakdown limit over a larger area. One can observe
the damaged area in figure \ref{Fig:ExperimentSetup} in a sample
with GaAs as substrate.

\section{Screening for a relativistic electron bunch}
\sectionmark{Screening}
The screening of the electric field and magnetic field of a
relativistic electron bunch by a metallic sample can be visualized
intuitively with the method of image charges.

\begin{figure}[!hbt]
  \begin{center}
  \includegraphics[width=\textwidth]{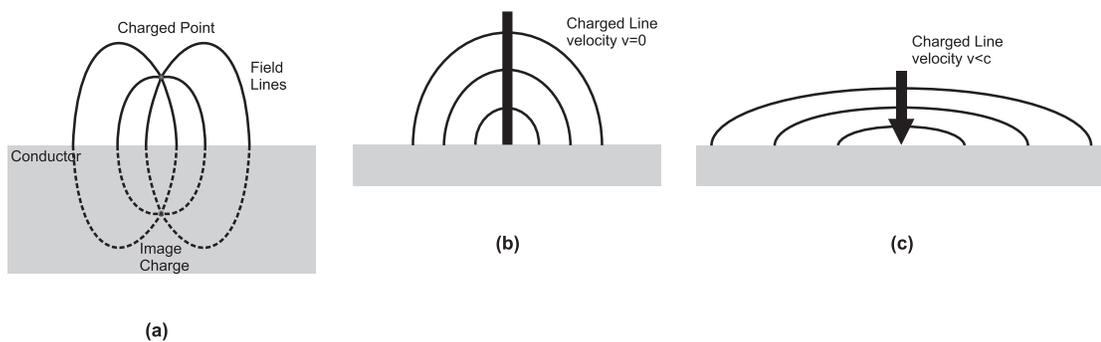}\\
  \caption[Method of Image Charges]{Method of image charges applied to
  (a) point charge
  (b) line charge (c) moving line charge close to a conductive surface.
  In all cases the electric field lines are perpendicular to the
  conductive surface. The conductor behaves as an opposite image charge.}
  \label{Fig:ImageCharge}
  \end{center}
\end{figure}

As a electron bunch approaches of the conductive sample the
electric field associated with the bunch will start to move
electrons inside the sample. Very quickly they reach such a
position as to make a zero total electric field inside the sample.
The same electric field at the surface of the sample is achieved
if one removes the sample and replaces it with distribution of
charge that is a mirror image of the electron bunch. Figure
\ref{Fig:ImageCharge} shows some model cases of charge
distribution close to a conductive surface and their field lines.

\begin{figure}[!hbt]
  \begin{center}
  \includegraphics[]{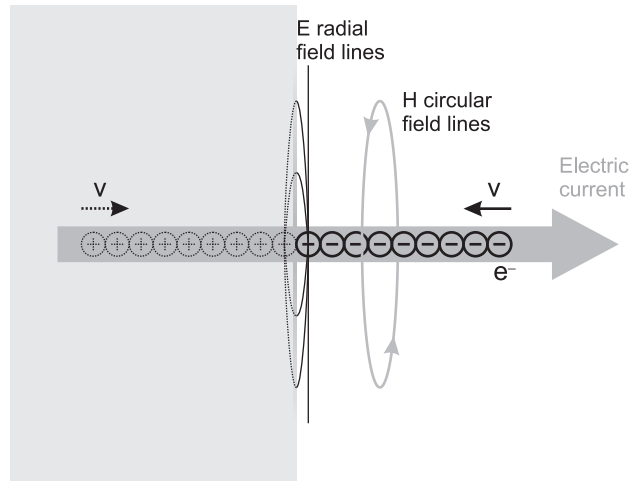}\\
  \caption[An Illustration of Electric and Magnetic Field Lines
  near the Surface of a Conductor]{An illustration of Electric and magnetic field lines
  near the surface of a conductor for a beam of relativistic
  electrons. The moving positive image charges are equivalent to
  a current in the same direction as that produced by the incoming
  electrons.}
  \label{Fig:SlacScreening}
  \end{center}
\end{figure}

An electron bunch has the electric field perpendicular to the
sample because the field lines are spread over a large area, and
so its surface charge associated with this field. The magnetic
field cannot be screened by these image charges because both the
moving electron bunch and its image produce a current in the same
direction, see figure \ref{Fig:SlacScreening}. Only the surface
eddy currents can screen the magnetic field.

When an experiment calls for a large electric field perpendicular
to a conductive sample the electron bunch has to fly over the
surface of the sample. This can also understood from the fact that
in figure \ref{Fig:ElectronFields} the electric field in the
direction of motion is $1/\gamma^3$ the electric field transversal
to the direction of motion.

\section{Energy loss of the electron beam in matter}
\sectionmark{Electron Energy Loss}
As high energy electrons pass through a sample they interact with
the matter and loose energy. Part of that lost energy is dumped
into the damaged area at the impact point.

The total stopping power (energy loss per unit length of path) of
electrons in matter be written as the sum of two terms
\cite{AIPdesk:book}, one due to ionization by collision and the
other due to radiation (bremsstrahlung):
\begin{equation}
    -\frac{dE}{dx} = \left(-\frac{dE}{dx}\right)_c + \left(-\frac{dE}{dx}\right)_r
    \label{Eq:StoppingPower}
\end{equation}.

The energy loss due to collisions is given by the Bethe-Bloch
equation:
\begin{align}
    \left(-\frac{dE}{dx}\right)_c = \quad & \frac{Z e^4 n}{8 \pi \epsilon_0^2 m_e v^2}
    \bigg[ ln\frac{m_e v^2 K}{2I^2(1-\beta^2)}-\left(2\sqrt{1-\beta^2}-1+
    +\beta^2\right)ln2 \nonumber\\
    & + 1 - \beta^2 + \frac{1}{8}\left(1-\sqrt{1-\beta^2}\right)^2 \bigg]
    \label{Eq:CollisionLoss}
\end{align}
where $Z$ is the atomic number of the absorber, $e$ the charge of
electron, $n$ the number of atoms per unit volume, $m_e$ the
static mass of electron, $\nu=\beta c$ the speed of electrons, $K$
the kinetic energy of electrons, $I$ the mean ionization potential
of the absorber.

The energy loss due to radiation is
\begin{equation}
    \left(-\frac{dE}{dx}\right)_r = \frac{\alpha n E Z(Z+1) e^4}{4\pi^2\epsilon_0^2 (m_e c^2)^2}
    \left[ln\frac{2E}{m_e c^2}-\frac{1}{3}\right]
    \label{Eq:RadiationLoss}
\end{equation},
where $E=K+m_e c^2$ is the energy of the electron and
$\alpha\approx 137$ is the fine constant.

\begin{figure}[!hbt]
  \begin{center}
  \includegraphics[]{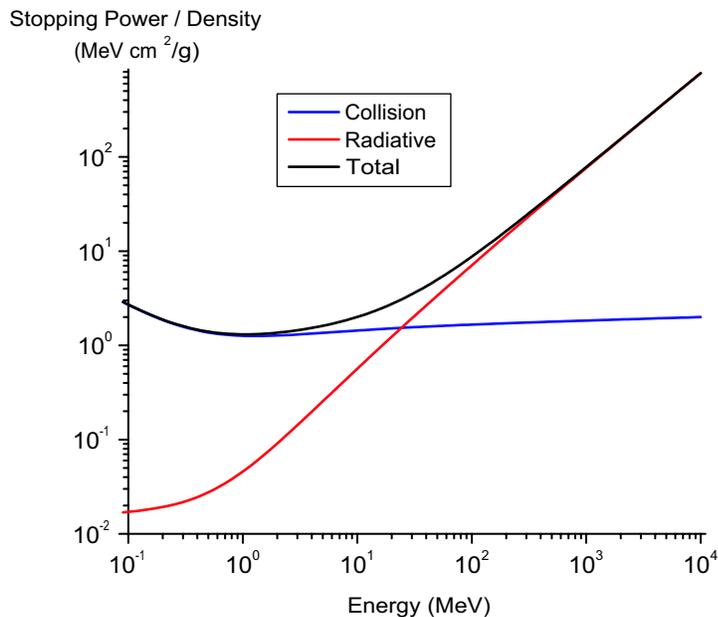}\\
  \caption[Stopping Power for Electrons of Copper ]{Stopping power of copper for electrons is dominated by radiative
  loss at high energies. The collision loss remains approximately at the same level.
  Note that here we have divided stopping power by density. }
  \label{Fig:StoppingPower}
  \end{center}
\end{figure}

The radiative contribution to the energy loss becomes important
only at high energies, typically above 10 MeV. The ratio between
two contributing stopping powers is approximately
\begin{equation}
    \frac{\left(-\frac{dE}{dx}\right)_c}{\left(-\frac{dE}{dx}\right)_r}
    =\frac{E Z}{1600 m_e c^2}
    \label{Eq:LossRatio}
\end{equation}.
At large energies the radiative loss processes dominates, see
figure \ref{Fig:StoppingPower} where the stopping power is a
measure of energy lost per unit length.
.
However, in general, matter is much more transparent to
electromagnetic radiation than to charged particles and the
radiation part of the energy loss does not remain in the sample
area. Thus the main contributing factor to the damage area is the
absorption of the collision part of the energy loss.

For copper with density $\rho = 9 g/cm^3$ and an electron energy
of 30 GeV the total stopping power is $\approx 20 GeV/cm$, while
the collision part is $\approx 2.4 MeV/cm$. An electron beam with
$10^10$ electrons will deposit an energy of $\approx 4 mJ$ in a
path of $1 cm$. These are only approximate numbers because they
neglect the effect of multiple scattering in altering the path.

If the electron beam is focused to spots of the order of microns
size the collision loss of the electrons will raise the
temperature of the impact point on the sample, and most of the
energy will be lost by radiation. A small part of it will be
conducted to adjacent area but, since conduction is slow, it does
not transport much energy before the impact point is cooled off by
radiation emission. Thus the damaged area is restricted around the
impact point.

\chapter{Experimental Setup}

The velocity electromagnetic field is short-range. In order to
observe its effects on the thin film sample one has to focus the
beam both to reduce the damage to a small area and to increase the
intensity of the electromagnetic field close to that area. For the
amount of charge in the electron bunch of the order of several
nanocoulombs, the magnetic field is useful on a length scale of
several hundred microns ($B\propto q/r$ see
\ref{Eq:WireMagneticField}). It is then desirable to keep the
focus below tens of microns.

\begin{SCfigure}[2][hbt]
  \includegraphics[height=2in]{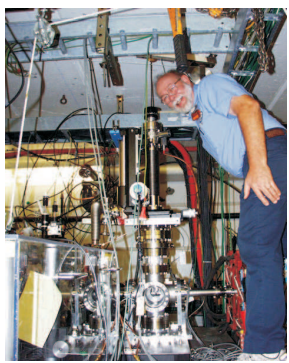}\\
  \caption[Experimental Setup and Location]{Experimental setup and location. (Jerry Collet, here standing
  near the whole apparatus, helped in the chamber design and assembly.)}
  \label{Fig:ChamberJerry}
\end{SCfigure}
For such a fine focus a good manipulator is needed for moving the
sample into the beam in the right position. The one used on our
experiment was an \textit{Omniax} type translator with an XY
table. The stepper motors had 20,000 steps per revolutions with
the motion transmitted as 1 rotation per mm for XY table while the
vertical Z translator was driven by a screw with 5 rotations per
mm. The manipulator (figure \ref{Fig:ChamberJerry}) was commanded
by a \textit{Compumotor} controller that communicated with an
outside computer through an Ethernet cable. During the actual run,
all personnel has to be outside the accelerator tunnel for
radiation safety reasons. The user interface was written in
\textit{Labview} and was structured as a state machine, the states
corresponding to different stages of the experiment (focusing of
the beam, exposure of the samples park in storage position etc.)

\begin{figure}[!hbt]
  \begin{center}
  \includegraphics[]{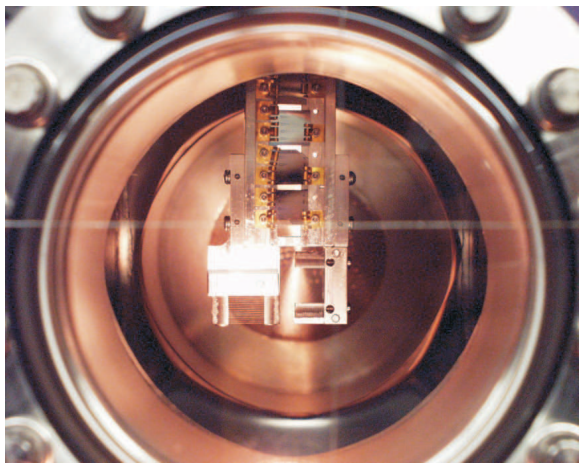}\\
  \caption[Wire Scanners for Monitoring the Beam Size and
  Position]{Wire scanners for monitoring the beam size and
  position. There are two groups of wires for vertical and horizontal
  directions. Some of the samples are visible in the upper part of the
  picture. The picture is taken from a window into the vacuum chamber
  situated on the accelerator line.}
  \label{Fig:WireScanner}
  \end{center}
\end{figure}
Also there is a need to find where the electron beam is with
respect to the sample holder. Once an absolute reference point is
found the sample can be exposed easily at the desired points using
incremental steps from that reference point. The size and position
of the beam are found by using wire scanners, see figure
\ref{Fig:WireScanner}. Carbon and tungsten wire were used with
diameters between $5-35\;\mu m$. Some of the thinner wires broke
after on or two shots under a short pulse of $100 fs$ and a focus
of $20\;\mu m$. The size of the beam is estimated after the
intensity curve of the gamma ray radiation generated when the
electron beam is swept across a wire.

The vacuum inside the whole chamber has to be very good (better
than $10^{-7}-10^{-8} torr$) otherwise the electron beam will
ionize the gas inside making a plasma which is not good for the
sample because the sample is heated and etched by the plasma. The
first experiments were unable to see a good magnetic pattern
precisely because of this damaging effect of a poor vacuum.
Although a magnetic pattern was recorded, the magnetic film was
destroyed by the interaction with ionized gases.

\begin{figure}[!hbt]
  \begin{center}
  \includegraphics[]{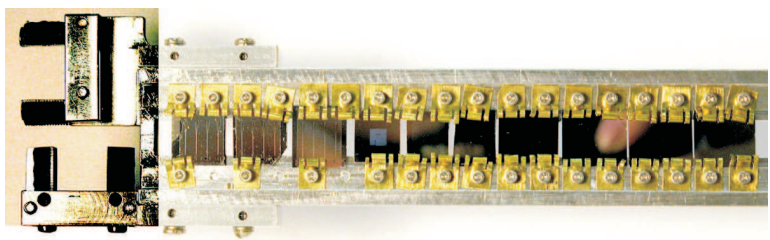}\\
  \caption[Overview of a Typical Sample Holder]{Overview of a typical sample holder with 10 samples
  loaded and the wire scanners attached to left side of the picture.}
  \label{Fig:SampleHolder}
  \end{center}
\end{figure}

The vacuum chamber is a six-way cross vacuum fitting. Two of the
side ports were connected to the accelerator, the top port was
used for loading the samples on a holder (figure
\ref{Fig:SampleHolder}) and two other side ports were windows.

As the electron bunch goes through the sample it generates
secondary electrons that exit the sample leaving it positively
charged. Grounding the sample holder is essential to prevent the
charging of the samples and the subsequent damaging sparks.
Fragile samples can even break under stresses caused by excessive
charges. Some magnetic films deposited on insulating glass
substrate were rendered unusable because the substrate fractured
in pieces. From this perspective a metallic substrate would be
better.

\chapter{Magnetic Imaging}

\section{Magneto-optic Effects}

The magneto-optic effects are due to the interaction of light and
electrons of the media through which light propagates. Visible
light does not penetrate far into the metals and thus the
magneto-optic methods are surface methods suitable for the study
of thin films. Although the magneto-optic effects are usually
associated with visible light they may be seen in the whole
spectrum of electromagnetic radiation.

\begin{figure}[!hbt]
  \begin{center}
  \includegraphics[]{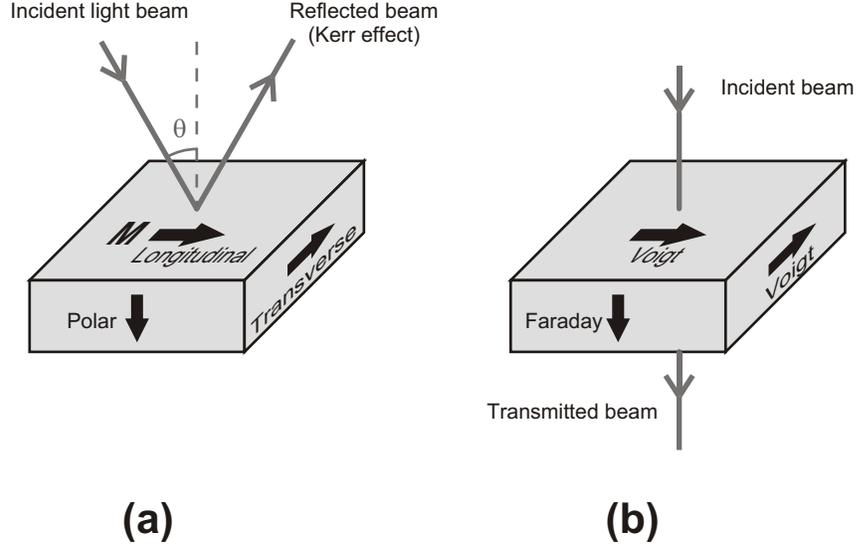}\\
  \caption[Magneto-Optic Effects]{Magneto-optic effects. (\textbf{a})
  Kerr effect in different geometries
  (orientation of magnetization with respect to the surface and plane of
  incidence). (\textbf{b}) Faraday and Voigt effects for transmitted light}
  \label{Fig:Magneto-optic_Effects}
  \end{center}
\end{figure}

These effects reflect the influence of an applied magnetic field
or a spontaneous magnetization on the polarization state and
intensity of light. The changes of the emergent light with respect
to incident light can be in amplitude (\textit{dichroism}) or in
phase (\textit{birefringence}). The emergent light may be
transmitted (Faraday and Voigt effects) or reflected (Kerr
effect), shown in figure \ref{Fig:Magneto-optic_Effects}.

All types of materials show magneto-optic effects: diamagnetic,
paramagnetic and ferromagnetic materials. These effects can be
attributed to off-diagonal terms in the dielectric or magnetic
susceptibility tensor. In first order the effects are due to the
interaction of the electric field of the light with the electrons.
The orbital motion of the electrons under the electric field is
influenced by the spin-orbit coupling and in the case of a
ferromagnet all spins are aligned, leading to a large
magneto-optic effect.

For example, to first order in magnetization $\mathbf{M}$, a cubic
or isotropic material has the following dielectric susceptibility
tensor $\epsilon$:
\begin{equation}
\epsilon = \epsilon_0 \left[
\begin{array}{rrr}
  1 & -iQM_z & iQM_y \\
  iQM_z & 1 & -iQM_x \\
  -iQM_y & iQM_x & 1 \\
\end{array}
\right] \label{Eq:DielectricTensor}
\end{equation}
where $\epsilon_0$ is the dielectric constant in the absence of
magnetization that has $M_i$ components. $Q$ is a parameter that
reflects the different behavior of right and left circularly
polarized light. When $\mathbf{M}$ reverses the off-diagonal terms
change sign and thus, the magneto-optic effects also change sign.
A detailed calculation involves solving the Maxwell equations with
the appropriate boundary conditions (for reflected and transmitted
light), but the main idea is that an incident electric field
$\mathbf{E}\nparallel\mathbf{M}$ produces a dielectric
polarization $\mathbf{P}\nparallel\mathbf{E}$ changing the
polarization of the emergent radiation.

\begin{figure}[!hbt]
  \begin{center}
  \includegraphics[]{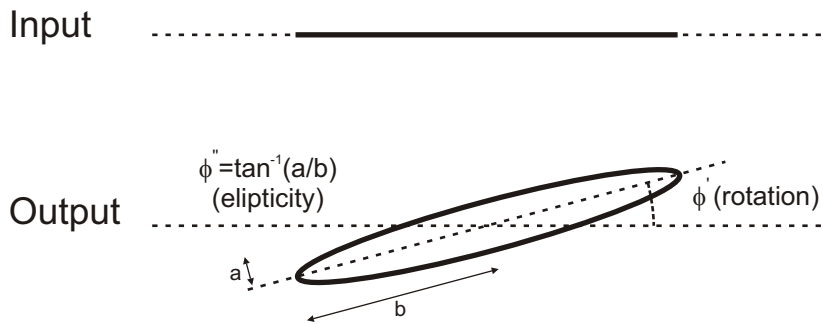}\\
  \caption[Change in Polarization State due to Magneto-Optic Effects]
  {Change in polarization state due to magneto-optic effects.}
  \label{Fig:Polarization_Change}
  \end{center}
\end{figure}

In general, reflection or transmission of light from a surface
changes the polarization of light even without the presence of
magnetization. However, if the light is linearly polarized in the
plane of incidence (p-polarization) or perpendicular to it
(s-polarization), there is no change of polarization due to the
surface. Then, if the material is magnetic, the output may be
elliptical ($\phi'$) and rotated ($\phi''$) with respect to the
initial polarization direction as shown in figure
\ref{Fig:Polarization_Change}.
The rotation and ellipticity are usually taken to be the real and
imaginary parts of a complex rotation $\phi=\phi'+i\phi''$. As an
example, Kerr effect in iron at room temperature and $633\,nm$
light has $\phi=0.14^\circ+i0.07^\circ$ in longitudinal geometry
and $45^\circ$ incidence while $\phi=0.63^\circ-i0.47^\circ$ in
polar geometry and normal incidence. Measurement conditions must
be specified since the measured angle varies with refractive
index, the angle of incidence, the temperature and the wavelength.

\section{Scanning Electron Microscopy}

Another method of imaging the magnetization of ferromagnetic thin
films is by detecting the polarization of secondary electrons. A
scanning electron microscope is well suited for this as it can
make a small scanning spot size of the incident electrons that
generate the secondary electrons. Scanning Electron Microscopy
with Spin Analysis (SEMPA) images the magnetization by measuring
the spin polarization of secondary electrons that is directly
related to the magnetization of the sample. Measurements of the
magnetization are intrinsically independent of the topography, but
the magnetic and topographic maps are measured simultaneously.
Because of the small (nanometers) secondary electron escape depth
the method is surface sensitive.
\begin{figure}[!hbt]
  \begin{center}
  \includegraphics[]{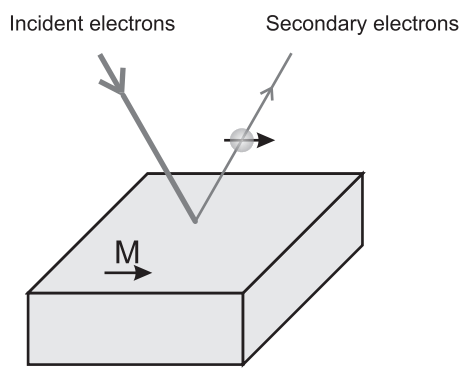}\\
  \caption[The Principle of Scanning Electron Microscope]{The principle
  of scanning electron microscope with
  polarization analysis: An unpolarized monoenergetic electron beam, focused by
  electromagnetic lenses, scans across the ferromagnetic surface.
  At each scanned position, spin-polarized secondary electrons
  are created near the surface, emitted into vacuum and transferred
  to the spin analyzer by electrostatic lenses.}
  \label{Fig:Sempa}
  \end{center}
\end{figure}

\section{Photoemission electron microscopy}

\begin{SCfigure}[][!hbt]
  \includegraphics[]{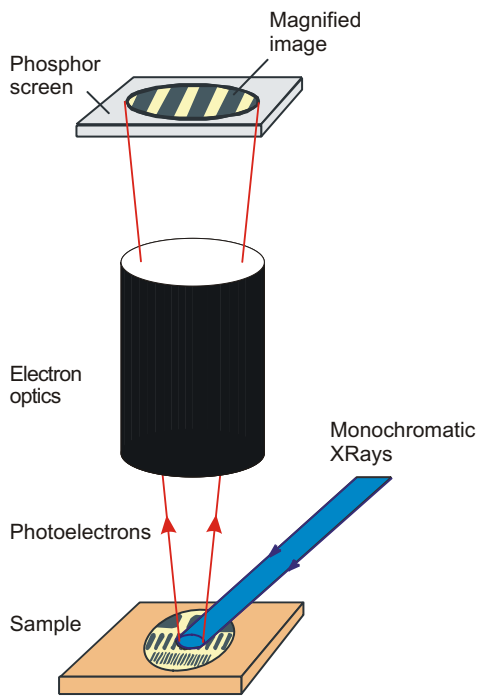}\\
  \caption[The Principle Of Photoemission Electron Microscope]{The
  principle of photoemission electron microscope with
  X-rays: An X-ray photon beam is shine on the sample and the
  secondary electrons pass through electron optics giving a magnified
  image on a phosphor screen.}
  \label{Fig:Peem}
\end{SCfigure}

X-ray photons impinging on a sample cause secondary electrons to
be emitted from the surface. One can then use regular electron
optics to image the place where these electrons originated (figure
\ref{Fig:Peem}). Absorption of photons and subsequent generation
of electrons can tell a lot of things about sample. The contrast
in the image can be elemental, chemical, magnetic or
topographical.

\begin{figure}[!hbt]
  \begin{center}
  \includegraphics[width=\textwidth]{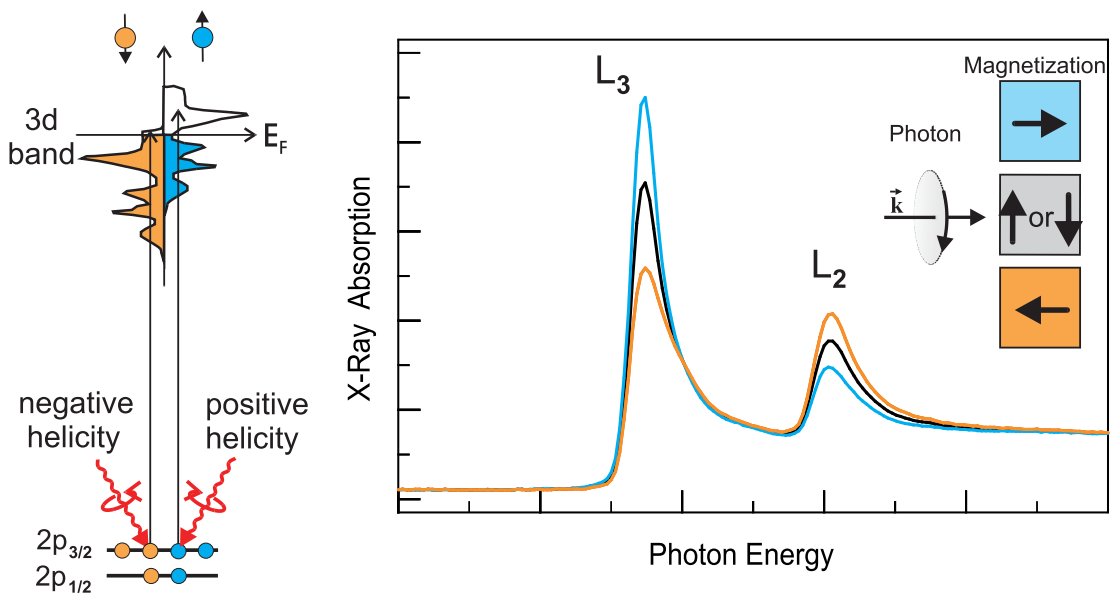}\\
  \caption[Illustration of the XMCD Effect]{Illustration of the XMCD
  effect: The angular momentum
  and the energy carried by the photon is transferred to a core level
  electron which is promoted to an empty valence band level. There are
  more empty levels of one spin orientation due to the exchange interaction
  and so the absorption of the photon is dependent on the orientation of
  the magnetization.}
  \label{Fig:XMCD_Effect}
  \end{center}
\end{figure}

When studying ferromagnetic samples the phenomenon of X-ray
magnetic dichroism is employed. Circularly polarized photons are
shined on the sample and absorbed by atoms promoting the electrons
in the core level to an empty valence band level. The subsequent
decay of the core level hole generates the secondary electrons
imaged by the electron optics. The absorption is sensitive to the
magnetization orientation because the electronic transition
conserves the electron spin while the valence band is spin split
by the exchange interaction in the ferromagnetic sample and the
core level is also split by spin-orbit interaction (figure
\ref{Fig:XMCD_Effect}). The absorption process is a miniature spin
polarizer-analyzer experiment (made possible by the conservation
of angular momentum).

\chapter{Magnetization Dynamics Equation}
\label{Ch:MagDynamics}

Magnetic materials contain a collection of magnetic moments
(dipoles) and the physics of magnetism tries to explain the
behavior of this collection of moments under a variety of
conditions and excitations. The most important behavior of a
magnetic moment is the precession in a magnetic field.

The magnetic moment $\mathcal{M}$ feels a torque from the magnetic
field $\mathbf{B}$ acting on it, torque given by
\begin{equation}
    \mathcal{T}=\mathcal{M}\times \mathbf{B}
    \label{Eq:Torque}
\end{equation}
An angular momentum $\mathbf{L}$ is always associated with the
magnetic moment, one being proportional to another (with
magneto-mechanical ratio $\gamma$)
\begin{equation}
    \mathcal{M}=\gamma\mathcal{L}
    \label{Eq:MagnetoMechanicalRatio}
\end{equation}
and since we know that $\mathcal{T}=d\mathcal{L}/dt$ the equation
of motion for the magnetic moment is
\begin{equation}
    \frac{d\mathcal{M}}{dt}=\gamma\mathcal{M}\times \mathbf{B}
    \label{Eq:MotionEquation}
\end{equation}
This is the basic equation of motion for a magnetic moment, but
the hard part is to find the magnetic field acting on the magnetic
moment. The field of micromagnetics simulates the dynamics of an
ensemble of magnetic moments, modelling the interaction of the
magnetic moment with the environment as an effective field.

When one has a lot of magnetic moments it is easier to use the
continuum approximation when one deals with the magnetic moment
density, also known as the magnetization
$\mathbf{M}=\mathcal{M}/\mathbf{V}$. With magnetic induction
expressed as $\mathbf{B}=\mu_0\mathbf{H}_{eff}$ the equation
\ref{Eq:MotionEquation} becomes
\begin{equation}
    \frac{d\mathbf{M}}{dt}=\gamma\mu_0\mathbf{M}\times \mathbf{H}_{eff}
    \label{Eq:MotionEquationHeff}
\end{equation}

The dissipation of angular momentum is usually modelled as a
phenomenological damping term and a simple way is to take it
proportional to $d\mathbf{M}/dt$ \cite{miltat:2002} like in ohmic
dissipation
\begin{equation}
    \mathbf{H}_{eff}=\mathbf{H}-\frac{\alpha}{\mu_0\gamma M_s}
                                \frac{d\mathbf{M}}{dt}
    \label{Eq:DampingTerm}
\end{equation}
The remaining part $\mathbf{H}$ includes the anisotropy, exchange,
dipolar (from neighboring magnetic moments) and external magnetic
fields which will be addressed later. The equation
\ref{Eq:MotionEquationHeff} then becomes Landau-Lifshitz-Gilbert
equation
\begin{equation}
    \frac{d\mathbf{M}}{dt}=\gamma\mu_0\mathbf{M}\times\mathbf{H}+
    \frac{\alpha}{M_s}(\mathbf{M}\times\frac{d\mathbf{M}}{dt})
    \label{Eq:LLGequation}
\end{equation}
which can be transformed into an explicit equation for
$d\mathbf{M}/dt$, known as Landau-Lifshitz equation.
\begin{equation}
    (1+\alpha^2)\frac{d\mathbf{M}}{dt}=\gamma\mu_0\left[\mathbf{M}\times\mathbf{H}+
    \frac{\alpha}{M_s}\mathbf{M}\times(\mathbf{M}\times\mathbf{H})\right]
    \label{Eq:LLequation}
\end{equation}
Historically equation \ref{Eq:LLequation} was derived first, the
damping term being introduced in such a way as to relax the
precessing magnetization into the field direction as shown in
figure .
\begin{figure}[!htb]
  \begin{center}
  \includegraphics[]{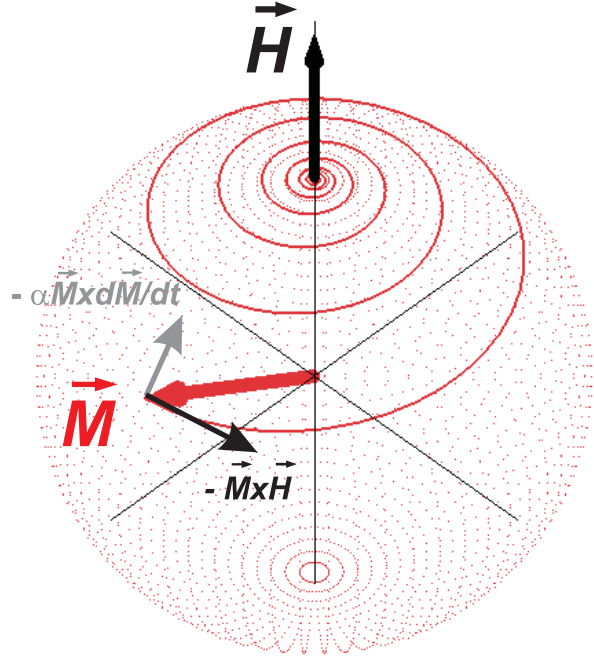}\\
  \caption[Directions of the Precessional and Damping Torques]{Directions
  of the precessional and damping torques for magnetization
  in a constant magnetic field. The damping torque eventually brings the
  magnetization into the direction of the field.}
  \label{Fig:LLGTorque}
  \end{center}
\end{figure}
The dimensions can be taken out of the equation
\ref{Eq:LLequation} by normalizing to $M_s$
$\mathbf{m}=\mathbf{M}/M_s$, $\mathbf{h}=\mathbf{H}/M_s$ and using
a time normalized to $\tau=t/(|\gamma|\mu_0 M_s)^{-1}$
\begin{equation}
    (1+\alpha^2)\frac{d\mathbf{m}}{d\tau}=-\mathbf{m}\times\mathbf{h}-
    \alpha(\mathbf{m}\times(\mathbf{m}\times\mathbf{h}))
    \label{Eq:LLequationNormalized}
\end{equation}
For an electron $\gamma$ is negative hence the minus sign in the
above equation.

The magnetic field $\mathbf{H}$ can be obtained if one interprets
the force acting on a magnetic moment $\mathcal{M}$ as the
negative gradient of a potential energy
$\mathcal{U}=-\mu_0\mathcal{M}\cdot \mathbf{H}$. The torque is
then $\mathcal{T}=d\mathcal{U}/d\theta$, where $\theta$ is the
angle between $\mathcal{M}$ and $\mathbf{H}$. This torque is the
precessional torque $\propto \mathbf{M}\times\mathbf{H}$ in figure
\ref{Fig:LLGTorque}. The damping torque just dissipates the excess
potential energy to bring the system to the lowest energy
configuration, i.e. the magnetization parallel to the magnetic
field

The variation of the free energy density $\mathrm{U}$ with respect
to magnetization direction $\mathbf{M}$ will give the magnetic
field $\mathbf{H}$.
\begin{equation}
    \mathbf{H}=-\frac{1}{\mu_0}\frac{d\mathrm{U}}{d\mathbf{M}}
    \label{Eq:FreeEnergyH}
\end{equation}
The part of free energy density $\mathrm{U}$ where $\mathbf{M}$
enters consists of the \textit{magnetic energy} in the external
field, the \textit{magnetostatic energy} from interaction between
different parts of the magnetized body, the \textit{anisotropy
energy} with respect to crystallographic axes and the
\textit{exchange energy} between neighboring regions with
different orientation of the magnetization. Thus the field
$\mathbf{H}$ is
\begin{equation}
    \mathbf{H}=\mathbf{H}_{extern}+\mathbf{H}_{magnetostatic}+
    \mathbf{H}_{anisotropy}+\mathbf{H}_{exchange}
    \label{Eq:SumH}
\end{equation}
The corresponding energy densities are \cite{miltat:2002} :
\begin{eqnarray}
  \mathrm{U}_{extern} &=& -\mu_0 \mathbf{M}\cdot\mathbf{H}_{extern}
                      = -\mu_0 M_s^2 \mathbf{m}\cdot\mathbf{h}_{extern}\\
                      \nonumber\\
  \mathrm{U}_{magnetostatic} &=& -\frac{\mu_0}{2} \mathbf{M}\cdot\mathbf{H}_{demag}
                             = -\frac{\mu_0}{2} M_s^2 \mathbf{m}\cdot\mathbf{h}_{demag}\\
                             \nonumber\\
  \mathrm{U}_{anisotropy} &=& -K_u(\mathbf{m}\cdot\mathbf{u}_{uniaxial})^2 +
                            K_{cubic}(m_x^4+m_y^4+m_z^4)\\
                            \nonumber\\
  \mathrm{U}_{exchange} &=& -A(\nabla\mathbf{m})^2
\end{eqnarray}
where $K_u$, $K_{cubic}$ are the uniaxial and cubic anisotropy
energy densities, $A$ the exchange constant and the $H_{demag}$ is
the demagnetizing field calculated from the magnetic potential of
all other parts of the magnetic body:
\begin{equation}
    \mathbf{H}_{demag}=-\nabla_{\mathbf{r}'}\left[\frac{1}{4\pi\mu_0}
    \int{\frac{\rho(\mathbf{r}')}{|\mathbf{r}-\mathbf{r}'|}d\mathbf{r}'}\right]
    \label{Eq:Hdemag}
\end{equation}
Most of the computation time in micromagnetics is spent in
computing \ref{Eq:Hdemag} because the integrand is long range and
slow varying. The convolution integral is more efficiently
calculated using Fast Fourier Transform techniques.

The dynamics described by the Landau-Lifshitz equation conserves
the magnitude of magnetization, an approximation which is true in
many systems. When spin-waves are generated the magnetization is
reduced in amplitude and one needs an additional parameter to
describe this process. The Bloch-Bloembergen equation
\cite{bloembergenn:1950} has such a term but only applies to small
angle perturbations as in ferromagnetic resonance.
\begin{eqnarray}
  \frac{M_{x,y}}{dt} &=& \gamma\mu_0 \big[\mathbf{M}\times\mathbf{H}\big]_{x,y}-
                \frac{M_{x,y}}{T_2} \\
  \nonumber\\
  \frac{M_z}{dt} &=& \gamma\mu_0 \big[\mathbf{M}\times\mathbf{H}\big]_z-
                \frac{M_z-M_s}{T_1}
\end{eqnarray}
There is a direct relation \cite{callen:1958} between the
transversal relaxation time $T_2$ and $\alpha$ in equation
\ref{Eq:LLequation}
\begin{equation}
    \frac{1}{T_2}=\alpha|\gamma|\mu_0 M_s\mathbf{h}\cdot\mathbf{m}
    \label{Eq:T2alpha}
\end{equation}
There is no equivalent in the Landau-Lifshitz equation for the
longitudinal relaxation time $T_1$, however the Landau-Lifshitz
equation is better suited to describing the process of switching
the magnetization direction as it is valid for large angle
perturbations.

\chapter{Experiments with In-plane Magnetic Thin Films}
\chaptermark{Model Thin Films}

\section{Introduction}
The bottleneck of magnetization dynamics is established by the
necessity to conserve angular momentum whenever the magnetization
$M$ changes direction or magnitude. After an external excitation
the spin system will ultimately equilibrate with the lattice on a
time scale of several hundred picoseconds (1 ps = 10$^{-12}$s), as
measured through the line width of ferromagnetic resonance (FMR).
Experiments based on precessional switching of $M$ are compatible
with the FMR derived dissipation \cite{schumacher:2003,
silva:2003, gerrits:2002, kaka:2002}. Applying the fastest
conventional magnetic field pulses of $\approx 10^4\ A/m$
amplitude and $\approx 100\ ps$ duration \cite{schumacher:2003,
silva:2003, gerrits:2002,kaka:2002, freeman:2002}, $M$ will switch
once performing a complex motion induced by the simultaneous
action of the pulse and the anisotropy fields. It is difficult to
evaluate the energy and angular momentum dissipated in such a
single complex switching process.

Our experiment separates the initial deposition of energy and
angular momentum into the spin system from the ensuing slower
magnetic switching and dissipation process, by using the extremely
fast and powerful magnetic field pulse generated by highly
relativistic electrons \cite{back:1998,back:1999,tudosa:2004}. The
field pulse amplitude varies across the sample thereby producing a
large magnetic pattern. This pattern, revealed by magnetic
microscopy, consists of a number $\nu$ of regions where $M$ has
switched its original direction. The location of the boundaries
between the regions reveals the energy required for the switching,
but owing to the internal clock provided by the precession of $M$
also corresponds to well defined times $t_{\nu}$ at which the
switching occurred. We find that the dissipation of the spin
angular momentum increases strongly after the first switch,
exposing the opening of a new dissipation channel, which we
associate with transfer of energy and angular momentum from the
uniform magnetization precession mode to higher spin wave modes.
In agreement with the recently developed quantitative theory
\cite{dobin:2003}, we find that this channel becomes less
effective with decreasing film thickness due to reduction of the
phase space for suitable spin waves. Our experimental results
reveal a larger dissipation than predicted with state-of-the-art
numerical simulations, but they do not exhibit the ultrafast
dissipation claimed in pulsed laser excitation.

\section{Experimental details}
\begin{figure}[hbt]
\centering
\includegraphics[width=\textwidth]{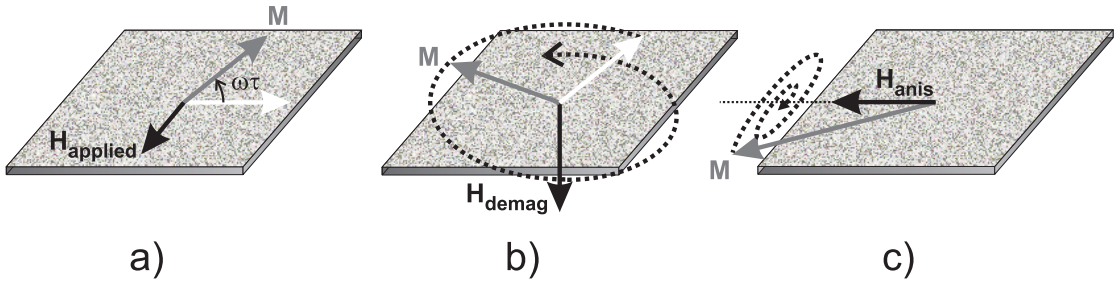}
\caption[Three Step Model of the Dynamics in In-plane Media]{An
illustration of the three step model for the in-plane media. (a)
Magnetization vector is brought out the sample plane by the
magnetic field of the electron bunch passing through. (b) The
demagnetization field created in the previous step makes the
magnetization precess around an axis perpendicular to the sample.
(c) As the precession around the demagnetization field damps out,
the anisotropy takes charge of the motion and the magnetization
relaxes into the easy axis.} \label{Fig:3step}
\end{figure}

Prior to the field pulse, the magnetization of the film, $M$, is
oriented along the easy direction which we assume to lie in the
$xy$ plane of the film, along the $x$-axis. A model of the
dynamics is illustrated in picture \ref{Fig:3step}. In step one,
the sample is excited by the magnetic field pulse generated with a
bunch of highly relativistic electrons from the linear
accelerator, see table \ref{Table:InPlaneBeam}.
\begin{table}[htb]
\begin{center}
\begin{tabular}{|l|c|}
  \hline
  \textbf{Property} & \textbf{Value} \\
  \hline
  Energy of electrons & $28\ GeV$ \\
  Number of electrons & $(1.0\ -\ 1.4)\times10^{10}$ \\
  Cross section (gaussian profile with $\sigma_x\times\sigma_y$) &
  $9\times6\ \mu m$ \\
  Pulse length (gaussian profile with $\sigma_z$) & $0.63\ mm$ \\
  Pulse duration ($\sigma_t=\sigma_z/c$) & $2.3\times10^{-10}\ s$ \\
  Peak magnetic field $B_p(Tesla)$ at distance $R(\mu m)$ &
  $B_p=(59.1\ -\ 82.1)/R$\\
  \hline
\end{tabular}
\end{center}
\caption[Electron Beam Used with In-plane Media ]{Some
characteristics of the electron beam used in the experiment using
in-plane media.} \label{Table:InPlaneBeam}
\end{table}
The electron beam travels along $z$, perpendicular to the film
plane and as it traverses the metallic film, its electric field is
screened at the $fs$-time scale. The magnetic field $B$ penetrates
the film because the skin depth is much larger than the film
thickness. It is oriented perpendicular to the beam axis,
resembling the familiar circular field generated by a straight
current carrying wire. The torque exerted by $B$ on the magnetic
moments makes them precess out of the film plane during the field
pulse. The angle of precession is
\begin{equation}
\varphi = \int{\omega dt} = \int{\gamma B_y(t)\, dt} \propto Q\ x
/ r^2
\end{equation}
where $\omega$ is the precession frequency, $r=[x,y,z]$ the
position vector with respect to the impact point, $Q$ the total
charge of the electron bunch. $\varphi$ lies in the range
$10^\circ-25^\circ$ in the present experiments.

In the next step, starting at the end of the field pulse, $M$
precesses around the demagnetizing field $H_{demag}$ generated
along the $z$-axis by the out-of-plane rotation of $M$ during
excitation. In this precession, a large angle $90^\circ - \varphi$
is enclosed between $H_{demag}$ and $M$, distinguishing it from
that in FMR where the precession angle is very small. In the large
angle precession, the in-plane component $M_x$ oscillates
periodically between the two easy directions. Owing to the damping
of the precession, $M $ spirals back into the plane of the film
until it can no longer overcome the anisotropy barrier imposed by
the uniaxial, in-plane crystalline anisotropy energy $K_u$. Then,
lastly $M$ oscillates about the in-plane uniaxial anisotropy field
$H_{ani,x}=2K_u/M $ until it comes to rest in either the initial
direction along $x$, or the direction opposite to it.

\begin{figure}[hbt]
\centering
\includegraphics[width=\textwidth]{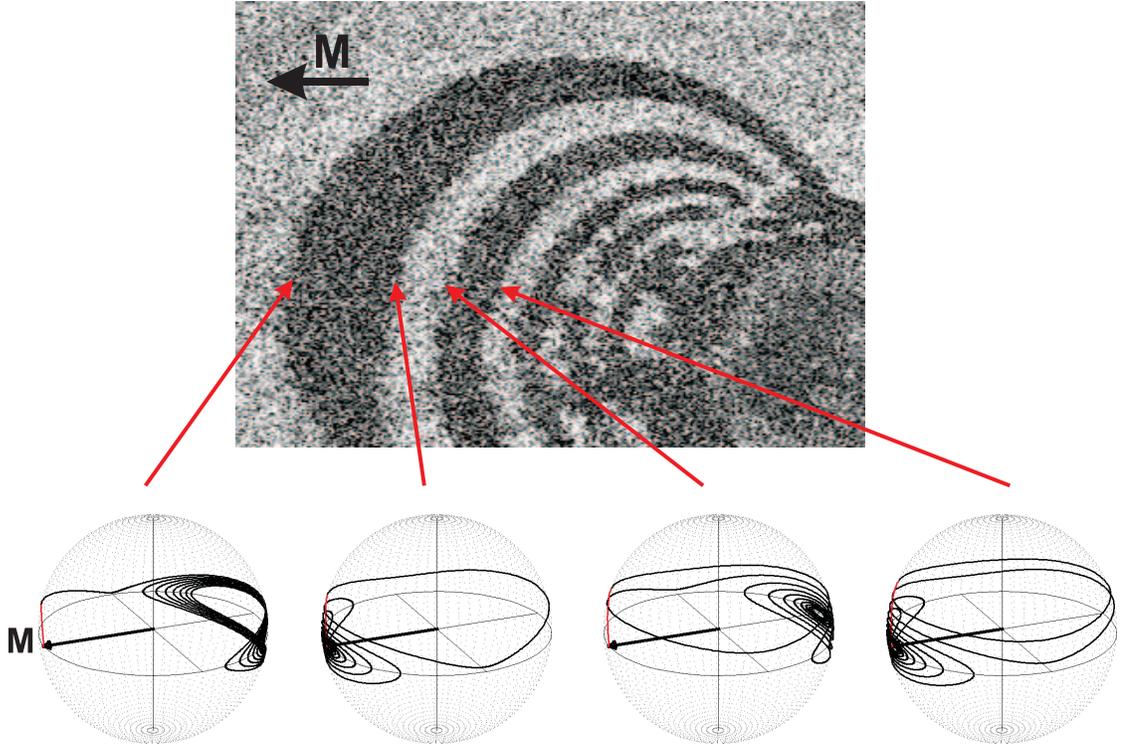}
\caption[Dynamics at the Boundaries in In-plane Media]{Dynamics at
the Boundaries in In-plane Media.} \label{Fig:Boundaries}
\end{figure}

If $K_{\perp}$ is the energy density of the total perpendicular
anisotropy, then the Zeeman energy density deposited in the spin
system by the magnetic field pulse is given by
\begin{equation}
E = K_{\perp}\ sin^2\varphi \label{Eq:EnergyDensity}
\end{equation}
where we have neglected higher order anisotropies because they
turn out to have a negligible effect. The energy $E(\varphi_1)$ to
induce the first switch is given by the energy $K_u$ to surmount
the anisotropy barrier and the damping loss in the precession of
$M$ about $H_{demag}$ by $90^\circ$ to reach the anisotropy
barrier. After that, the magnetization relaxes into the new
direction in the last step of the switching without consuming any
additional energy. To switch $M$ back, a higher energy
E($\varphi_2$) is needed to account for the damping loss in the
additional precession by $180^\circ$. Each additional switching
requires an energy increment $\Delta E_{\nu} = E(\varphi_{\nu}) -
E(\varphi_{\nu-1})$ to compensate for the damping loss in the $\nu
th$ large angle precession by $180^\circ$. This is a truly ideal
situation to measure the dissipation development in large angle
precession. The boundaries along which $M$ has switched are
contour lines of constant Zeeman excitation energy
$E(\varphi_{\nu}) = const$. This is shown in picture
\ref{Fig:Boundaries}.

\begin{figure}[hbt]
\centering
\includegraphics[width=\textwidth]{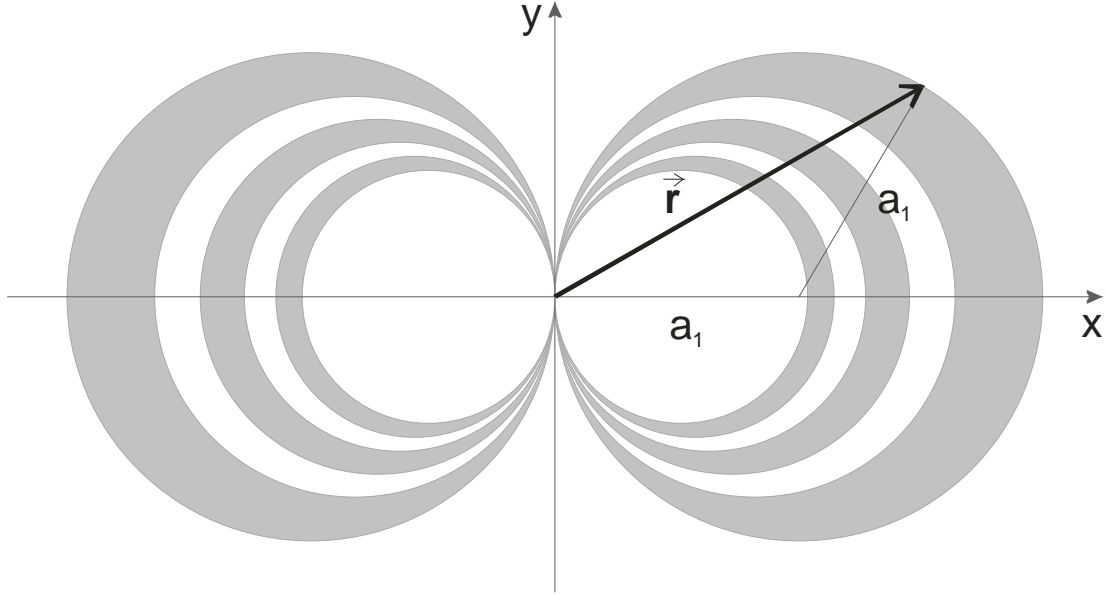}
\caption[Lines of Constant Excitation for very Short Pulses]{Lines
of constant excitation for very short magnetic field pulses.
Because the torque acting on magnetization is proportional to the
sine of the angle between the applied magnetic field and the
magnetization, the lines of constant torque (excitation) are
circles, which are drawn here for several values of the torque.}
\label{Fig:ConstantTorque}
\end{figure}

The contour lines $E=const$ for a small damping and very short
excitation pulse are calculated from $\varphi \propto x/r^2 =
const$ yielding:
\begin{equation}
(\frac{x - a_{\nu}}{a_{\nu}})^2 + (\frac{y}{a_{\nu}})^2 = 1
\label{Eq:ConstantExcitation}
\end{equation}
The contour lines are thus circles of radius $a_{\nu}$ whose
origin is shifted by $\pm a_{\nu}$ on the x-axis, as shown in
figure \ref{Fig:ConstantTorque}.

\begin{figure}[hbt]
\centering
\includegraphics[width=\textwidth]{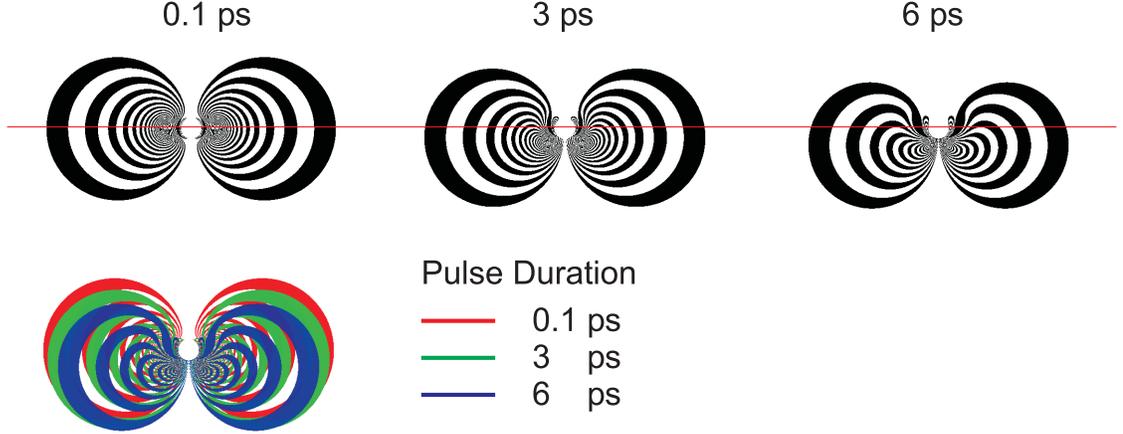}
\caption[Effect of the Pulse Length on the Magnetic
Pattern]{Effect of the Pulse Length on the Magnetic Pattern. The
red line indicates the center of the electron bunch that passes
through the sample. It is also the orientation of the easy axis.
As the bunch length gets longer the anisotropy field becomes
comparable to the magnetic field created by the bunch and the
magnetization rotates also around the easy axis, hence the pattern
gets shifted down. One could imagine using this effect to measure
the bunch length.} \label{Fig:DurationEffect}
\end{figure}

When the excitation pulse is longer, comparable with the
precession time in the anisotropy field, the magnetization not
only goes out of the sample plane but also rotates around the easy
axis. This has the effect of shifting the whole pattern to one
side as shown in figure \ref{Fig:DurationEffect}.

The validity of (\ref{Eq:ConstantExcitation})is proven by figure
\ref{Fig:VA1Circles} and \ref{Fig:VA3Circles} displaying examples
of magnetic switching patterns obtained with ultrathin films
premagnetized along the easy direction along the horizontal
x-axis. We used single crystalline bcc Fe-films grown epitaxially
on a GaAs(001) surface, protected by a capping of 10 ML Au. The
films have been characterized by FMR (yielding g=2.09), and other
techniques \cite{georg:thesis}. The width of the FMR resonance is
found to be independent of film thickness, and increases linearly
with FMR-frequency from 9-70 Ghz, corresponding to a damping
$\alpha=0.004$. Up to ten switches induced by a single electron
bunch can be distinguished, as opposed to at most 4 switches in
previous experiments with thicker Co-films \cite{back:1999}. The
constants $a_{\nu}$ are obtained by fitting circles (table
\ref{Table:InplaneMediaResults}) to the patterns as indicated in
the figures.
\begin{table}[hbt]
\centering \resizebox{\textwidth}{!}{
\begin{tabular}{|c|c|c|c|c|c|c|c|c|}
  \hline
  Switching boundary & 1 & 2 & 3 & 4 & 5 & 6 & 7 & 8 \\
  \hline
  Diameter($\mu m$) for 10 ML Fe/GaAs & 292 & 227 & 200 & 172 & 159 & 139 & - & - \\
  Diameter($\mu m$) for 15 ML Fe/GaAs & 325 & 270 & 238 & 209 & 185 & 167 & 152 & 139 \\
  \hline\multicolumn{9}{l}{}
\end{tabular}
} \caption[Results of Experiments Run on In-plane Magnetic
Media]{Circle diameters of the switching boundaries for the two
samples shown in figures \ref{Fig:VA1Circles} and
\ref{Fig:VA3Circles}.} \label{Table:InplaneMediaResults}
\end{table}

The pattern of figure \ref{Fig:VA1Circles} is produced with an
electron bunch of charge Q = 1.73 nC in a 15 ML Fe-film with
$H_{ani} = 4.72\times10^4\ A/m$ and $H+{demag} = 128\times10^4\
A/m$. The pattern of figure \ref{Fig:VA3Circles} is generated with
Q = 2.1 nC in a 10 ML Fe film. The thinner film exhibits a larger
uniaxial anisotropy field of $H^{ani}$ = 8.21x10$^4$ but a smaller
$H^{demag}$ = 109x10$^4$ A/m compared to the 15 Ml Fe-film. The
magnetic patterns have been imaged 12 weeks after exposure of the
samples to the field pulse by sputtering away the capping layers
of 10 ML Au and then imaging the direction of $M$ in spin resolved
scanning electron microscopy (Spin-SEM). $M$ is either parallel
(light grey) or antiparallel (black) to the horizontal easy
direction.

\section{Potential problems}
\begin{SCfigure}[][hbt]
\includegraphics[]{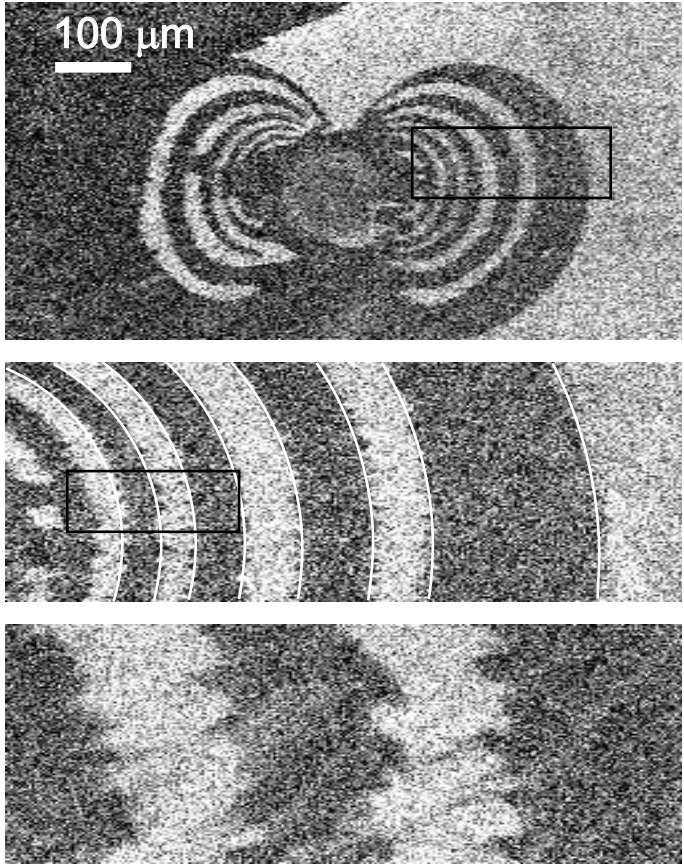}
\caption[Magnetic Pattern Generated in 15 ML
Fe/GaAs(001)]{Magnetic pattern generated with a single electron
bunch in a 15 ML Fe/GaAs(001) epitaxial bcc Fe-film. The magnetic
image is obtained by Spin-SEM after sputtering off the capping
layer of 10 ML Au. Prior to the field pulse, $M$ is aligned
horizontally to the right shown in light grey. The regions were
$\vec M $ has switched to the left are shown dark. On the left and
lower left side, the pattern is disturbed by motion of domain
walls after exposure. In the center, a large spot due to beam
damage appears. The framed part is shown at greater magnification
in the middle with the fitted circles and at the bottom at still
larger magnification exposing zig-zag domain boundaries.}
\label{Fig:VA1Circles}
\end{SCfigure}

\begin{SCfigure}[][hbt]
\includegraphics[]{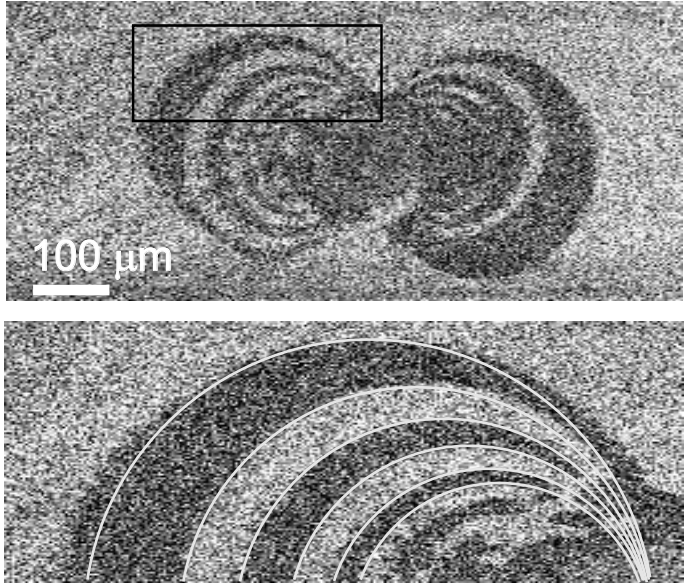}
\caption[Magnetic Pattern Generated in 10ML Au/10 ML
Fe/GaAs(001)]{Magnetic pattern generated with a single electron
bunch in an  10ML Au/10 ML Fe/GaAs(001) epitaxial Fe-film,
otherwise like figure \ref{Fig:VA1Circles}. No after-pulse motion
of domain walls occurred in this sample, but the pattern is less
regular than with 15 ML}\label{Fig:VA3Circles}
\end{SCfigure}

One type of problem encountered in determining the contour lines
\ref{Eq:ConstantExcitation} is is due to rugged
zig-zag-transitions between regions of opposite $M$. Such zig-zag
domain walls are displayed with high spatial resolution in the
bottom section of figure \ref{Fig:VA1Circles}. The switching leads
initially to the unfavorable "head-on" position of $M$ when a
contour line runs $\perp$ to the x-axis. As noted before
\cite{back:1999}, the head on-transitions relax later into the
longer, but more favorable zig-zag domain walls. The location of
the switching transition is the average over the zig-zag-walls.

A second type of uncertainty arises from the fact that the samples
are soft-magnetic with a coercivity of $1-2\ kA/m$ only. This
means that domains may easily shift, e.g., in accidental magnetic
fields. Apparently, domain wall motions occurred after exposure
and deformed the left side of the pattern of figure
\ref{Fig:VA1Circles} while on the right side, the pattern appears
to be undisturbed.

A third problem is the damage caused by the high energy electron
bunch in the sample. With the semiconducting GaAs substrate we
observe larger damage compared to metallic buffer-layer substrates
used in prior experiments \cite{back:1999, tudosa:2004}. The
damage may be attributed to the electric field $E_p = c\times B_p$
running perpendicular the magnetic field $B_p$ of the pulse. $E_p$
is not rapidly screened in a semiconductor, resulting in
electrostrictive deformation of the GaAs-template responsible for
the uniaxial magnetic anisotropy of the Fe-film. The permanent
beam damage is delineated by a halo around the location of beam
impact at $r \leq 50\ \mu m$. Although the halo is below the
distances of the measurable switching events, it cannot be
excluded that the magnetic anisotropy is affected transiently even
at larger distances by the electrostrictive chock of the template.

With the 10 ML Fe-film, the domain pattern is less regular
compared to the 15 ML film. This must be due to larger local
variations of the magnetic properties in the thinner film at the
length scale of $100\ \mu m$ as the irregularities repeat
themselves in different exposures.

\section{Interpretation of the results}
\begin{SCfigure}[1.2][hbt]
\includegraphics[]{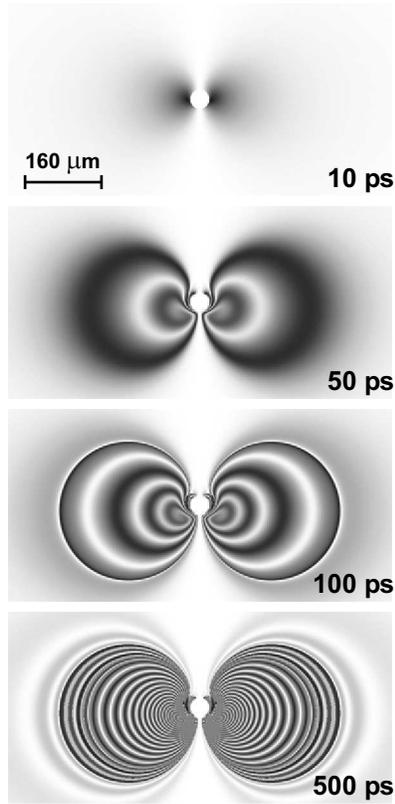}
\caption[Uniform Mode Pattern Snapshots]{Snapshots of the
magnetization projection on the anisotropy axis for $10ML Fe/GaAs$
sample using the damping measured with FMR. Time and distance
scale are indicated, time zero being the time the front of the
electron bunch enters the sample. Initially the sample is
uniformly magnetized (white) in one direction along the anisotropy
axis. The gray scale indicates the projection on that axis, white
being the initial direction and black the opposite. Although the
intermediate snapshots look similar to picture in figure
\ref{Fig:VA3Circles} the last snapshot at $500\ ps$ (close to the
end of the relaxation) has many rings not appearing in the
experimental image. Some other additional damping prevents the
sample to relax into the pattern shown in the last snapshot.}
\label{Fig:VA3FMRdampingDynamics}
\end{SCfigure}

\begin{figure}[hbt]
\centering
\includegraphics[]{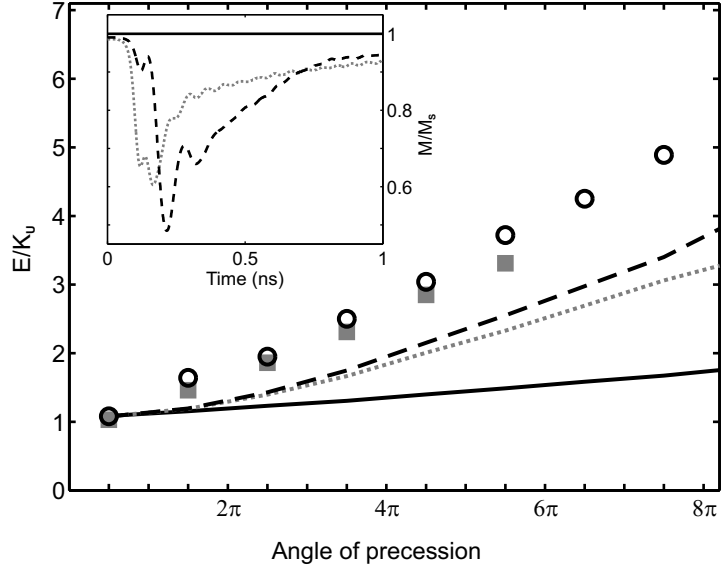}
\caption[Energy Deposited in the Spin System]{Energy deposited in
the spin system in units of the uniaxial in-plane anisotropy
constant $K_u$ vs polar precession angle. Data points are for 10
Fe-ML(squares) and 15 Fe-ML(circles). The simulations are with the
FMR Gilbert damping $\alpha=0.004$ and no magnon scattering (---),
and for 10 Fe-ML ($\cdots$) and 15 Fe-ML (-\,-) including magnon
scattering. The inset shows the relative saturation magnetization
$M(t)/M_s(0)$ where t is the time after an exciting field pulse of
amplitude $0.24\times10^6\ A/m$ for 10 Fe-ML ($\cdots$) and
$0.175\times10^6\ A/m$ for 15 Fe-ML (-\,-). But without magnon
scattering(---), $M(t)/M_s = 1$} \label{Fig:EnergySwitch}
\end{figure}

If the energy (\ref{Eq:EnergyDensity}) required for the onset of a
new switch is plotted in units of $K_u$ vs the angle of precession
of $M$, one obtains the universal switching diagram shown in
figure \ref{Fig:EnergySwitch}. The switching diagram is
independent of the magnetic parameters of the films, but depends
somewhat on film thickness as apparent with increasing number of
switches. The first switch requires the reduced energy $E \approx
1$, compatible with the small damping observed in FMR. The small
dissipation contribution in the first precessional switch explains
the difficulty to determine it with conventional magnetic field
pulses inducing only one switch. Yet already with the second
switch, the additional precession by $180^\circ$ requires much
more energy than what results from FMR damping, see figure
\ref{Fig:VA3FMRdampingDynamics}. The loss in the higher switches
is nearly an order of magnitude larger than the dissipation
extracted from FMR.

The increase of the energy loss after the first switch shows that
dissipation of spin angular momentum increases with time. Such
delayed dissipation is characteristic for the Suhl instability
\cite{suhl:1957}, which is the transfer of energy from the uniform
precession mode with wavevector $k=0$ to higher spin wave modes
with $k \neq 0$. The transfer of energy, induced by non-linear
interactions owing to $H_{demag}$ and $H_{ani}$, takes time
because the numbers of excited non-uniform spin waves grow
exponentially with time.

A quantitative theory for the dissipation caused by the Suhl
instability has been developed recently \cite{dobin:2003}. In the
inset of figure \ref{Fig:EnergySwitch} we show simulations for an
area of $1\ \mu m \times 1\ \mu m$ of the 10 and 15 ML Fe-film
with a respective pulse amplitude that completes the first switch.
It demonstrates one important consequence of the generation of
higher spin wave modes, namely the decrease of the space averaged
order parameter $M/M_s$ with time. It is seen that $M/M_s$
decreases sharply $\approx\ 50\ ps$ after the field pulse, and
recovers slowly through spin lattice relaxation of the spin waves.
Now, from the time $t_{\nu}$ after the field pulse at which the
last change of sign of $M_x$ occurs, we know the moment in time at
which the energy consuming part of the switch $\nu$ is terminated.
With 15 ML-Fe we obtained $t_1,\ldots,t_8$ = 40, 115, 155, 195,
235, 270, 310, 360 ps respectively. Large dissipation is observed
only \emph{after} the first switch. This agrees with the $50\ ps$
delay seen in the development of spinwave scattering. Furthermore,
the fluctuations of $M/M_s$ in time and space manifest themselves
through increasingly random switching as the angle of precession
grows. Another characteristic of the Suhl instability concerns the
film thickness. To conserve energy and momentum, the effective
scattering of the uniform mode requires the excitation of low
energy spin waves. The phase space for such low energy, long
wavelength modes decreases with film thickness, and this explains
the experiment as well as the simulation both showing smaller
dissipation as the number of ML is reduced. Hence there is no
reasonable doubt that the Suhl instability contributes
significantly to the dissipation observed in the experiment.

However, as apparent from figure \ref{Fig:EnergySwitch}, the
simulations are short by a factor 2 to fully account for the
observed damping. Surface roughness is known to contribute to the
damping . However, the detailed analysis based on
\cite{dobin:2004} shows that the surface roughness measured on the
present films \cite{georg:thesis} is not enough to explain the
observations, and furthermore should show up in FMR as well. We
therefore have to conclude that additional, so far unknown
relaxation mechanisms must be active in large angle precession of
the magnetization as well.

\chapter{Experiments with Perpendicular Magnetic Media}
\chaptermark{Magnetic Media}

\section{Introduction}
The dynamic response of a spin system reveals itself in the image
of the final magnetic switching pattern generated by very short,
intense and gaussian shaped magnetic field pulses. Our experiment
shows that under these extreme conditions, precessional switching
in magnetic media supporting high bit densities no longer takes
place at well-defined field strengths; instead, switching occurs
randomly within a wide range of magnetic fields. We attribute this
behavior to a momentary collapse of the ferromagnetic order of the
spins under the load of the short and high-field pulse.

\section{Experimental details}

\begin{table}[htb]
\begin{center}
\begin{tabular}{|l|c|}
  \hline
  \textbf{Property} & \textbf{Value} \\
  \hline
  Energy of electrons & $28\ GeV$ \\
  Number of electrons & $(1.05\pm0.05)\times10^{10}$ \\
  Cross section (gaussian profile with $\sigma_x\times\sigma_y$) &
  $10.8\times7.4\ \mu m$ \\
  Pulse length (gaussian profile with $\sigma_z$) & $0.7\ mm$ \\
  Pulse duration ($\sigma_t=\sigma_z/c$) & $2.3\times10^{-10}\ s$ \\
  Peak magnetic field $B_p(Tesla)$ at distance $R(\mu m)$ &
  $B_p=54.7/R$\\
  \hline
\end{tabular}
\end{center}
\caption[Characteristics of the Electron Beam]{Some
characteristics of the electron beam used in the experiment.}
\label{Table:BeamCharacteristics}
\end{table}

The experiment is based on the unique magnetic field pulses
generated in solids by a passing-through finely focused high
energy electron bunch as shown previously in chapter
\ref{Ch:EMeffects}. The magnetic field lines are the circular
field lines generated by a straight electric current. The peak
field strength is calculated from Amp\`eres law valid for
distances greater than the size of the beam. A summary of the
electron beam characteristics is shown in table
\ref{Table:BeamCharacteristics}. The electron beam has a gaussian
density profile in all three directions and we use the notation
$B_p$ for its peak magnetic field intensity.

With the films magnetized perpendicular to the film plane, the
magnetic field $B$ and magnetization $M$ are orthogonal
everywhere. This is the optimum geometry to induce a precessional
motion of $M$ about the magnetic field.
\begin{figure}[!htb]
  \begin{center}
  \includegraphics[width=\textwidth]{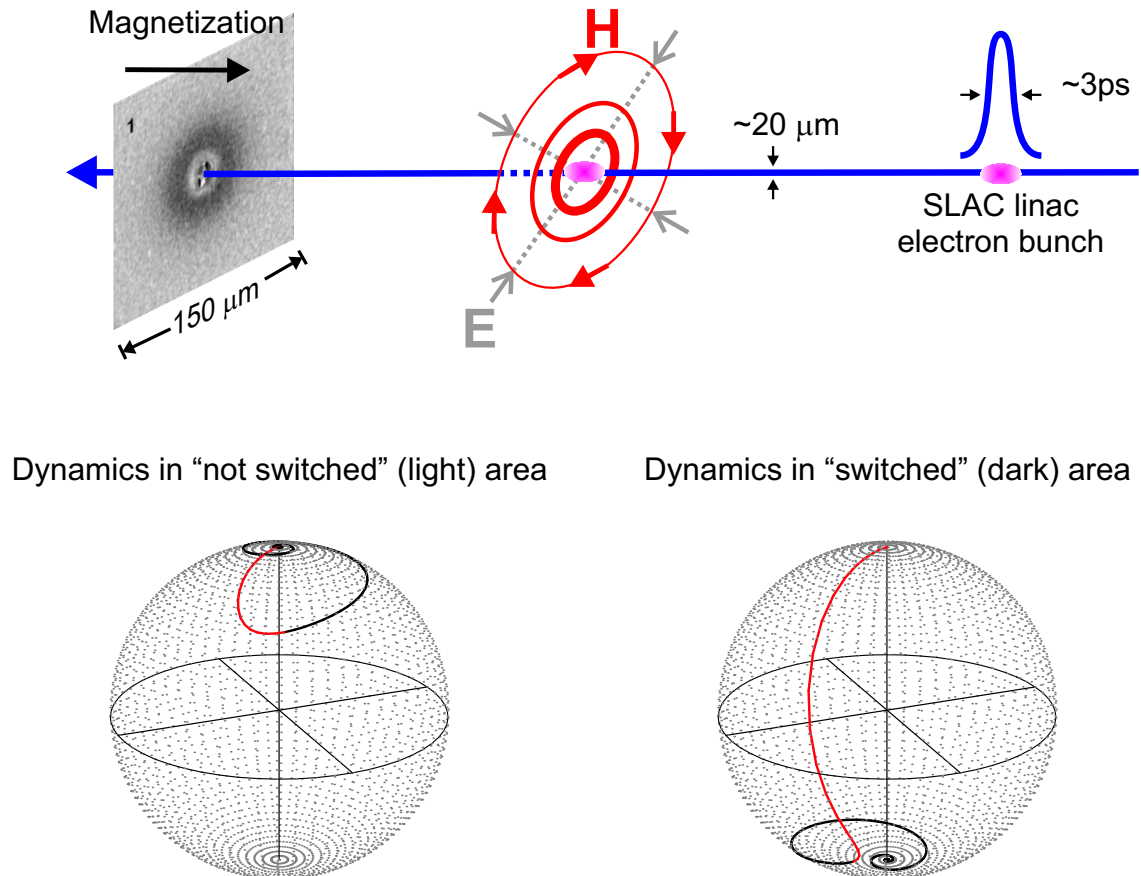}\\
  \caption[Illustration of Experiment with Perpendicular Media]{
  An illustration of the experiment with with the dynamical path of the magnetization
  for two locations on the sample with different magnetization switching results. Outer
  areas that are light-colored are not exposed to field of enough intensity
  to switch them, whereas the dark areas precess during the field pulse( red color)
  over the magnetically hard plane( the equator), which is also the sample plane. North and south
  pole are the equilibrium directions of the magnetization.}
  \label{Fig:PerpendicularExperiment}
  \end{center}
\end{figure}

Once M has precessed about $B$ by an angle large enough to cross
the magnetically hard plane of the sample, it will continue to
relax (spiral) by itself into the opposite direction. In the end
it has switched from one easy direction into the opposite easy
direction. If the magnetic field terminates before $M$ has reached
the hard plane, $M$ is expected to relax back to its original
perpendicular direction, hence no switch is observed. The
condition for switching is that the angle of precession
$\varphi\geq\pi/2$, where $\varphi \propto B_p\sigma_t$,
$\sigma_t$ describing the gaussian temporal profile of the beam.
We thus obtain the switching condition $B_p\sigma_t\geq const$.

For our case of uniaxial perpendicular magnetocrystalline
anisotropy, no demagnetization field is generated that could
continue to provide a torque propelling $M$ uphill over the hard
plane, both demagnetization and anisotropy fields having the same
direction. This is different with uniaxial in-plane magnetized
materials where the field pulse causes $M$ to precess out of the
plane of the film and a demagnetization field along the surface
normal is generated, resulting in demagnetization being
perpendicular to the anisotropy field. Precession about this
demagnetization field can lead to multiple in-plane reversals of
$M$ after the field pulse has ceased to exist \cite{back:1999}.
Therefore, perpendicular magnetized materials are unique in
providing the opportunity to observe any fields different from the
anisotropy field that might have been generated by the interaction
of the spins with the field pulse.

Generally, precessional magnetization switching is interesting
because it is faster by an order of magnitude and uses much less
energy to reverse the magnetization $ \vec M $ compared to the
traditional methods of switching in use today
\cite{gerrits:2002,schumacher:2003,kaka:2002}. Furthermore, $ \vec
M $ may be switched ``back" (by rotating another $180^\circ$ in
the same direction) without changing the polarity of the magnetic
field pulse simply by applying the switching pulse again. This
characteristic is used here to test the reversibility of the
switching by applying identical pulses several times. If the
switching is deterministic, the second pulse should restore the
magnetic state present before the first pulse, and so on. All
patterns produced by an odd number of pulses should be identical
and so should be all even pulse patterns. In contrast, we show
here that the switching probability is independent of whether or
not a particle has switched before and is therefore dominated by a
stochastic process. As a consequence, only the average motion of $
\vec M $ can be described by the conventional coherent precession
model. This remained hidden in previous work \cite{back:1998}
where reversibility was not tested.

To test this switching, films of perpendicular granular magnetic
recording media of the CoCrPt-type developed for high-density
magnetic recording \cite{weller:2003} were used. The main
condition for high-density recording is that the grains are
decoupled so that the medium can sustain narrow transitions
between "up" and "down" bits.
\begin{figure}[!htb]
  \begin{center}
  \includegraphics[]{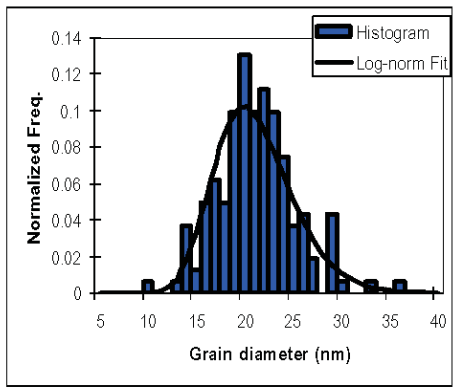}\\
  \caption[Grain Size Distribution of Perpendicular]{ Grain size
  distribution of perpendicular granular
  magnetic recording media as determined by X-ray diffraction for
  sample $CoCrPt\ I$ in table \ref{Table:PerpMediaResults}}
  \label{Fig:SizeDistribution}
  \end{center}
\end{figure}
The decoupling of the grains occurs through segregation of Cr to
the grain boundaries induced by deposition at elevated
temperature. The grain size was determined by X-ray diffraction to
be $20 \pm 5\ nm$, while the film thickness was $14\ nm$. This
grain size is so small that the magnetic field is homogeneous over
the grain size to better than $0.1\%$. We can then assume that the
switching of a grain occurs in a homogeneous applied field. The
films were protected from atmospheric corrosion by a cover of
$1.5\ nm$ Pt and used glass substrate with appropriate thin buffer
layers, with and without adding a soft magnetic underlayer such as
needed in perpendicular recording \cite{weller:1999}.

\section{Magnetic Patterns}
\begin{SCfigure}[][!htb]
  \includegraphics[]{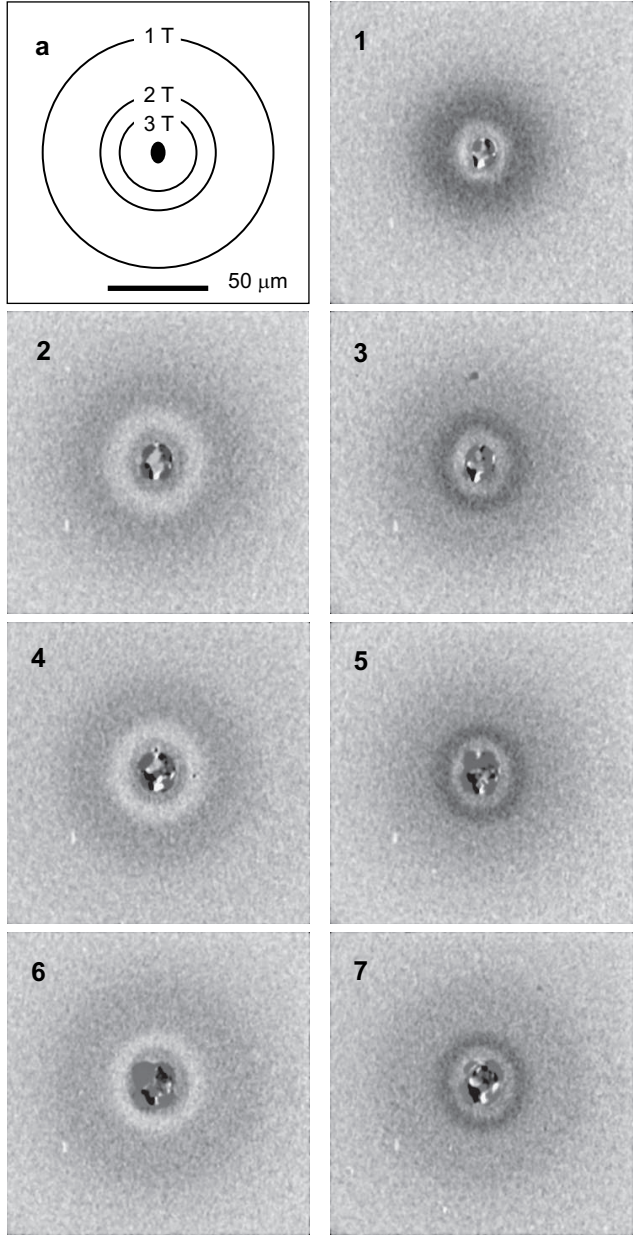}\\
  \caption[Magneto-optic Patterns of Magnetization]{ Magneto-optic
  patterns of magnetization for $CoCrPt\ I$ in table \ref{Table:PerpMediaResults}. Diagram at
  top left shows contour lines of constant peak magnetic field $B_p$
  with area of the electron beam focus in the center around which
  there is localized beam damage. The numbers on subsequent panels
  indicate the number of electron bunches (shots) that passed
  through the sample. Grey contrast is such that the outer light
  region corresponds to $M$ in the initial "up" state. As darkening
  intensifies, $M$ has switched increasingly to the "down" direction.
  The contrast in the central region at $R < 10\ \mu m$ is due to beam damage.}
  \label{Fig:MultipleShots}
\end{SCfigure}

Before exposure, the samples were magnetized perpendicular to the
film plane into what we shall call the "up" direction. We recorded
patterns on the same sample corresponding to a single shot
(pulse), a pattern corresponding to two shots at the same
location, and so on, up to seven consecutive shots per pattern.
The time separation of consecutive shots was $1\ s$. Three weeks
after exposure, the perpendicular component of $M$ was imaged by
polar magneto-optic Kerr microscopy. A set of such images is shown
in figure \ref{Fig:MultipleShots}. The spatial resolution of Kerr
microscopy is $1\ \mu m$, so that we integrate over $\approx
2,500$ grains. The spot in the center of the pattern is due to
beam damage. It extends to roughly twice the beam focus. The
increase of the damaged area with the number of shots is due to
beam jitter, which is estimated to be $\pm 2\ \mu m$ per shot
only. The grey scale of the images is such that the light regions
near the edge of the frames correspond to the initial "up" state.
Darkening indicates that particles in the regions have
increasingly switched to the "down" direction.

It is evident that switching occurs along circular contour lines
$B_p = const$, with the contrast changing gradually over a
distance of tens of $\mu m$, rather than abruptly. The switching
is not reversible, because the second pulse does not return M to
the initial "up" direction. For odd shot numbers, the dark ring
where $M$ has switched from "up" to "down" narrows with increasing
number of shots, whereas the outer grey zone corresponding to
partially switched $M$ expands. With an increasing number of even
shots, the central light ring narrows and the outer grey zone
expands. This switching behavior is characteristic of a stochastic
process. Starting with a homogeneous magnetic "up" state, it takes
only seven shots to create a random distribution of magnetization
directions throughout the large grey zone, where $M$ is "up" in
some grains but "down" in others.

\section{Interpretation of the results}
To describe the stochastic switching, consider that the angle
$\varphi$ of precession could be short of $\pi/2$ needed for
switching. The lacking precessional angle may be supplied by
random torques or by random initial conditions. If we assume that
the probability $p$ of such stochastic events is gaussian and can
be expressed as the probability density of an additional magnetic
field $\Gamma$:
\begin{equation}
p(\Gamma) = \frac {1}{\Delta B \sqrt{2
\pi}}\,e^{-\frac{\Gamma^2}{2 (\Delta B)^2}}
\end{equation}
The relative magnetization $M$ (the magnetic order parameter) then
depends on the pulse peak amplitude $B_p$ and is given by the
fraction of particles that switch minus the fraction of particles
that do not switch:
\begin{equation}
M(B_p)= \int\limits_{-\infty}^{B_1-B_p} p(\Gamma)\, d\Gamma -
\int\limits_{B_1-B_p}^{+\infty} p(\Gamma)\, d\Gamma
\end{equation}

\begin{SCfigure}[][!htb]
  \includegraphics[]{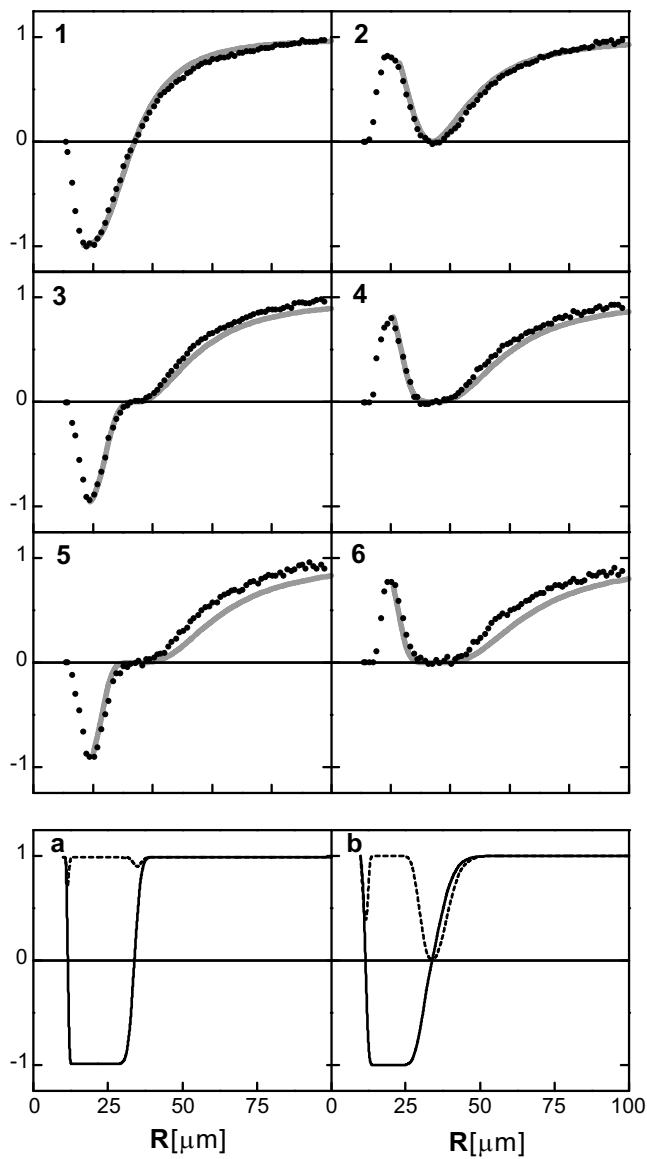}\\
  \caption[Radial Profile of Magnetic Order
  Parameter]{Magnetization( order parameter) radial profile $M_n(R)$ for figure
  \ref{Fig:MultipleShots}. \textbf{[a]} Data points
  are from averaged radial cuts of the Kerr images of figure \ref{Fig:MultipleShots}, setting
  $M(100\ \mu m) = +1$ and $M(20\ \mu m) = -1$. The grey lines are generated
  from $M_n(R) = M_1^n(R)$. \textbf{[b]} Calculated $M_1(R)$ (---) and
  $M_2(R)$ (-\,-\,-) with the observed easy-axis dispersion of $5.5^\circ$ \cite{weller:2003},
  showing $M_2(R)\,\simeq\, M_0=+1$ in gross contradiction to the
  experiment. \textbf{[c]} Calculated $M_1(R)$ (---) and $M_2(R)$ (-\,-\,-)
  assuming the excitation of the uniform precession mode (figure )
  corresponding to $K_uV/kT = 40$ (anisotropy/thermal energy ratio), where $K_u$ is the uniaxial crystalline
  anisotropy constant, $V$ the volume of a grain, $k$ the Boltzmann
  factor and $T$ the temperature \cite{weller:1999}.}
  \label{Fig:MultipleShotsProfile}
\end{SCfigure}

By choosing the average switching pulse amplitude $B_1 = 1.70\ T$
and its variance $\Delta B = 0.59\ T$ we obtain the order
parameter $M_1(B_p)$ after the first shot generated from the
uniform initial state $M_0 = +1$. Substituting the variable $B_p$
by $R$ yields the radial dependence $M_1(R)$ of the magnetization
after the first shot. $M_1(R)$ is plotted for $R > 20\ \mu m$ as a
solid line on top of the data points in panel 1 of figure
\ref{Fig:MultipleShotsProfile}. It agrees with the experimental
data. The increase of the observed $M_1(R)$ at $R < 20 \mu m$
indicates the onset of the second switch where $M$ precessed by
$\varphi \geq 3\pi/2$. We do not analyze this second switch
because it occurs close to the beam damage area. The easy axis
dispersion has been determined with X-ray diffraction to be
$5.5^{\circ}$ (full width half maximum). This generates a
distribution of switching fields which is much smaller than
$\Delta B$.

The other distributions of $M_n(R)$ shown as solid lines in panels
$2-6$ of figure \ref{Fig:MultipleShotsProfile} are generated by
multiplication $M_n(R) = M_{n-1}(R)\times M_1(R) = M_1^n(R)$. This
accounts very well for the features observed in the experimental
data shown as points, despite the fact that raising $M_1(R)$ to
the n-th power enhances errors. Experimental errors are caused by
beam jitter, variations in the number of electrons per bunch, and
the uncertainty in the extrapolation of $M_1(R)\rightarrow +1$. At
any rate, multiplicative probabilities are the signature of a
random variable. Therefore, this analysis reveals that a
memoryless process, usually referred to as a Markov uniform
stochastic process, dominates the switching.

Different samples have been studied within a range of magnetic
parameters such as saturation magnetization and magnetic
anisotropy. Information is also available
\cite{siegmann:1995,back:1998} about switching with older media
types such as $Co/Pt$ magnetic multilayers and $CoPt$ alloys,
designed for thermomagnetic recording and exhibiting exchange
coupled grains and larger anisotropy. The relative broadening $
\Delta B/B_o $ of the first switching transition has been found to
be of similar magnitude than observed here with the largely
uncoupled grains of $CoCrPt$. The conclusion emerging from all
experiments is that stochastic switching is a general feature of
ultrafast precessional magnetization reversal. The results of the
experiments are presented in the table
\ref{Table:PerpMediaResults}.

\begin{table}[htb]
\centering \resizebox{\textwidth}{!}{
\begin{tabular}{|c|c|c|c|c|c|c|c|c|}
  \hline
  $\mathbf{Media Type}$ & $\mathbf{M_s}$ & $\mathbf{H_c}$ & $\mathbf{H_{eff}}$ &
  $\mathbf{Dispersion\ of}$ & $\mathbf{Grain Size}$ & $\mathbf{B_0}$ & $\mathbf{\Delta B}$ & $\mathbf{2\Delta B/B_0}$ \\
  & $(T)$ & $(kA/m)$ & $(kA/m)$ & $\mathbf{easy\ axis}({}^\circ)$ & $(nm)$ & $(T)$ & $(T)$ & \\
  \hline\hline
  CoCrPt I alloy& $0.65$ & $483$ & $684$ & $5.5$ & $20.6\pm4$ & $1.7$ & $0.59$ & $0.7$ \\
  \hline
  CoCrPt II alloy& $0.58$ & $171$ & $398$ & $11$ & $19.9\pm5.1$ & $1.55$ & $0.57$ & $0.74$ \\
  \hline
  CoCrPt II alloy& $0.58$ & $203$ & $398$ & $12.7$ & $19.9\pm5.1$ & $1.55$ & $0.51$ & $0.66$\\
  on soft underlayer& & & & & & & & \\
  \hline
  CoPt(95) alloy\cite{siegmann:1995} & $0.47-0.58$ & $117-178$ & $1380-1600$ & $\approx 0$ & - & - & - & $0.6$ \\
  \hline
  CoPt(98) multilayers\cite{back:1998}& - & - & $1274-2548$ & $0-20$ & $15$ & - & - & $0.2-0.8$ \\
  \hline
  \multicolumn{9}{c}{}\\
  \multicolumn{9}{l}{$M_s$ - saturation magnetization; $H_c$ - coercive field; $H_{eff}$ - effective anisotropy
  field}\\
  \multicolumn{9}{l}{$B_0$ - average switching field; $\Delta B$ - variance of switching field}\\
  \multicolumn{9}{l}{$2\Delta B/B_0$ - relative width in switching fields}\\
  \multicolumn{9}{l}{$SUL$ - soft magnetic underlayer ($FeCoB$), see figure \ref{Fig:SUL}}\\
  \multicolumn{9}{l}{}
\end{tabular}
} \caption[Results of Experiments Run on Perpendicular Magnetic
Media]{Some older and present results of running the electron beam
through different magnetic media. In all cases the relative
transition width is similar} \label{Table:PerpMediaResults}
\end{table}

\begin{figure}[htb]
  \centering
  \includegraphics[]{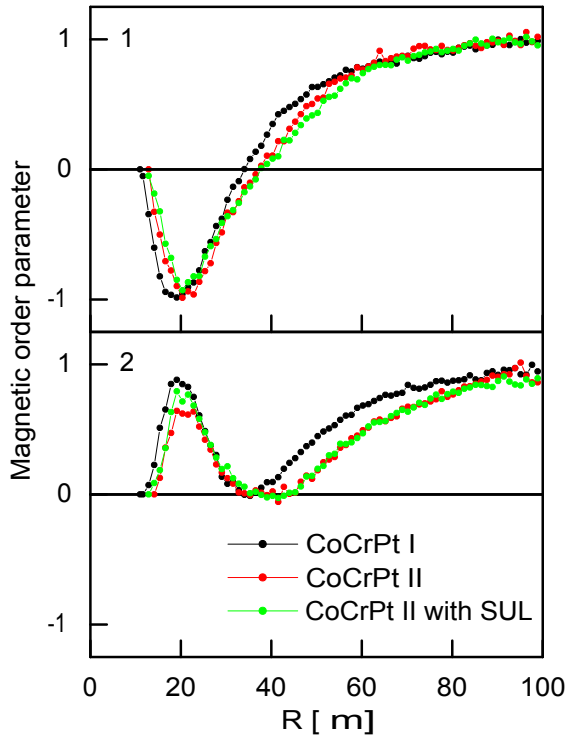}\\
  \caption[Radial Profile of One and Two Shot Locations]{Radial
  profile of one and two shot locations for different types of samples.}
  \label{Fig:OneTwoShotsProfile}
\end{figure}

The magnetic signal obtained by Kerr microscopy for samples
$CoPtCr\ I-II$ is shown in figure \ref{Fig:OneTwoShotsProfile}.
With samples $CoPt(95)$ and $Co/Pt(98)$ the magneto-optic patterns
are not of sufficient quality to permit similar quantitative
analysis. The earlier patterns are asymmetric due to oblique
illumination, unspecified magneto-optic contrast enhancement, and
moreover, the contour lines of constant magnetic field were quite
elliptical.
The difference in figure \ref{Fig:OneTwoShotsProfile} between
samples $CoCrPt\ I$ and $CoCrPt\ II$ may be accounted for by the
difference in the values of the anisotropy. It should be noted,
that the magnetic anisotropy has a reduced influence in
precessional reversal. Increasing it by a factor of $1.72$ on
going from $CoCrPt\ II$ to $CoCrPt\ I$ increases the average
switching pulse amplitude by a factor of $1.10$ only. In the
hysteretic switching commonly practiced one expects that the
switching field is proportional to the anisotropy.

\begin{figure}[htb]
  \centering
  \includegraphics[width=0.5\textwidth]{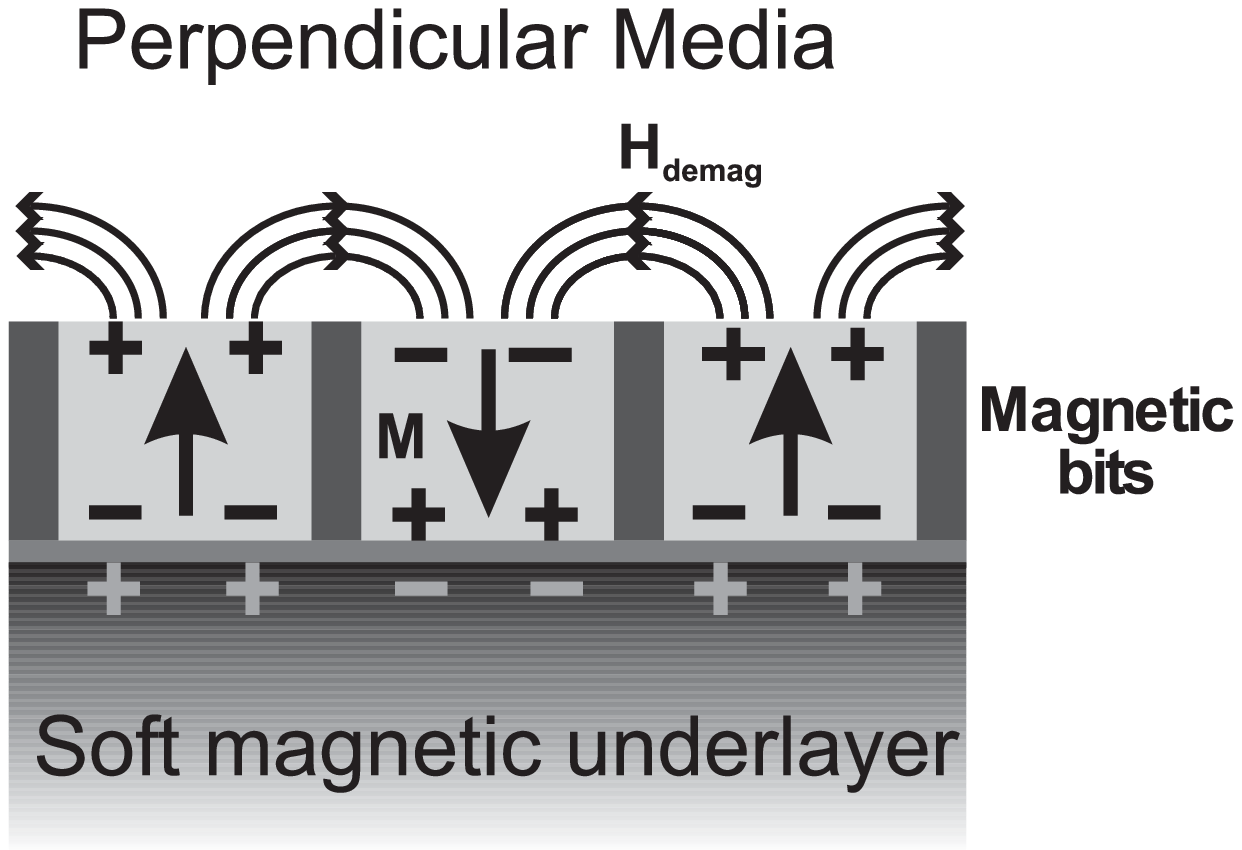}\\
  \caption[Soft Magnetic Underlayer in Perpendicular Media]{Illustration of
  the role of soft magnetic underlayer in perpendicular media. Magnetization (M)
  interacts with its image in the soft magnetic underlayer increasing
  the magnetostatic interaction of the grains by a factor of about two.}
  \label{Fig:SUL}
\end{figure}
The soft-magnetic underlayer( SUL) is needed in perpendicular
media to enhance the perpendicular field of the writing head pole.
This happens by inducing an image magnetic pole in the underlayer.
Without SUL, there is no high-density perpendicular magnetic
recording. The magnetization also interacts with its own induced
magnetic image in the soft magnetic underlayer (see figure
\ref{Fig:SUL}), yet this additional magnetostatic interaction does
not have a significant effect on the width of the transition. The
curves in figure \ref{Fig:OneTwoShotsProfile} for the sample
$CoCrPt\ II$ with and without SUL are nearly identical, confirming
the theoretical conjecture that magnetostatic interactions are not
responsible for the transition width.

\begin{figure}[hbt]
  \centering
  \includegraphics[]{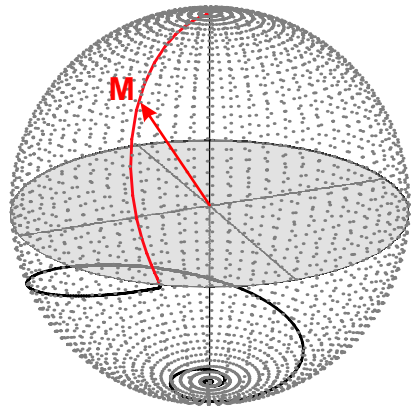}\\
  \caption[Average Magnetization Dynamics at the Switching Boundary]{The mean
  path followed by the magnetization with a pulse amplitude $B_p$ close to $B_0$.
  The portion in red represents the pathway during the magnetic field pulse.}
  \label{Fig:DynamicsSwitchingBoundary}
\end{figure}

The angular velocity of magnetization precession is independent of
grain size and it is proportional to the effective magnetic field
at the grain location. If all grains precess with the same angular
velocity, they will arrive at the hard plane simultaneously and
the dipolar coupling field vanishes just at the critical moment in
time. The mean path followed by the magnetization, calculated from
the Landau-Lifshitz-Gilbert equation assuming a damping parameter
of the precession equal to 0.3, is shown in figure
\ref{Fig:DynamicsSwitchingBoundary}.

\begin{figure}[!h]
  \begin{center}
  \includegraphics[width=0.5\textwidth]{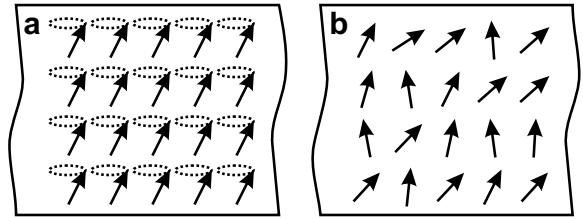}\\
  \caption[Spin Motion in a Magnetic Grain]{Spin motion in a magnetic grain. \textbf{[a]} The uniform precession
  mode of the spins with wave-vector $q = 0$. The excitation of this mode
  determines the long-term stability of the magnetization direction in
  the grain \cite{weller:1999}. \textbf{[b]} A moment in time with non-uniform excitation of the
  spins. At ambient temperature, these excitations have small amplitude,
  which however dramatically increases after the field pulse has been
  applied. Sizeable exchange fields are generated by the angles between
  neighboring spins that can account for the random torques operating
  after the magnetic field pulse.}
  \label{Fig:UniformNonuniformMode}
  \end{center}
\end{figure}

To explore causes of the randomness, we carried out various
calculations using the Landau-Lifshitz-Gilbert equation,
introduced in chapter \ref{Ch:MagDynamics}. Theoretical results
that explore two hypothetical sources of the randomness are shown
in figure \ref{Fig:MultipleShotsProfile}. Panel \textbf{b.}
demonstrates that static dispersion of the easy axis of
magnetization in the decoupled grains cannot produce anything but
deterministic switching reversing M to the original state $M_0 =
+1$ in the second shot. This is in gross contradiction to the
experiment. Panel \textbf{c.} explores thermal excitations in
terms of the uniform precession mode (illustrated in figure
\ref{Fig:UniformNonuniformMode}). The degree to which the uniform
mode is excited is known from the long-term stability of the
magnetic bits \cite{weller:1999}. It induces randomness in the
direction of $M$ before the arrival of the field pulse and indeed
generates dispersion of $M(R)$, but the dispersion is much too
small to explain the data.

The grain size distribution is important for thermal fluctuations
of the magnetization direction. The amplitude of such fluctuations
depends on the ratio between anisotropy energy and thermal energy
$K_uV/kT$ ($K_u$ is the anisotropy energy density and V is the
volume of the grain). All our samples have similar grain size
distribution and figure \ref{Fig:SizeDistribution} shows it for
$CoCrPt I$ sample.

The effect of heating of the sample by the electron pulse can be
asserted without calculation. The supersonic heat wave emerging
from the point of beam impact requires $10^{-9}\ s$ to travel $1\
\mu m$. However, the switching at $R> 20\ \mu m$ is already
completed at that time. Similarly, magneto-static coupling between
the grains cannot explain the variance of the switching fields
because it is small at the end of the field pulse when the spins
have precessed close to the hard plane.

\begin{figure}[!h]
  \begin{center}
  \includegraphics[width=0.8\textwidth]{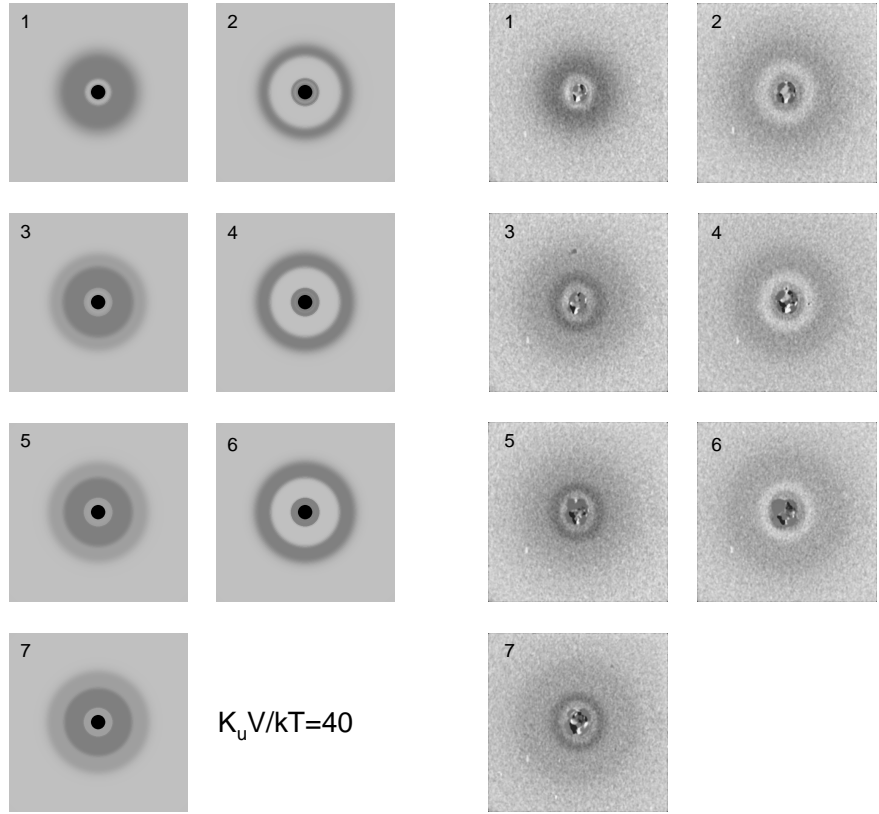}\\
  \caption[Broadening due to thermal fluctuation for $K_uV/kT = 40$]
  {Transition broadening due to thermal fluctuation for $K_uV/kT = 40$. The left side
  shows the simulation and the right side the experimental images. }
  \label{Fig:PerpendicularComparison}
  \end{center}
\end{figure}
The thermally activated homogeneous precession mode generates
deviations of $ \vec M$ from the ideal initial direction. Using
the resulting variable initial conditions prior to the arrival of
the pulse one can explain the width of the switching transition by
assuming $ K_uV/kT = 8 $. Such particles would be
super-paramagnetic \cite{weller:1999}. Yet the media are evidently
not super-paramagnetic, because they have retained the patterns
written by the electron bunches for three weeks, and in fact were
designed to carry magnetic information for years. Therefore, the
thermal fluctuations prior to the arrival of the field pulse
cannot be the main reason for the broadening of the switching
transition. The broadening due to thermal fluctuations for a
$K_uV/kT = 40$, a value close to the lower limit for stability of
the information stored, is illustrated in figure
\ref{Fig:PerpendicularComparison}.

The thermal fluctuations within a grain also include higher modes
in which the spins are not parallel to each other (illustrated in
figure \ref{Fig:UniformNonuniformMode}). At ambient temperature,
these fluctuations have small amplitude. Calculations show that
the sharply rising field pulse greatly amplifies the pre-existing
thermal randomness. The amplification of the thermal fluctuations
leads to the observed variance of the switching fields.

The crystalline magnetic anisotropy field $H_A = 1.2\times 10^6\
A/m$, the saturation magnetization $M_s = 0.652\ T$, and the
average switching field $B_1 = 1.7\ T$ are compatible with the
Landau-Lifshitz-Gilbert equation if the damping of the
magnetization precession in a grain is assumed to be $\alpha =
0.3$. This extremely large damping shows that torque is lost at a
high rate to the spin system, proving indeed the excitation of
spin fluctuations. It is well known that the spin system is pushed
easily into auto-oscillation and chaos as the absorbed power
increases \cite{hillebrands:book,safonov:2001}. At the end of the
field pulse, the non-equilibrium modes illustrated in figure
\ref{Fig:UniformNonuniformMode} exert the postulated random
torques. Our experiment reveals a "fracture of the magnetization"
under the load of the fast and high field pulses, putting an end
to deterministic switching at shorter time scales.

\section{In-plane magnetic media}

Beside the perpendicular media, in-plane media were also studied,
although lately the magnetic recording industry moves to
perpendicular magnetized media as it enables a denser packing of
bits per surface area.

\begin{figure}[hbt]
  \centering
  \includegraphics[width=0.7\textwidth]{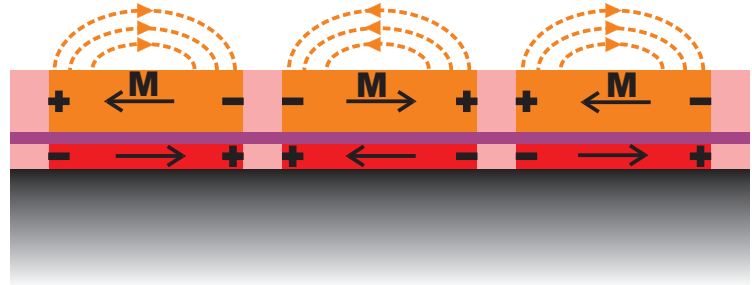}\\
  \caption[Synthetic Antiferromagnet - In-plane media]{The principle of
  synthetic antiferromagnet underlies most of the in-plane magnetic
  recording media. Two magnetic layers are antiferromagnetically coupled
  by a very thin nonmagnetic mettalic layer, usually Ru. The stray magnetic
  field of the transition between domains at the surface is not
  significantly changed but the thermal stability of the whole structure is enhanced.}
  \label{Fig:SAF}
\end{figure}

The magnetic anisotropy of the in-plane media has to be quite
large to preserve the orientation of the bit against the thermal
fluctuations over a long period of time. The saturation
magnetization is just large enough to give a readable magnetic
field at the magnetic transition, see fig \ref{Fig:SAF}. Since the
magnetic thermal stability is given by the volume of the magnetic
bit, one can enhance it by antiferromagnetically coupling an
underlayer that doesn't contribute to readable signal. This is a
so called synthetic antiferromagnet (SAF).

\begin{figure}[hbt]
  \centering
  \includegraphics[width=0.7\textwidth]{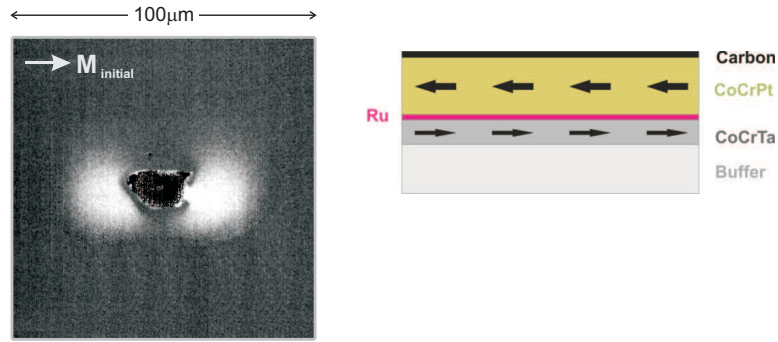}\\
  \caption[Typical magnetic pattern for in-plane media]{A typical magnetic
  pattern for in-plane media with the usual chemical composition of the layers.
  An initially premagnetized sample is exposed to the circular magnetic
  field of the electron beam. The beam damage can be seen as the dark area in the middle
  of the left picture. The picture was taken with Photo-Electron Emission Microscopy
  (PEEM)  using circularly polarized x-rays in order to be sensitive to the magnetic signal.}
  \label{Fig:Komag}
\end{figure}

\begin{figure}[hbt]
  \centering
  \includegraphics[width=0.5\textwidth]{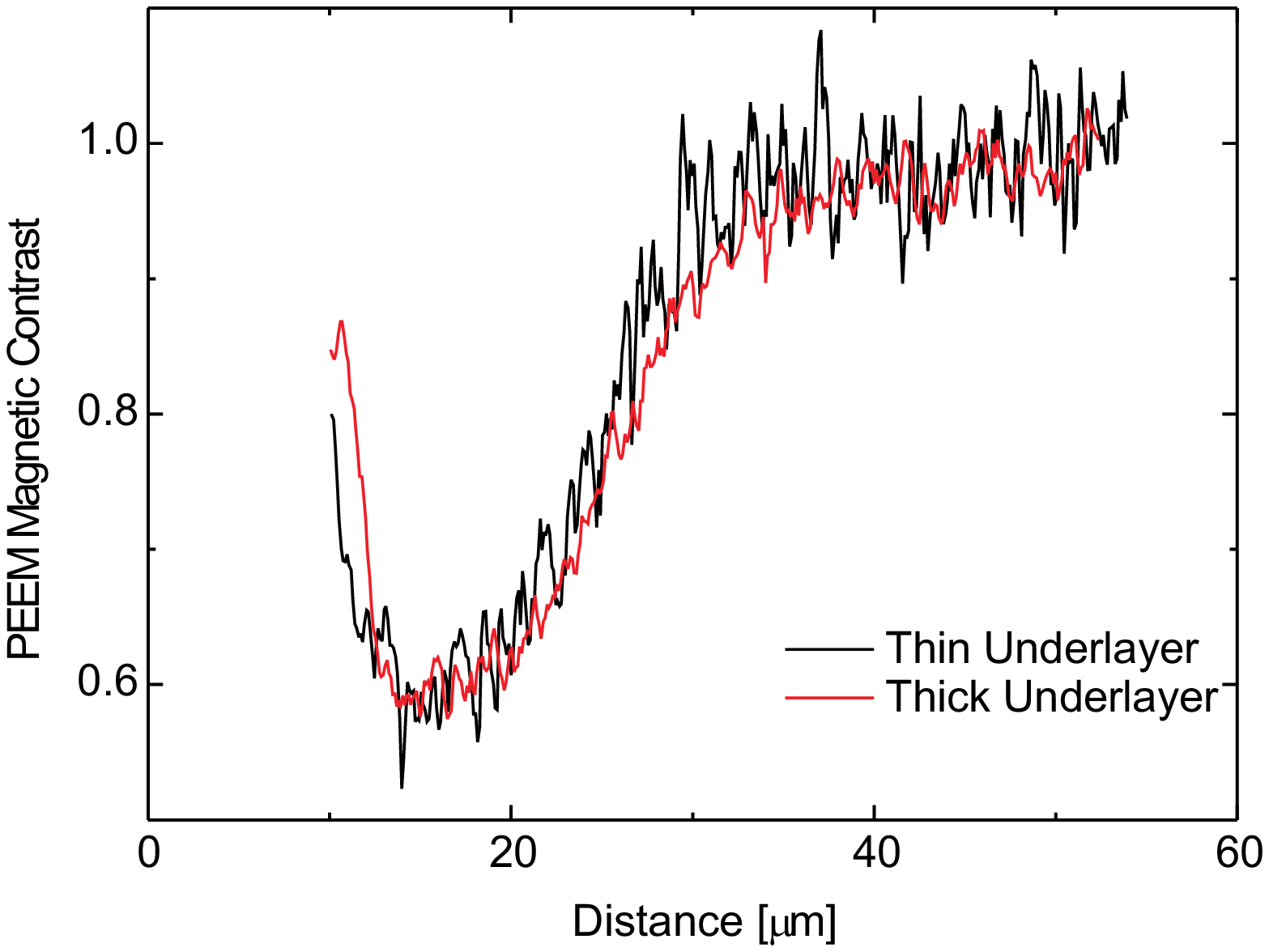}\\
  \caption[Line scans for different thickness of the underlayer
  in the in-plane magnetic media]{Line scans for different thickness of the
  underlayer in the in-plane magnetic media. In the precessional switching
  the underlayer has the same motion of the magnetization as the overlayer
  and does not contribute any additional demagnetization field.}
  \label{Fig:SAFunderlayer}
\end{figure}

We run the same electron beam, see table
\ref{Table:BeamCharacteristics}, through our samples and saw
patterns that looks like the one in figure \ref{Fig:Komag}.
Samples with different thickness for the antiferromagnetically
coupled underlayer were used but no significant difference was
observed as shown in the line scans in figure
\ref{Fig:SAFunderlayer}. The main reason is that the two layers
precess in unison and no additional demagnetizing field is
generated. Thus one can use the macrospin approximation to explain
reasonably well the shape and size of the pattern.

\begin{figure}[hbt]
  \centering
  \includegraphics[width=0.8\textwidth]{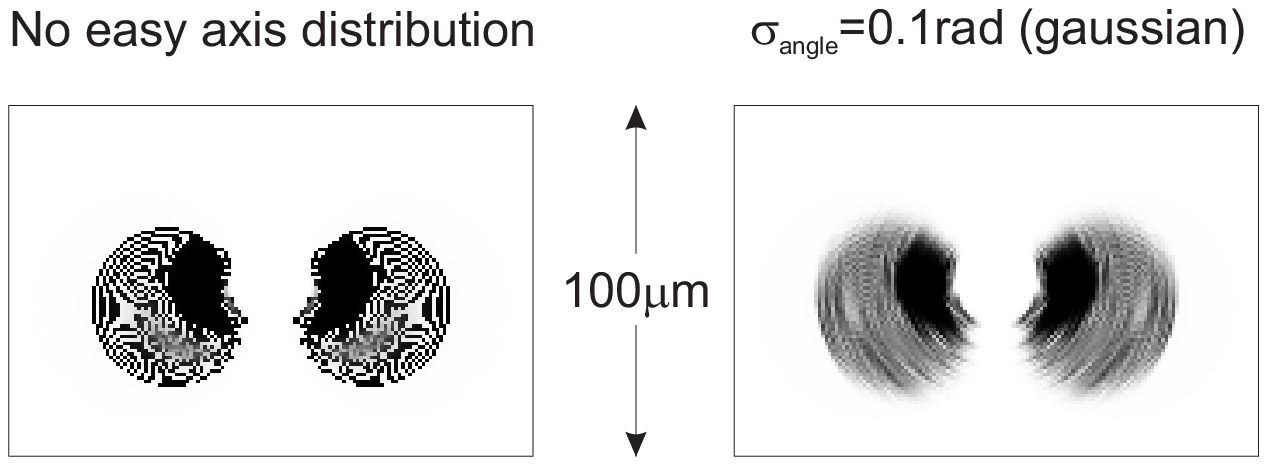}\\
  \caption[In-plane media simulation of the magnetic pattern]{In-plane
  media simulation of the magnetic pattern using the macrospin approximation. The left
  picture shows the pattern when all grains have the same easy axis direction, while on
  the right the average over a gaussian distribution of the direction is taken. }
  \label{Fig:EasyAxisDispersion}
\end{figure}

A simulation using the macrospin approximation is shown in figure
\ref{Fig:EasyAxisDispersion}, where the in-plane uniaxial
anisotropy field is $H_{anis}=0.89\times10^6\,A/m$, the
demagnetizing field is $H_{demag}=0.33\times10^6\,A/m$ and a
damping of $\alpha=0.01$ is used. The magnetic material usually
has a random anisotropy direction as sputtered on the magnetic
disk but these samples have a preferential distribution along the
circular direction on the disk to enhance the magnetic signal. If
all magnetic grains had the same orientation of the easy axis one
would see details inside the magnetic pattern, however, because of
the random distribution the pattern becomes diffuse.

A comparison with the in-plane magnetic thin films reveals that,
for the magnetic recording media, the out of plane demagnetization
energy is much less than the uniaxial in-plane anisotropy energy
($H_{anis}=0.89\times10^6\,A/m\quad < \quad
H_{demag}=0.33\times10^6\,A/m$). The applied magnetic field cannot
store much energy by bringing the magnetization out of the sample
plane and the dynamics looks very much like a swing from one
direction to the opposite as shown in figure  for a point situated
$20\,\mu m$ from the electron beam center.
\begin{figure}[hbt]
  \centering
  \includegraphics[width=0.5\textwidth]{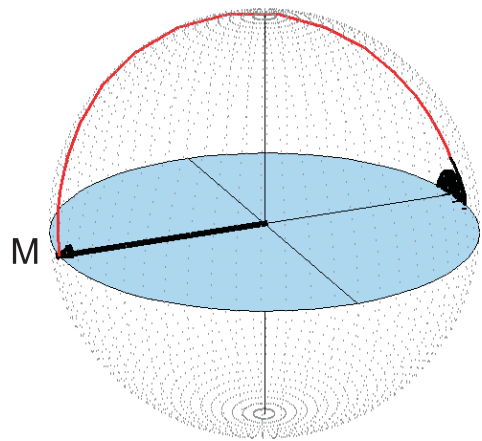}\\
  \caption[Magnetization dynamics for a point $20\mu m$ from the beam center]{Magnetization
  dynamics for a point $20\mu m$ from the beam center. The applied field makes the
  magnetization precess across the hard plane from the initial easy direction (red curve). The
  demagnetizing energy as the magnetization is brought out of the sample (equatorial) plane is small
  and not effective in storing energy from the applied field.}
  \label{Fig:SAFdynamics}
\end{figure}

\chapter{Conclusions}

This thesis describes new results in precessional dynamics of the
magnetic materials using the magnetic field pulse associated with
an electron bunch from a linear accelerator. The dynamics was
inferred from the shape and size of the magnetic pattern imprinted
on the initially uniformly magnetized sample. Although we don't
have real time information as that obtained in pump-probe
experiments, each location on the pattern sees a different applied
field and we get a good statistics not easily obtained in
pump-probe experiments. Additionally, the precession of the
magnetization around the demagnetization field of the in-plane
samples provides a stopwatch for the switching boundaries.

The perpendicular magnetic media patterns can be explained if we
assume a random magnetic field in addition to the usual one caused
by the thermal fluctuations. This field might come from the energy
stored in the exchange interaction between the spins. The samples
show enhanced damping of the uniform precession mode, consistent
with the above hypothesis. A parallel can be made with
ferromagnetic resonance where high power rf excitation leads to
chaotic dynamics. The Fourier transform of our magnetic field
pulse does contain large amplitude, high frequency components. The
additional randomness found in our experiment could impose
limitations on the reliability of magnetic switching at very high
speed if the magnetic recording industry will be able to go from
around 1 ns to about 10 ps bit writing time.

The model thin magnetic films developed big patterns with many
rings due to the low in-plane uniaxial anisotropy. The position of
the switching boundaries tell how much energy is needed to push
the magnetization over the anisotropy barrier to the other
direction. By subtracting this from the energy pumped by the
magnetic field of the pulse we get the energy dissipated by the
system. A peculiar feature of the pattern is that the first
outside ring is thicker than the inner ones. A micromagnetics
simulation shows that the spin wave scattering of the uniform mode
into higher modes. This takes time to develop and the it is not
active for the first switching boundary. The additional
dissipation due to this mechanism is still not sufficient (by a
factor of two) to explain location of the rings. Here again there
might be some additional mechanism for dissipation of angular
momentum and energy.

Two main features of the magnetic field pulses are important to be
simultaneously present for meaningful precessional switching
experiments: high intensity and short duration. The tabletop
experiments parameters, although improving, are still a factor of
10 behind in either of them and a factor of 100 combined. Thus
they are limited to study of soft magnetic materials. On the other
hand accelerator electron bunches of 100fs duration are currently
available and experiments exploring this time scale are scheduled
to be run shortly after the submission of this thesis. A
preliminary test experiment showed a chaotic magnetic pattern and
perhaps unexpected results will come out.

\appendix

\chapter{Li\'{e}nard-Wiechert Potentials}

\section{Charge in uniform motion} \label{UniformMotion}
\begin{figure}[!hbt]
  \begin{center}
  \includegraphics[]{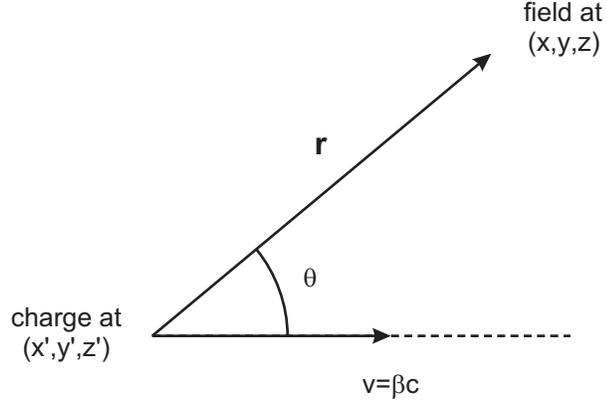}\\
  \caption[Angles and Distances for Li\'{e}nard-Wiechert formula]{The angle $\theta$ in the text is defined as the the angle between
  the velocity direction and the distance vector pointing from the charge location
  to the location of the electromagnetic fields.}
  \label{Fig:LienardWiechertUniformMotion}
  \end{center}
\end{figure}

Instead of describing the electromagnetic field by $\mathbf{E,B}$
we'll describe it by the scalar potential $\phi$ and vector
potential $\mathbf{A}$, a totally equivalent approach, with
\begin{equation}
    \mathbf{E}=-\mathbf{\nabla} \phi-\dt{\mathbf{A}}{}\hspace{1em},\hspace{1em}
    \mathbf{B}=\mathbf{\nabla}\times \mathbf{A}
    \label{Eq:FieldsPotentials}
\end{equation}
In free space and using the Lorentz gauge condition $\nabla\cdot
A+\frac{1}{c^2}\dt{\phi}{}=0$ the Maxwell's equations are
equivalent to wave equations
\begin{equation}
\begin{aligned}
  \nabla^2\phi-\frac{1}{c^2}\dt{\phi}{} &= -\mu_0\mathbf{j} \\
  \nabla^2\mathbf{A}-\frac{1}{c^2}\dt{\mathbf{A}}{} &= -\rho/\epsilon_0
\end{aligned}
\label{Eq:WaveEq}
\end{equation}
The field of an electron moving with uniform velocity must be
carried convectively along with the electron which implies that
the time and and space derivatives are not independent, i.e.
\begin{equation}
    \frac{\partial}{\partial t}=-\mathbf{v\cdot\nabla}
\end{equation}
For example, if the electron moves along x direction, the wave
equation becomes
\begin{equation}
    (1-\frac{v^2}{c^2})\frac{\partial^2\phi}{\partial x^2}+
    \frac{\partial^2\phi}{\partial y^2}+
    \frac{\partial^2\phi}{\partial z^2}=-\rho/\epsilon_0
\end{equation}
Changing the variables to
$\underline{x}=x/\sqrt{1-v^2/c^2},\underline{y}=y,\underline{z}=z$
we get the electrostatic Poisson equation
\begin{equation}
    \underline{\nabla}\phi=-\rho(\sqrt{1-v^2/c^2}\underline{x},\underline{y},\underline{z})/\epsilon_0
\end{equation}
with the solution
\begin{equation}
    \phi(\underline{x},\underline{y},\underline{z})=
    \frac{1}{4\pi\epsilon_0}\int{\frac{\rho(\underline{x}',\underline{y}',\underline{z}')}
    {\sqrt{(\underline{x}-\underline{x}')^2+(\underline{y}-\underline{y}')^2+(\underline{z}-\underline{z}')^2}}}d\underline{v}'
\end{equation}
Changing back to the original variables the solution becomes
\begin{equation}
    \phi(\underline{x},\underline{y},\underline{z})=\frac{1}{4\pi\epsilon_0}\int{\frac{\rho(\underline{x}',\underline{y}',\underline{z}')}{s}}d\underline{v}'
\end{equation}
with
\begin{equation}
\begin{aligned}
    s&=\sqrt{(x-x')^2+(1-v^2/c^2)[(y-y')^2+(z-z')^2]}
    &=r\sqrt{1-v^2/c^2\sin^2\theta}
\end{aligned}
\label{Eq:sDefinition}
\end{equation}
The angle theta is defined in the figure
\ref{Fig:LienardWiechertUniformMotion}.
In particular for an electron with velocity $\mathbf{v}$,
\begin{equation}
\label{Eq:LienardWiechertUniformMotion}
\begin{aligned}
  \phi &= \frac{e}{4\pi\epsilon_0 s} \\
  \mathbf{A} &= \frac{e\mathbf{v}}{4\pi\epsilon_0c^2s}
\end{aligned}
\end{equation}
These are also called Li\'{e}nard-Wiechert potentials, derived
here for uniform motion, although in the general case the velocity
can vary in time. Now we are in a position to calculate the
Lorentz force
\begin{equation}
\begin{aligned}
  \mathbf{F} &=& \frac{e^2}{4\pi\epsilon_0}\left[-\nabla\left(\frac{1}{s}\right)+
  (\mathbf{v\cdot\nabla})\frac{\mathbf{v}}{c^2s}+
  \frac{\mathbf{v}}{c^2}\times\left(\mathbf{\nabla}\times\frac{\mathbf{v}}{s}\right)\right] \\
   &=& -\frac{e^2}{4\pi\epsilon_0}\mathbf{\nabla}\left(\frac{1-v^2/c^2}{s}\right)
\end{aligned}
\end{equation}
The result is that the force can be expressed as gradient of a
potential, called the convective potential.
\begin{equation}
    \mathbf{F}=-\mathbf{\nabla}\psi,\qquad where \qquad
    \psi=\frac{e^2(1-v^2/c^2)}{4\pi\epsilon_0s}
\end{equation}
The scalar convective potential does not have spherical symmetry
about the electron, see how $s$ is defined in
\ref{Eq:sDefinition}.

The electric field of a moving electron, using
\ref{Eq:LienardWiechertUniformMotion} in \ref{Eq:FieldsPotentials}
is
\begin{equation}
    \mathbf{E}=\frac{e\mathbf{r}(1-v^2/c^2)}{4\pi\epsilon_0s^3}
        =\frac{e\mathbf{r}}{4\pi\epsilon_0r^3}
        \frac{1-v^2/c^2}{(1-v^2/c^2\sin^2\theta)^{3/2}}
\end{equation}

\section{Charge in arbitrary motion}
\begin{figure}[!hbt]
  \begin{center}
  \includegraphics{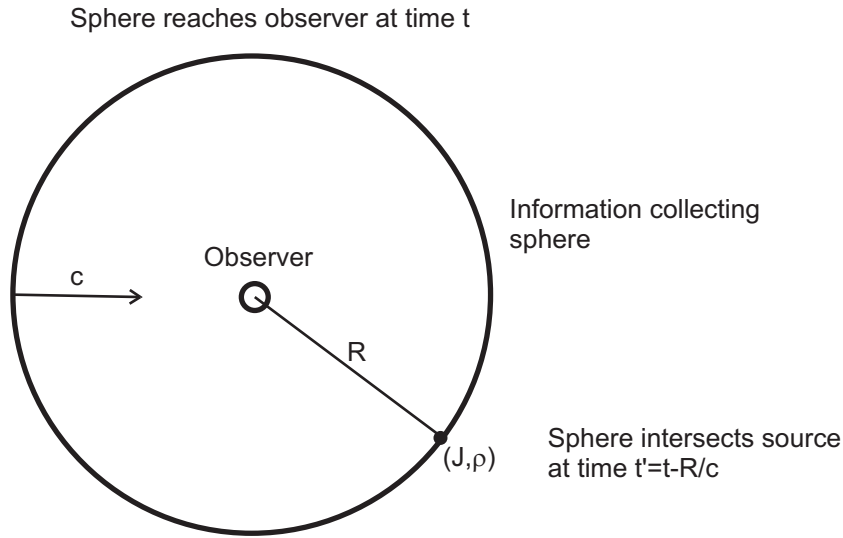}\\
  \caption[Retarded Potentials and a Moving Sphere]{A helpful
  instrument in keeping track of the time is to
  imagine a sphere collecting information for retarded potentials.
  As the sphere moves towards the observer with the speed of light, it
  carries information about the field at the observer that is generated
  by the charges that intersect the sphere.}
  \label{Fig:RetardedPotentialsSphere}
  \end{center}
\end{figure}

A general solution for \ref{Eq:WaveEq} for a charge in arbitrary
motion, not just uniform motion is
\begin{equation}
\begin{aligned}
    \phi(x,y,z,t)&=\frac{1}{4\pi\epsilon_0}\int{\frac{[\rho(x',y',z')]}{R}}dv'\\
    \mathbf{A}(x,y,z,t)&=\frac{\mu_0}{4\pi}\int{\frac{[\mathbf{j}(x',y',z')]}{R}}dv'\\
    R&=\sqrt{(x-x')^2+(y-y')^2+(z-z')^2}
\end{aligned}
\end{equation}
where square brackets are evaluated at a retarded time $t'=t-R/c$.
These are known as \textit{retarded potentials} and they can be
visualized as follows \cite{panofsky:book}. Consider an observer
situated at the point $r$, see figure
\ref{Fig:RetardedPotentialsSphere}, and let a sphere centered at
$r$ contract with radial velocity $c$ such that it converges on
the point at the time of observation $t$. The time at which this
information-collecting sphere passes the source at point $r'$ is
then the time at which the source produced the effect which is
felt at $r$ at time $t$.

Care must be taken when applying the concept of retarded potential
because the integrand $[\rho(\mathbf{r'})d\mathbf{r'}]$ is
evaluated at different times. As the information collecting sphere
sweeps over the charge distribution, the charges may move so as to
appear more or less dense. If the charges move in the same
direction as the converging sphere the volume integral of the
\emph{retarded charge} density will be more than the total charge
and if charges move in opposite direction the integral will give
less than the total charge. The retarded potential of an
approaching charge will be greater than that of a receding charge
at the same distance from an observer, since the approaching
charge stays longer within the information collecting sphere.
\begin{figure}[!hbt]
  \begin{center}
  \includegraphics[]{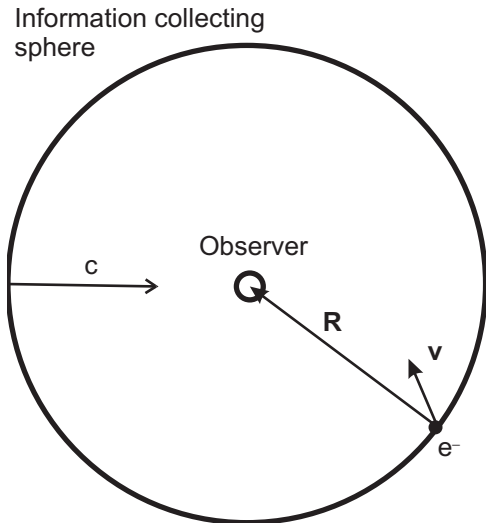}\\
  \caption[Moving Sphere in Case of an electron]{The information
  collecting sphere and a moving electron.}
  \label{Fig:RetardedPotentialsElectron}
  \end{center}
\end{figure}

In the case of an electron approximated as a point charge, we can
express the total charge in terms of the retarded charge in the
following way, see figure \ref{Fig:RetardedPotentialsElectron}:
\begin{equation}
    e=[e]-[e]\frac{\mathbf{v}}{c}\cdot\frac{\mathbf{R}}{R}
\end{equation}
where $\mathbf{v}$ is the speed of electrons. The retarded
potentials are then
\begin{equation}
\begin{aligned}
    \phi(\mathbf{r},t)&=\frac{1}{4\pi\epsilon_0}\;\frac{e}{[R-\mathbf{v}\cdot\mathbf{R}/c]}\\
    \\
    \mathbf{A}(\mathbf{r},t)&=\frac{\mu_0}{4\pi}\;\frac{e\mathbf{v}}{[R-\mathbf{v}\cdot\mathbf{R}/c]}
\end{aligned}
\end{equation}
also called Li\'{e}nard-Wiechert potentials for an electron (the
square brackets are evaluated at $t'=t-R/c$).

\begin{figure}[!hbt]
  \begin{center}
  \includegraphics[]{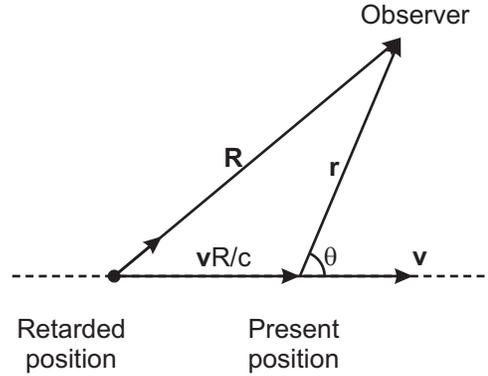}\\
  \caption[Retarded and Present Position of a Moving Charge]{Retarded
  and present position of a moving charge.}
  \label{Fig:UniformMotionDistances}
  \end{center}
\end{figure}

For an electron in uniform motion one can show with the help of
figure \ref{Fig:UniformMotionDistances} that the quantity
$[R-\mathbf{v}\cdot\mathbf{R}/c]$ in the denominator is exactly
$s$ in \ref{Eq:sDefinition} (see \cite{jackson:book}).

From the figure, with the help of the relations
$\mathbf{r}=\mathbf{R}-\mathbf{v}R/c$ and
$\mathbf{r}\times\mathbf{v}=\mathbf{R}\times\mathbf{v}$, we get
\begin{equation}
\begin{aligned}
    (R-\mathbf{v}\cdot\mathbf{R}c)^2&=r^2-\left(\frac{\mathbf{r}\times\mathbf{v}}{c}\right)^2\\
    &=r\sqrt{1-\frac{v^2}{c^2}\sin^2\theta}\\
    &=s^2
\end{aligned}
\end{equation}

\chapter{Labview code for the sample manipulator}

 As the manipulator has to be quite versatile and is used for many
 different procedures, the software written in Labview 6.1 was
 structured as a state machine, similar to how a computer works. A
 state machine can be in any of a collection of states. At every
 step the machine evaluates its input and decides what state it will be in
 the next step. A schematic of the code implementing the state
 machine is given in figure below. The hierarchy of the code and one of the
 monitoring front panel are shown next.
\begin{figure}[!hbt]
  \begin{center}
  \includegraphics[]{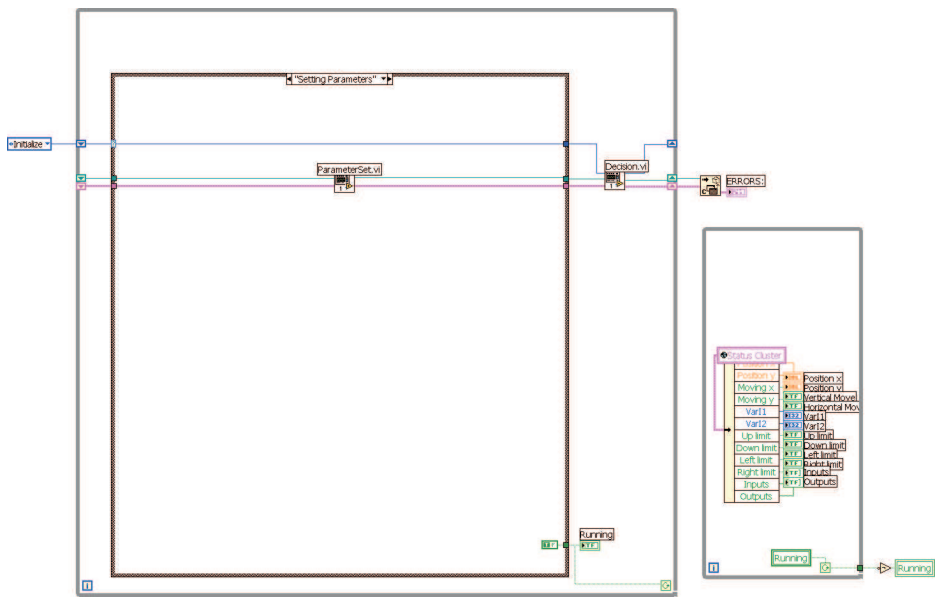}\\
  \caption[Labview Diagram Implementing the Manipulator Control]
  {Labview diagram implementing the manipulator control as
  a state machine. The states are shown pasted inside the
  \textit{Case} structure. One can see the monitoring cluster on the right.
  The two \textit{While} structures run in parallel, both being terminated by the
  variable \textit{Running}.}
  \label{Fig:LabvieDiagram}
  \end{center}
\end{figure}

\begin{figure}[!hbt]
  \begin{center}
  \includegraphics[height=3in]{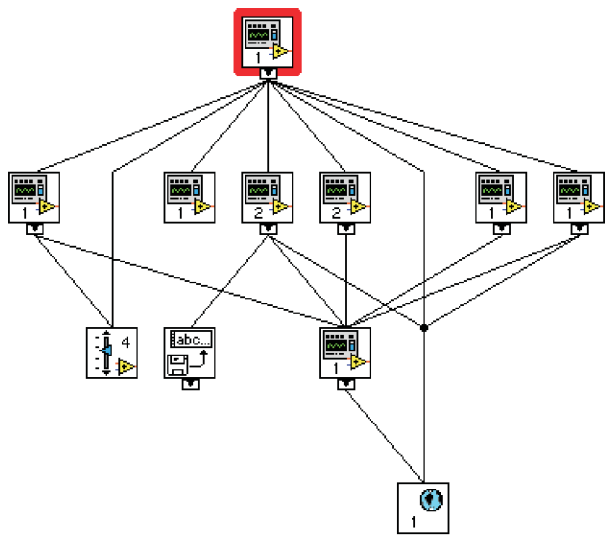}\\
  \caption[Labview Hierarchy of the Program that Controls the Manipulator]
  {Labview hierarchy of the program that controls the manipulator.
  The programs spends most of its time in one of the states on the
  second level. \textit{UpdateStatus} displays the information contained in
  the \textit{GlobalStatus} variable. The list of exposure is read from a file
  in \textit{HitPlan}.}
  \label{Fig:LabviewHierarchy}
  \end{center}
\end{figure}

\begin{figure}[!hbt]
  \begin{center}
  \includegraphics[height=2.5in]{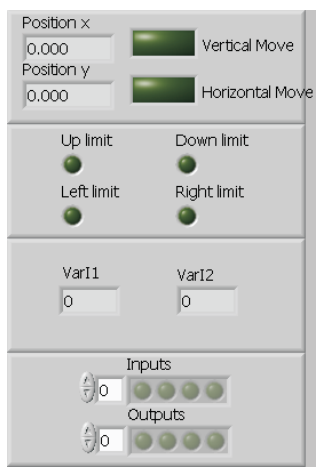}\\
  \caption[Labview Monitoring Front Panel]{Labview monitoring front panel. VarI1 and VarI2 are used for
  monitoring multiple shots. The lower inputs and outputs can be used
  for different triggers}
  \label{Fig:LabviePanel}
  \end{center}
\end{figure}
All the Labview components of the file send text commands to a
controller for the stepper motors that move the manipulator.

\chapter{Comparison: Electron Bunch and Half Cycle Pulses}
\chaptermark{Half Cycle Pulses}

A gaussian shaped electron bunch has an associated EM field that
travels along with it with the same gaussian shape. To a
stationary observer, this field very much resembles to a half
cycle of an oscillating electromagnetic field. However, this is
not a radiation field as it doesn't travel to infinity but is
localized around the electron bunch.

\begin{figure}[hbt]
\centering
\includegraphics[]{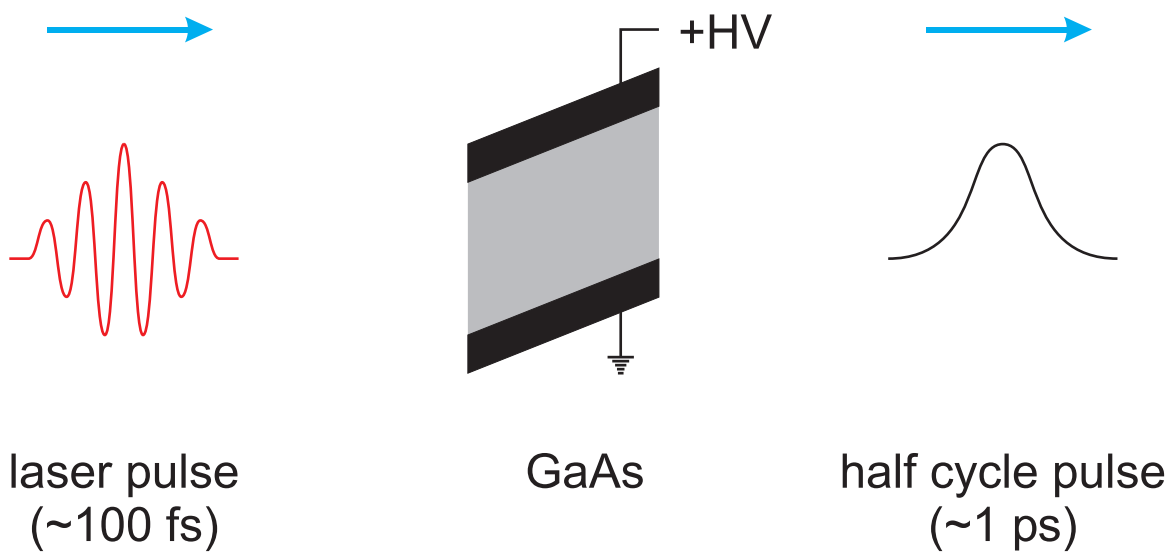}
\caption[Schematic for Generation of Half Cycle Pulses]{Schematic
showing the method to produce half cycle pulses of radiation
field. A large amount of conduction electrons are produced in a
short time by the laser pulse. These electrons are accelerated in
the high voltage (+HV) bias and radiate. The radiated field shape
follows the shape of the acceleration curve.}
\label{Fig:HCPProduction}
\end{figure}

There is a method to generate a true radiation field that also
resembles a half cycle most of the time. By illuminating a wafer
of biased GaAs semiconductor with a short pulse of laser light
electrons are promoted to the conduction band. While in the
insulating band, each GaAs wafer can hold off several kilovolts
across $1\ cm$ without breaking down.

A schematic of the half cycle pulse (HCP) production process
\cite{zeibel:thesis} is shown in figure \ref{Fig:HCPProduction}.
When the laser pulse hits the wafer, electrons within the wafer
quickly accelerate due to the bias field. The accelerating
electrons radiate a short pulsed field which propagates away from
the wafer. The radiated field is polarized along the bias axis,
its amplitude is proportional to the amplitude of the bias field
and its rise time is about $0.5\ ps$. The HCP field produced by
the GaAs wafer scales linearly with the area of the wafer up to an
area of approximately $1\ cm^2$.

\begin{figure}[hbt]
\centering
\includegraphics[]{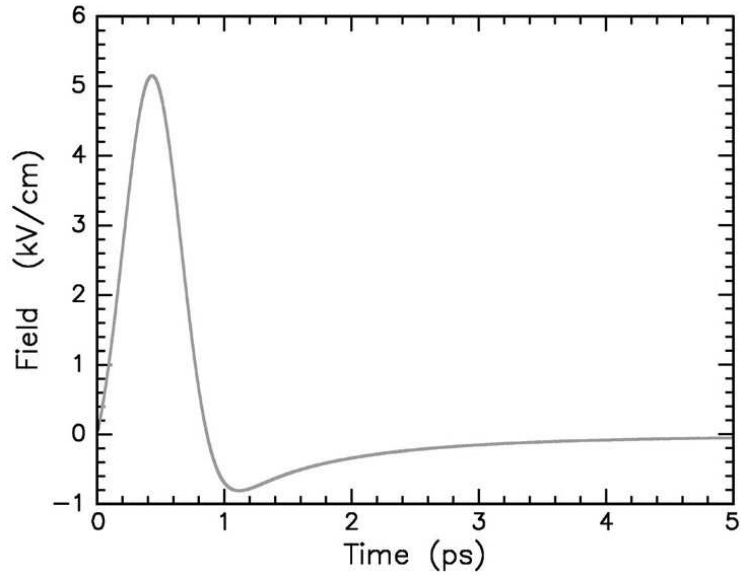}
\caption[Schematic of a Half Cycle Pulse Shape]{The shape of the
electric field from a typical half cycle pulse. The initial
positive pulse is about $1\ ps$ in duration and the negative tail
persists for hundreds of picoseconds \cite{zeibel:thesis}.}
\label{Fig:HCPShape}
\end{figure}

The wafer returns to insulating state after the laser pulse has
past, but this transition is much slower and electrons decelerate
over a period of hundreds of picoseconds \cite{zeibel:thesis}. The
deceleration produces a field in the opposite direction to that of
the initial field, however the maximum strength of the "recoil"
field is typically a factor of $5-10$ smaller than the initial
pulse height. A graphical representation of a typical HCP temporal
profile is shown in figure \ref{Fig:HCPShape}.
\begin{figure}[hbt]
\centering
\includegraphics[]{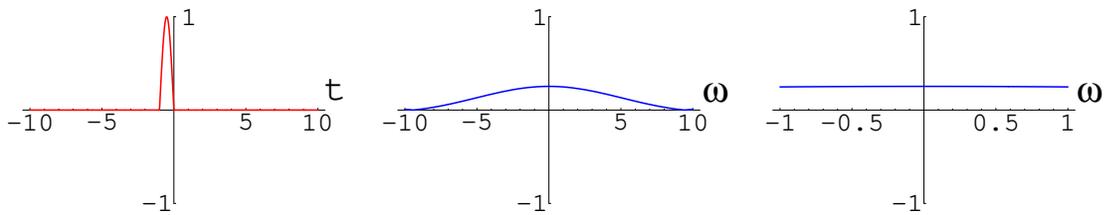}
\caption[Fourier Transform of a Half Cycle of a Sine Wave]{The
Fourier transform of a half cycle of a sine wave with a zoom in
around zero frequency. There is a zero frequency component
proportional to the area under the pulse.} \label{Fig:HCPHalfSine}
\end{figure}

\begin{figure}[hbt]
\centering
\includegraphics[]{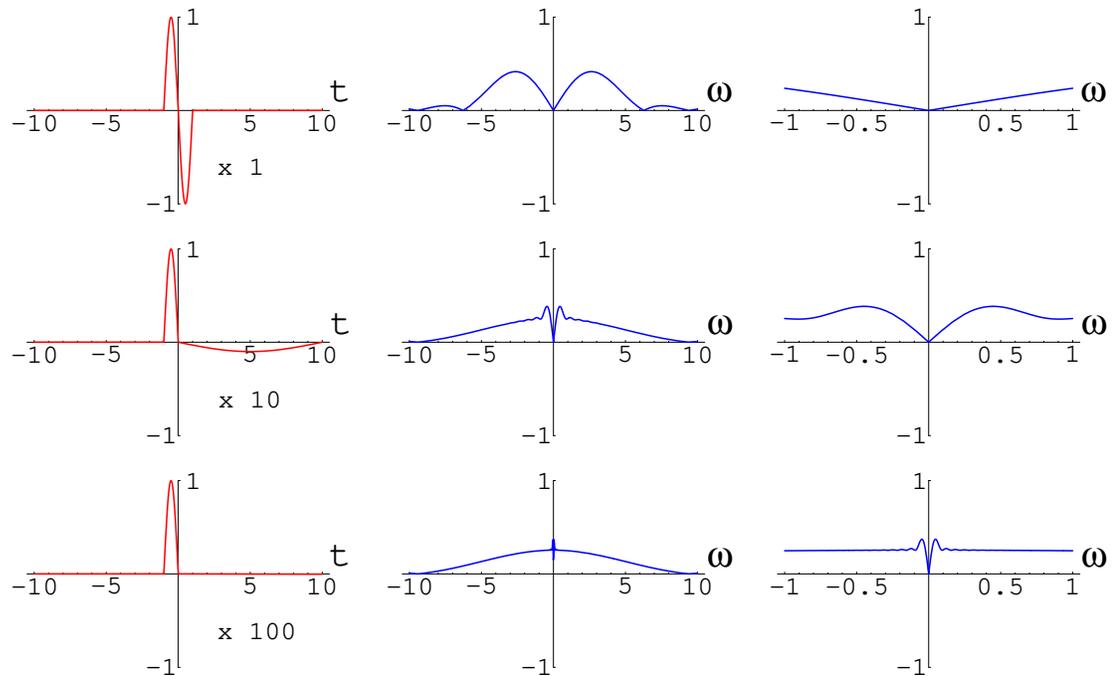}
\caption[Fourier Transform of a Full Cycle of a Sine Wave with the
Negative Part Stretched]{The Fourier transform of a full cycle of
a sine wave having the negative part stretched, with a zoom in
around zero frequency. There is no zero frequency component.}
\label{Fig:HCPFullSine}
\end{figure}

Although the pulse is called "half cycle" it has a negative part
that is about 100 times longer in duration and lower in amplitude.
This can be seen clearly by comparing the Fourier transform of a
half sine cycle \ref{Eq:HalfSine}, \ref{Fig:HCPHalfSine} with that
of a full cycle having the negative part stretched
\ref{Eq:FullSine}, \ref{Fig:HCPFullSine}.

\begin{equation}
    \mathbf{E}(t)=
    \begin{cases}
    \mathbf{E_0}\sin(\omega t) & \text{$-\pi/2<\omega t<0$},\\
    0 & \text{elsewhere}
    \end{cases}
    \label{Eq:HalfSine}
\end{equation}
\begin{equation}
    \mathbf{E}(t)=
    \begin{cases}
    \mathbf{E_0}\sin(\omega t) & \text{$-\pi/2<\omega t<0$},\\
    \mathbf{E_0}/a\quad\sin(\omega t/a) & \text{$0<\omega t<\pi/2$, $a\in\{1,10,100\}$},\\
    0 & \text{elsewhere}
    \end{cases}
    \label{Eq:FullSine}
\end{equation}

\begin{figure}[hbt]
\centering
\includegraphics[]{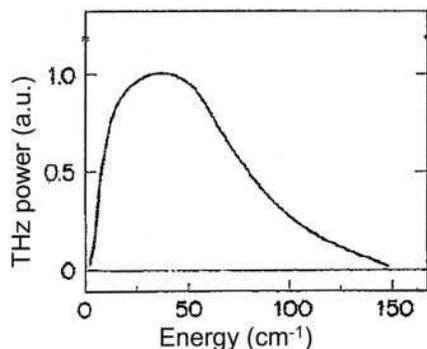}
\caption[Spectrum of a Half Cycle Pulse]{Spectrum of a half cycle
pulse determined by Fourier-transform detection \cite{tsen:book}.
Compare it with figure \ref{Fig:HCPFullSine}.}
\label{Fig:HCPSpectrum}
\end{figure}

The pure half cycle has a zero frequency component whereas the
stretched full cycle doesn't have it. This zero frequency
component is directly proportional to the total integral of the
electric field. Fourier-transform detection of HCP has been done,
see \ref{Fig:HCPSpectrum}, and its spectrum has a shape similar to
the full cycle with the negative part stretched \cite{tsen:book}.

\begin{figure}[hbt]
\centering
\includegraphics[]{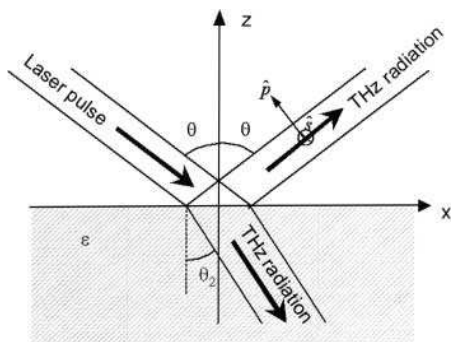}
\caption[Direction of a Half Cycle Pulse]{Direction of a half
cycle pulse is determined by Snell's law \cite{tsen:book}.}
\label{Fig:HCPDirection}
\end{figure}

The direction of the THz radiation can actually be controlled by
the incident laser as shown in the picture \ref{Fig:HCPDirection}.
The angles are given by the nonlinear Snell law, with frequency
dependent refraction index. Because of the tilt of the laser beam
there is a phase delay in the radiation from different parts of
the emitter that in the end make the radiation concentrate along
the angle given by the Snell Law. The generation of THz radiation
can also be thought as a nonlinear effect where the mixing of
different frequencies in the incoming beam gives the THz radiation
as a difference.

\begin{figure}[hbt]
\centering
\includegraphics[]{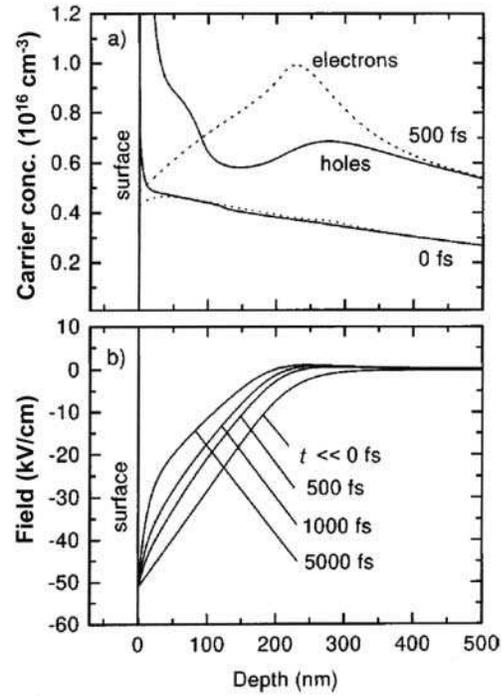}
\caption[Charge Dynamics in a Device Biased by a Surface Depletion
Field]{Charge dynamics in an emitter biased by a surface depletion
field. The changes in the field over time is due to space charge
screening \cite{tsen:book}.} \label{Fig:HCPChargeDynamics}
\end{figure}

The acceleration of the charge carriers in the electric bias field
is the most critical aspect of THz because this is the process
that determines the temporal evolution of the radiated electric
field. It can be viewed as involving an initial ballistic
acceleration of the carriers on a time scale shorter than the
carrier scattering time, followed by an approach to drift
velocity. For high enough density of charge carriers the initial
acceleration can lead to a space charge that can screen the bias
field reducing the efficiency of the THz generation. A simulation
for the dynamics of the charge in an emitter biased perpendicular
to the surface by doping is shown in figure
\ref{Fig:HCPChargeDynamics}.

\chapter{Coherent Magnetic Switching via Magnetoelastic Coupling}
\chaptermark{Magnetoelastic Switching}

\section{Stress Induced, Dynamically Tunable Anisotropy}
\sectionmark{Tunable Anisotropy}
A magnetic moment is intrinsically linked to an angular momentum,
either of the orbital or the spin type. I like to think of the
magnetic storage as storage of information in little spinning
things. To change a magnetic moment one has to supply angular
momentum. The usual method is to use a magnetic field or, more
recently, circularly polarized light that carries one quantum of
angular momentum. Another way of angular momentum transfer is to
use the coupling between spin and orbital angular momentum.

The magnetic moment in the ferromagnetic materials commonly used
in information storage consists of mainly spin magnetic moment.
The orbital magnetic moment is "quenched" by the crystal lattice,
that is the electric interactions between the electron and the
lattice prefers orbits that are equal combination of right and
left circling electron. This quenching is not complete and there
is a few percent orbital magnetic moment contribution to the total
magnetic moment.

The important fact we care about here is that there is a very
strong coupling of the orbital moment to the crystal lattice.
There is also a weaker coupling of the orbital moment to the spin
moment. As the electron moves about a charged atomic core it sees
the moving core as an electric current which makes a magnetic
field acting on its spin moment. An idea I have thought about is
to coherently modify the crystal lattice, then the orbital moment,
being strongly coupled to the lattice will very fast adjust its
orientation to it after which the orbital moment will exert a
torque on the spin moment via spin-orbit coupling.

There is a whole branch of magnetism that studies the coupling
between the lattice and the magnetic moment, also called the
magnetostriction. A correspondent phenomenon with electric field,
the coupling between the crystal lattice and the electric dipole
moment, is the electrostriction. Suppose we generate a mechanical
stress in a piezoelectric material by applying an electric field
to it, and adjacent to it we have a magnetic material in full
mechanical contact. The stress transmitted to the magnetic
material changes its crystal lattice and if this magnetic material
has a large magnetoelastic coefficient it will change the
orientation of the magnetic moment. What we have done in fact is
generate angular moment in the crystal lattice by applying an
electric field, transfer it to the orbital angular moment very
fast and then transfer it more slowly to the spin moment via the
spin-orbit coupling.
\begin{figure}[!h]
\centering
\includegraphics[width=0.5\textwidth]{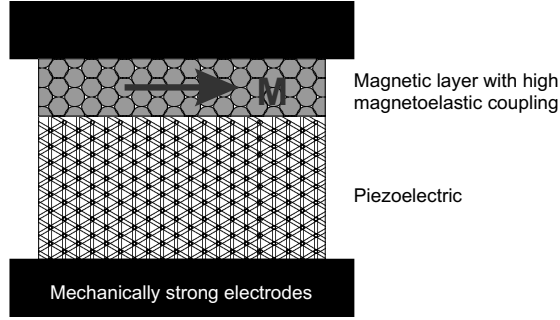}
\caption[Magnetoelastic Memory]{An illustration of a memory device
using the magnetoelastic effect. The piezoelectric layer converts
the electric voltage into mechanic stress that changes the
orientation of the magnetization $M$ in the magnetic layer.}
\label{Fig:MagnetoelasticMemory}
\end{figure}

A simple magnetic device functioning as magnetic memory can be
imagined using the above mechanism. A capacitor with electrodes
made of strong metallic material, such as tungsten, has a
dielectric made of good piezoelectric material. In between one of
the electrodes and the piezoelectric we have a layer of magnetic
material with high magnetoelastic coefficient. When we apply an
electric voltage to the electrodes the piezoelectric contracts or
expands in the fixed space provided by the mechanically strong
electrodes. The stress is transmitted to the magnetic material and
its magnetic moment will change its orientation. In principle one
could use precessional switching in the magnetic field generated
by the mechanical stress; the precise timing could be provided by
tuning the voltage pulse shape. With the simple ferromagnetic
materials the magnetic field generated by stress can reach $1000\
Oe$. Thus the time scale of precessional switching is around $100\
ps$.

The induced anisotropy due to stress can be illustrated in the
case of isotropic materials, i.e. with isotropic magnetoelastic
coefficient. In this case the magnetoelastic energy has the form:
\begin{equation}
    E_{mel}=\frac{3}{2}\lambda_s\sigma\sin^2\theta=K_u\sin^2\theta
    \label{Eq:MagnetoelasticEnergy}
\end{equation}
where $\lambda_s$ is the magnetoelastic coefficient, $\sigma$ is
the stress, $\theta$ is the angle between the applied stress and
the magnetization direction. $K_u=\frac{3}{2}\lambda_s\sigma$ is
the uniaxial anisotropy energy density due to the stress; it can
have both signs depending whether the stress is tensile
($\sigma>0$) or compressive ($\sigma>0$) \cite{tremolet:book}. For
a positive magnetoelastic coefficient ($\lambda_s>0$) a tensile
stress will favor the alignment of magnetic moments with the
direction of the stress. A compressive stress will favor the
settling of the magnetic moments in a plane perpendicular to the
stress direction, see figure . For a negative $\lambda_s$ the
behavior is opposite to the above.
\begin{figure}[!h]
\centering
\includegraphics[height=3in]{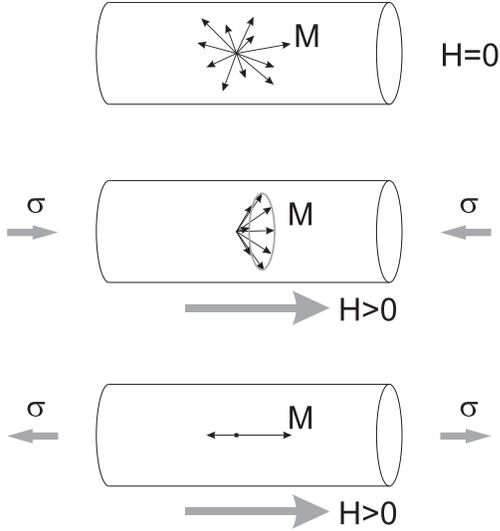}
\caption[The Effect of Stress Anisotropy]{The effect of stress
anisotropy on the orientation of magnetization for a magnetic
material with isotropic positive magnetostriction. With no
magnetic field and no stress applied the magnetization has no
preferred direction. A compressive stress will favor an
orientation perpendicular to the stress direction and with an
applied field the magnetization lies in a cone around the magnetic
field direction. A tensile stress will tend to align the
magnetization along its direction and an applied magnetic field
will favor domains aligned with it.} \label{Fig:StressAnisotropy}
\end{figure}

The stress induced in a structure like the one in figure
\ref{Fig:MagnetoelasticMemory} can be enhanced when there is a
resonance with the mechanical structure. The stress in the
magnetic material can be several times the applied stress
\cite{claeyssen:1997}, the multiplication factor being the
Q-factor of the mechanical resonator.

\section{Spin Amplifier}

An apparatus that uses the spin degree of freedom consists of
polarizers, analyzers and spin rotators. The whole field of spin
dependent electric conduction exploded with the use of the Giant
Magneto-Resistance (GMR) effect in the read heads of magnetic hard
drives. The system there consists of a polarizer and analyzer
ferromagnetic layers with a non-magnetic metallic layer to
transmit the spin polarized current. The stray magnetic field from
the bits in the magnetic media changes the direction of the
polarizer and one can see this with the help of the analyzer
layer.

Using the spin degree of freedom of conduction electrons means
performing operation on their spin polarization (the normalized
difference between the number of spin up and spin down electrons).
Making an analogy with regular charge electronics, which took off
after the invention of the transistor, one would need a spin
transistor to build a spin polarization amplifier. A regular
transistor is a current/voltage source which controlled by the
input current/voltage. A spin transistor is a spin polarization
source(a filter) controlled by the input spin polarization. One
has to employ a spin dependent effect to build such a transistor.

\begin{figure}[!h]
\centering
\includegraphics[]{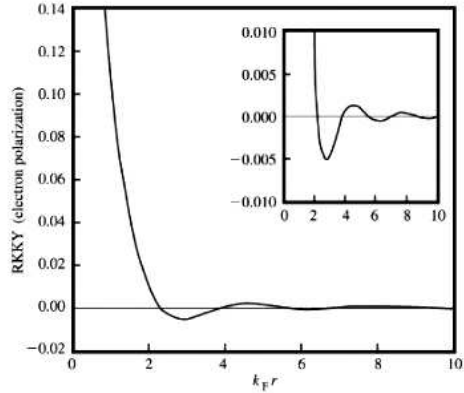}
\caption[The RKKY Coupling]{RKKY coupling between two magnetic
moments separated by a distance r, in three dimensions
\cite{jones:1998}. The figure can also be viewed as the negative
of the electron spin polarization as a function of distance around
a single magnetic moment. Positive value corresponds to
ferromagnetic coupling, the distance is scaled by the Fermi wave
vector and the inset shows the same plot with expanded vertical
scale.} \label{Fig:RKKYcoupling}
\end{figure}

An effect that is sensitive to the spin of the conduction
electrons is the RKKY exchange interaction mediated by the
conduction electrons. Localized magnetic moments are too far to
have a direct exchange interaction but they polarize the mobile
conduction electrons and through their polarization they interact
with neighboring magnetic moments. Because the cutoff of the
momenta near the Fermi surface the polarization of the conduction
electron, in fact a spin screening effect, has an oscillatory
behavior. The exchange coupling can be ferromagnetic or
antiferromagnetic depending on the distance between neighboring
magnetic moments \cite{jones:1998}, see figure
\ref{Fig:RKKYcoupling}.

\begin{figure}[!h]
\centering
\includegraphics[]{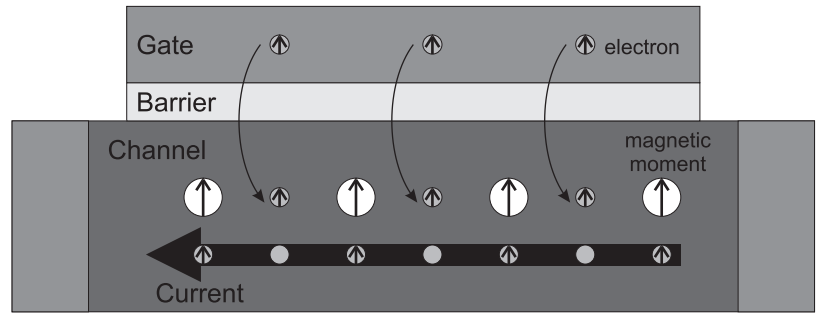}
\caption[A Spin Transistor Using RKKY Coupling]{A proposal for a
spin transistor using the RKKY exchange coupling. polarized
electrons are injected from the gate electrode across the barrier
into the conducting channel. Here the injected electrons change
the coupling between the magnetic moments. That in turn modifies
the spin filter effect of the channel and the whole structure
behaves like a spin-dependent spin filter, an important part of a
spin amplifier. } \label{Fig:RKKYtransistor}
\end{figure}

Imagine now a MOSFET type structure, see \ref{Fig:RKKYtransistor},
where the channel under the gate is an RKKY ferromagnetic material
and we inject spin polarized electrons from the gate across the
tunnel barrier. The exchange coupling constant in the channel
depends on the polarization of the conduction electrons, so if we
inject more we can shift the curve in figure
\ref{Fig:RKKYcoupling}, for instance change the ferromagnetic
coupling into an antiferromagnetic one. The channel will
spin-filter the electrons that go through it when it is
antiferromagnetic. This is now a spin-dependent spin source. There
are weaknesses in this model however. For one thing we do not know
if injecting polarized electrons does indeed affect the RRKY
coupling, also once the coupling is established maybe it will
self-sustain itself and be hard to turn off. A
ferromagnetic-antiferromagnetic phase transition will be sensitive
to temperature and it may be slow.

\begin{figure}[!h]
\centering
\includegraphics[width=0.7\textwidth]{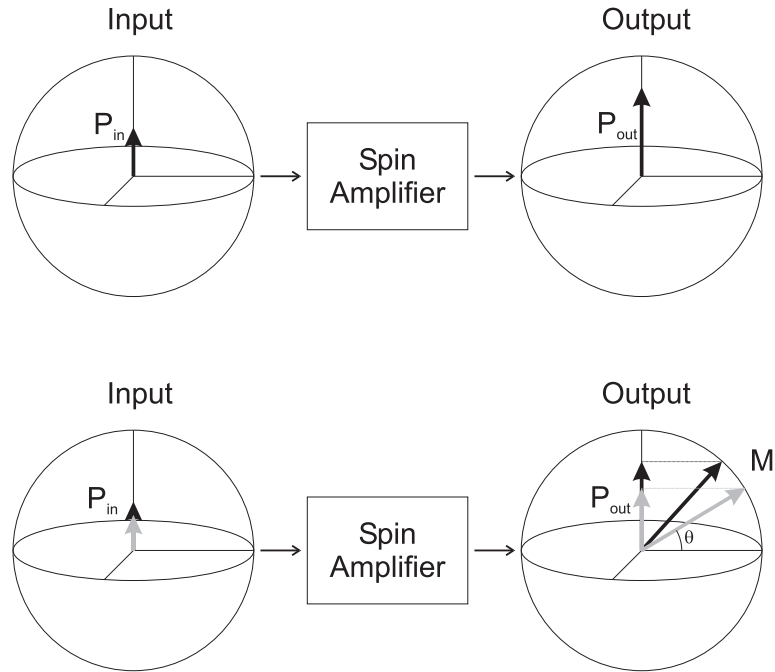}
\caption[Schematic of a Spin Amplifier]{A spin amplifier is
supposed to increase the input spin polarization, shown here on
the Poincar\'{e} sphere. It can do it either by increasing the
length of the output polarization along the z-axis, as in the
upper part of the picture, or it can rotate the unit polarization
on the Poincar\'{e} sphere, thus changing the projection along
z-axis, as in the lower part of the picture. The latter process is
potentially easier to implement as the magnetization (spin filter)
is easier to rotate than to increase.} \label{Fig:SpinAmplifier}
\end{figure}

A different idea for a spin amplifier comes up when one realizes
that what matters is actually the projection of the spin
polarization on some direction. Thus, by rotating a polarizer
(usually the magnetization of a ferromagnetic material) one can
increase or decrease that projection, see figure
\ref{Fig:SpinAmplifier}. The only thing remaining is to make this
rotation dependent on the input spin polarization.

\begin{figure}[!h]
\centering
\includegraphics[height=3in]{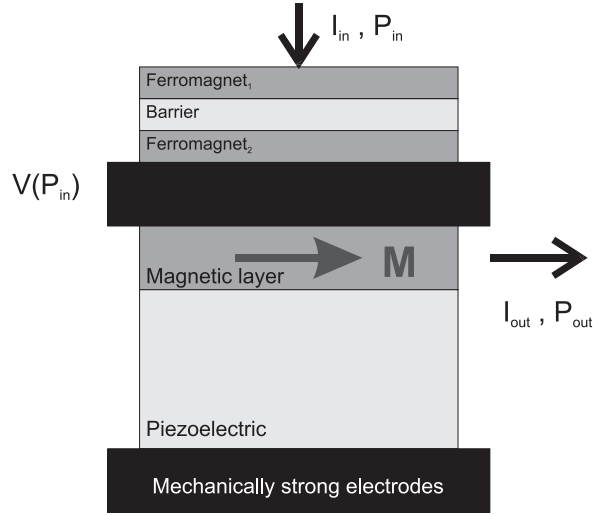}
\caption[Principle of a Magnetoelastic Spin Amplifier]{A spin
amplifier using the magnetoelastic effect and the tunnelling
magnetoresistance (TMR). The TMR sensor converts the input spin
polarization $P_{in}$ into voltage $V$ which in turn, through the
piezoelectric layer, generates a mechanical stress that changes
the orientation of the magnetization. The polarization $P_{out}$
of the output current will depend on the input polarization.}
\label{Fig:MagnetoelasticSpinAmplifier}
\end{figure}

A spin dependent rotation of the magnetization could be
implemented in several ways. We now show an implementation using
the magnetoelastic effect, see figure
\ref{Fig:MagnetoelasticSpinAmplifier}. The orientation of the
magnetization in \ref{Fig:MagnetoelasticMemory} is dependent on
the stress that is given by the applied voltage. We can make the
applied voltage spin dependent either by using a GMR structure or
a TMR (tunnelling magneto-resistance) structure. The TMR has a
larger relative resistance change, it involves electrons
tunnelling from one ferromagnetic layer to another across an
insulating barrier, and is probably more suitable for sensing the
input polarization in this case.

\begin{figure}[!h]
\centering
\includegraphics[]{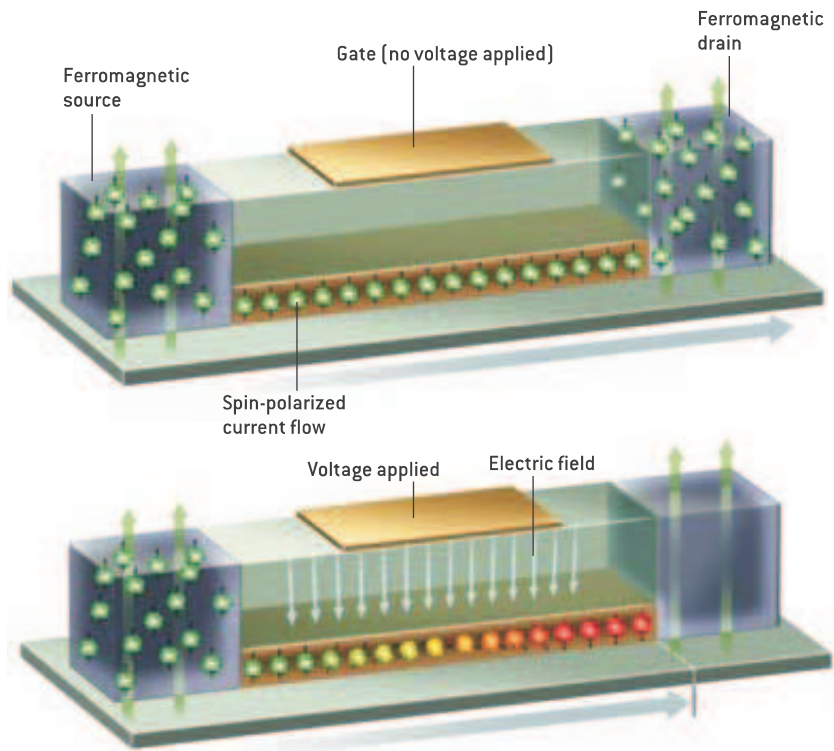}
\caption[Datta Das Transistor]{The principle of Datta Das
transistor. The electric field is transformed into a magnetic
field in the frame of reference of the spin polarized electrons in
the channel. This magnetic field rotates the spin polarization and
one can have a variable resistance between the two ferromagnetic
layers that act as polarizer and analyzer \cite{awschalom:2002}.}
\label{Fig:DattaDas}
\end{figure}

The spin polarized devices can be implemented using semiconductor
materials, in fact the spin decay length is hundreds of microns as
compared to hundreds of nanometers in non-magnetic metals
\cite{awschalom:2002}. However, it is difficult so far to inject
spin polarized electrons in semiconductors from metallic
ferromagnetic layers because of conductivity mismatch and
ferromagnetic semiconductors still have a long way to reach a room
temperature Curie point. Nevertheless a spin transistor has been
proposed, the Datta-Das transistor, see figure \ref{Fig:DattaDas}.
It is basically a MOSFET transistor with source and drain made of
ferromagnetic material, acting as polarizer and analyzer. The gate
creates an electric field in the channel and in some
semiconductors the moving electrons see the electric field as a
magnetic field in their frame of reference. This magnetic field
will make the spin of the electrons precess, this is the Rashba
effect, in essence a spin-orbit coupling similar to the
magnetoelastic coupling. We can make the electric field
spin-dependent by placing a spin sensor on top of the gate, a GMR
or TMR structure as above. This way we have a spin amplifier using
semiconductors.

\bibliographystyle{unsrt}
\bibliography{suthesis}

\end{document}